%% file: ms.tex
\newcommand\ForIEEE

\ifdefined\ForIEEE
    \newcommand\SubmitToComSoc
\fi

\ifdefined\SingleColumnDraft
    \documentclass[12pt,draftclsnofoot,onecolumn]{IEEEtran}
\else
    \ifdefined\ForIEEE
        \ifdefined\SubmitToComSoc
            \documentclass[journal,comsoc,twoside]{IEEEtran}
        \else
            \documentclass[journal,twoside]{IEEEtran}
        \fi
    \else
        \documentclass[journal]{IEEEtran}
    \fi
\fi

\usepackage[T1]{fontenc}
\usepackage{cite}
\usepackage[cmintegrals]{newtxmath}

\ifdefined\SubmitToComSoc
    \usepackage[cmintegrals]{newtxmath} 
\else
    \usepackage{amsmath,amsthm,amsfonts,amssymb,amsbsy,esint,cancel}
\fi

\usepackage{graphicx}
\usepackage{xcolor}
\usepackage{enumerate}

\usepackage{textcomp,gensymb,bbm,bm,dsfont,upgreek}
\usepackage[caption=false,font=footnotesize]{subfig}

\usepackage{url} 
\usepackage[bookmarks=false,hidelinks]{hyperref}

\usepackage[normalem]{ulem} 

\usepackage{tikz}
\tikzset{every picture/.style={font issue=\footnotesize},
	font issue/.style={execute at begin picture={#1\selectfont}}
}
\usepackage{pgfplots}
\pgfplotsset{compat=newest}
\usetikzlibrary{backgrounds,plotmarks}
\usetikzlibrary{positioning,calc} 
\usetikzlibrary{arrows,arrows.meta}
\usepackage{ulem} 

\usepackage{mathtools}

\newtheorem{theorem}{Theorem}

\usepackage[capitalize]{cleveref}
	\Crefname{figure}{Fig.}{Fig.}
	\Crefname{section}{Sec.}{Sec.}
	\Crefname{subsection}{Sec.}{Sec.}
	\Crefname{prop}{Proposition}{Proposition}
	\Crefname{lemma}{Lemma}{Lemma}
	\Crefname{equation}{}{}
	\Crefname{footnote}{Footnote}{Footnote}
	\Crefname{appendix}{Apdx.}{Apdx.}

\input{macros}

\newcommand\OurPaperTitle{%
Load Modulation for Backscatter Communication: Channel Capacity and Near-Capacity Schemes}
\title{\OurPaperTitle}

\author{%
Gregor~Dumphart,~Johannes~Sager,~and~Armin~Wittneben%
\thanks{%
Accepted to appear in the IEEE Transactions on Wireless Communications.
Manuscript received  July 6, 2022; revised January 31, 2023 and July 17, 2023; accepted August 22, 2023.
This article was presented in part at the IEEE Wireless Communications and Networking Conference,
Austin TX, USA, April 2022 \cite{DumphartWCNC2022}.
(\textit{Corresponding author: Gregor Dumphart})
}%
\thanks{The authors are with
ETH Zurich, Z\"urich, 8092 Switzerland, e-mail: gdumphart@gmail.com, sagerj@student.ethz.ch, wittneben@nari.ee.ethz.ch. G. Dumphart is now with u-blox, Thalwil, 8800 Switzerland.}%
\thanks{See the Matlab code at \url{https://github.com/GrDu/BackscatterAtCapacity}.}
\thanks{Digital Object Identifier 10.1109/TWC.2023.3313110}
}

\begin{document}

\maketitle

\begin{abstract}
In backscatter communication (BC), a passive tag transmits information by just affecting an external electromagnetic field through load modulation. Thereby, the feed current of the excited tag antenna is modulated by adapting the passive termination load. 
This paper studies the achievable information rates with a freely adaptable passive load.
As a prerequisite, we unify monostatic, bistatic, and ambient BC with circuit-based system modeling. We present the crucial insight that channel capacity is described by existing results on peak-power-limited quadrature Gaussian channels, because the steady-state tag current phasor lies on a disk. Consequently, we derive the channel capacity for the case of an unmodulated external field, for general passive, purely reactive, or purely resistive tag loads. We find that modulating both resistance and reactance is important for very high rates. We discuss the capacity-achieving load statistics, rate asymptotics, technical conclusions, and rate losses from value-range-constrained loads (which are found to be small for moderate constraints).
\ifdefined\SingleColumnDraft\else

\fi
We then demonstrate that near-capacity rates can be attained by more practical schemes: (i) amplitude-and-phase-shift keying on the reflection coefficient and (ii) simple load circuits of a few switched resistors and capacitors.
\ifdefined\SingleColumnDraft\else

\fi
Finally, we draw conclusions for the ambient BC channel capacity in important special cases.
\end{abstract}

\newcommand\OurKeywords{%
backscatter communication, load modulation, channel capacity, ambient backscatter, RFID} 


\newcommand\CasesTable{
\begin{table*}
\caption{Cases considered in the capacity analysis.}
\begin{center}
\begin{tabular}{c||c|l|c|c|c|c}
assumption on   & equiv.\,constraint & \multicolumn{1}{c|}{equivalent}   & \multicolumn{3}{c|}{channel capacity} & high-SNR\,($\,\SNR\!\gg\!1$) \\
modulated load  & on refl.\,coeff. & \multicolumn{1}{c|}{AWGN channel} & ref. & ref. here & symbol        & asymptote \\\hline
general passive & $|\r\Gamma| \leq 1$ & $\bbC$, limited peak-power&\cite{ShamaiTIT1995} & \Cref{sec:CapacityGeneral}&$\RateMaxGP(\SNR)$ & $\log_2(1+\SNR/e)$ \\
\!uniform distribution\! & \!$\r\Gamma \!\sim\! \calU$(unit\,disk)\! & $\bbC$, specific signaling &\cite{ShamaiTIT1995} & \Cref{sec:MaxEntropyRate}&$\RateME(\SNR)$ & $\log_2(1+\SNR/e)$ \\
purely reactive & $|\r\Gamma| = 1$    & $\bbC$, constant power    &\cite{WynerBSTJ1966} & \Cref{sec:CapacityReact}  &$\RateReact(\SNR)$ & $\f{1}{2}\log_2(1+4\pi\SNR/e)$ \\
purely resistive& $\r\Gamma \in [-1,1]$ & $\bbR$, limited peak-power&\cite{Smith1971}     & \Cref{sec:CapacityResist} &$\RateResist(\SNR)$& $\f{1}{2}\log_2(1+4\SNR/(\pi e))$
\end{tabular}
\end{center}
\label{tab:cases}
\end{table*}
}

\section{Introduction}
\label{sec:intro}
\input{01-Intro}

\section{System Modeling}
\label{sec:model}
\input{02-Model}

\section{Achievable Information Rates}
\label{sec:capacity}
\input{03-Capacity}

\subsection{Effect of Value-Range Constraints on the Load}
\label{sec:ValueRange}
\input{03f-ValueRange}

\section{Finite-Constellation Schemes Near Capacity}

The schemes in \Cref{sec:CapacityGeneral,sec:MaxEntropyRate,sec:CapacityReact} use continuous transmit distributions, which are of academic nature and an implementation nightmare. Most every practical scheme instead uses a finite constellation (a.k.a. symbol alphabet). It is thus worthwhile to explore if these more practical schemes can still attain near-capacity rates. This is the topic of this section.

Formally, a finite constellation means choosing $\r\Gamma$ from $\{ \Gamma_1 , \ldots , \Gamma_{M} \} \subset \bbC$ with certain probabilities $\SymbProb_m$.
The Euclidean symbol distance is capped by the unit-disk diameter, $|\Gamma_m - \Gamma_n| \leq 2$.
The information rate $I(\r{y};\r\Gamma)$ will obviously be below $\RateMaxGP$. However, for a finite constellation that resembles the capacity-achieving distribution, $I(\r{y};\r\Gamma)$ can approach channel capacity. This has been argued rigorously in \cite{HuleihelISIT2018,MericCL2015,WuACCC2010}.

We note that, here, the information rate is capped by the finite source entropy
$H(\r\Gamma) = -\sum_{m=1}^M \SymbProb_m \log_2(\SymbProb_m)$, i.e.
$I(\r y;\r\Gamma) \leq H(\r \Gamma) \leq \log_2(M)$ \cite{Cover2006}.
A good design should assert that $H(\r \Gamma) > \RateMaxGP(\SNR)$ for the target SNR range.

\subsection{Approaching Capacity with Finite APSK Constellations}
\label{sec:APSK}
\input{04a-APSK}

\subsection{High Rates from Simple Switched Loads}
\label{sec:Switch}
\input{04b-Switch}

\section{Implications for Ambient BC Capacity}
\label{sec:Ambient}
\input{05-Ambient}

\section{Summary \& Outlook}
\label{sec:summary}
\input{99-Summary}

\begin{appendices}
\crefalias{section}{appendix}
\crefalias{subsection}{appendix}
\crefalias{subsubsection}{appendix}

\section{Impedance Statistics}
\label{apdx:zStats}
\input{AP-Impedance}

\section{Rate Calculation Details}
\input{AP-RateCalc}

\section{Maximum Entropy of Complex Variables Constrained to Finite Area}
\label{apdx:MaxEntropy}
\input{AP-SubareaEntropy}

\section*{Acknowledgement}
We would like to thank Henry Schulten, Robert Heyn, Amos Lapidoth, Christoph Mecklenbr\"auker, and the reviewers for valuable suggestions.

\end{appendices}

\bibliographystyle{IEEEtran}
\bibliography{ref}

\pdfinfo{
/Author(Gregor Dumphart, Johannes Sager, Armin Wittneben)
/Title(\OurPaperTitle{})
/Subject(\OurKeywords{})
/Keywords(\OurKeywords{})
}

\end{document}

%% file: macros.tex
  \newcommand\f[2]{\frac{#1}{#2}} 
  \newcommand\fp[2]{\f{\partial #1}{\partial #2}} 
  \newcommand\re{\mathrm{Re}} 
  \newcommand\im{\mathrm{Im}} 
\renewcommand\Re{\re} 
\renewcommand\Im{\im} 
  \newcommand\reSub{_\mathrm{R}} 
  \newcommand\imSub{_\mathrm{I}} 

\renewcommand\H{^\mathrm{\normalfont{H}}}        

\newcommand\bbN{\mathbb{N}} 
\newcommand\bbZ{\mathbb{Z}} 
\newcommand\bbR{\mathbb{R}} 
\newcommand\bbC{\mathbb{C}} 

\newcommand\SymbolOfEV{\mathbb{E}} 
\newcommand\EVVar[2]{\SymbolOfEV_{#1}[\hspace{.2mm}#2\hspace{.2mm}]}
\newcommand\EV[1]{\EVVar{}{#1}}
\newcommand\EVVarLR[2]{\SymbolOfEV_{#1}\left[\hspace{.2mm}#2\hspace{.2mm}\right]}

\newcommand\iid{\overset{\text{i.i.d.}}{\sim}}

\newcommand\calU{\mathcal{U}}   
\newcommand\calN{\mathcal{N}}   
\newcommand\calCN{\mathcal{CN}} 
  
\newcommand\unit[1]{\,\mathrm{#1}}	
\newcommand\dB{\unit{dB}}

\newcommand\SymbolOfTx{\hspace{.2mm}\mathrm{T}}
\newcommand\SymbolOfRx{\hspace{.2mm}\mathrm{R}}
\newcommand\Txs[1]{_{\SymbolOfTx#1}} 
\newcommand\Rxs[1]{_{\SymbolOfRx#1}} 
\newcommand\RxTxs[1]{_{\hspace{.2mm}\mathrm{RT}#1}}
\newcommand\Tx{\Txs{}} 
\newcommand\Rx{\Rxs{}} 
\newcommand\RxTx{\RxTxs{}} 
\newcommand\SNR{\rho}

\newcommand\iTx{i\Tx}
\newcommand\Radius{a}
\newcommand\PM{_0}

\newcommand\RateReact{R_\mathrm{psk}}
\newcommand\RateResist{R_\mathrm{res}}
\newcommand\RateME{R_\mathrm{ud}}
\newcommand\RateMaxGP{R_\mathrm{max}}
\newcommand\RateMaxAmbient{\RateMaxGP^\mathrm{AmBC}}

\newcommand\CircProb{q}
\newcommand\SymbProb{p}

\renewcommand\L{}                
\newcommand\Ind{^\mathrm{ind}}   
\newcommand\N{_\mathrm{N}}
  
\renewcommand\r[1]{{\bm{#1}}}

\newcommand\myFigWidth{.9\textwidth}
\newcommand\myFigHeight{4cm}

\newcommand*{\QEDB}{\null\nobreak\hfill\ensuremath{\square}}%

%% file: 01-Intro.tex
Backscatter communication (BC) allows a simple passive tag to transmit data with essentially zero transmit power and no transmit amplifier.
Instead, the tag scatters a preexisting external electromagnetic field in an adaptive fashion, in order to modulate data onto this field.
The external field can stem from a dedicated unmodulated source or a modulated ambient source \cite{LiuSmithSIGCOMM2013,HuynhCST2018,WangTC2016}.
The receiver may decode the data after detecting the changes in the field.
BC has found widespread use in smart cards and radio-frequency identification (RFID) \cite{ChawlaCM2007,Finkenzeller2010} and is a promising approach to ultra-low-energy communication in wireless sensor networks and the Internet of Things (IoT) \cite{DuanCM2020,WangTC2016,HuynhCST2018,ZhaoNAT2020}.
The different basic setups are illustrated in \Cref{fig:Concepts}.
Promising BC systems that utilize ambient commodity signals (e.g., Wi-Fi) were presented in \cite{WangMercierISSCC2020,ZhaoNAT2020,YangTC2017,KimSPAWC2017}.
Other exciting BC applications are given in \cite{PengACM2018,Hessar2019,Brooker2013}.

In BC, the tag antenna is excited by an external field and terminated by a passive load.
Transmitting information requires to modulate the antenna feed current, for which load modulation is the only available option under the passive low-power paradigm of BC (with no transmit amplifier) \cite{LiuSmithSIGCOMM2013,ChawlaCM2007,Finkenzeller2010}.
Therefore, BC tags use load modulation to transmit data, at a chosen symbol rate.
For example, switching between a resonant load and an open circuit results in binary amplitude-shift keying (2-ASK) for the current.
Such modulation has been used, e.g., in \cite{ZhaoNAT2020}. %
\ifdefined\SingleColumnDraft
\renewcommand\myFigWidth{50mm}
\begin{figure}[!ht]
\centering
\subfloat[bistatic BC or ambient BC]{\label{fig:ConceptBistatic}
\includegraphics[width=\myFigWidth,trim=0 0 0 5]{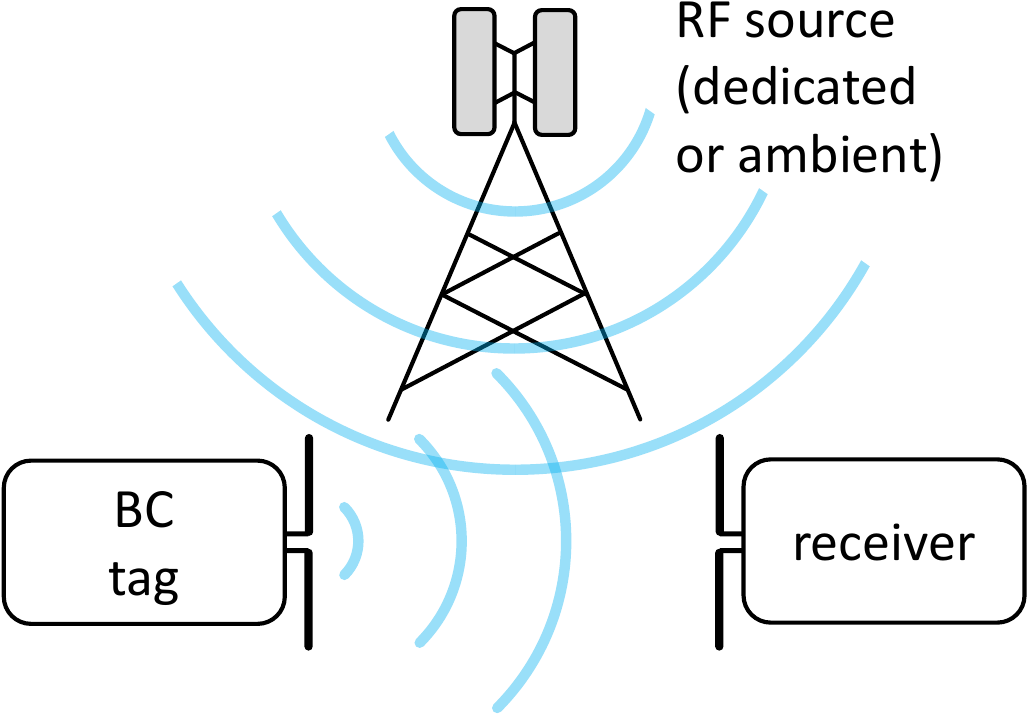}}
\ \ \ \ \ \
\subfloat[monostatic BC (e.g., RFID)]{\label{fig:ConceptMonostatic}
\includegraphics[width=\myFigWidth,trim=0 -2 0 5]{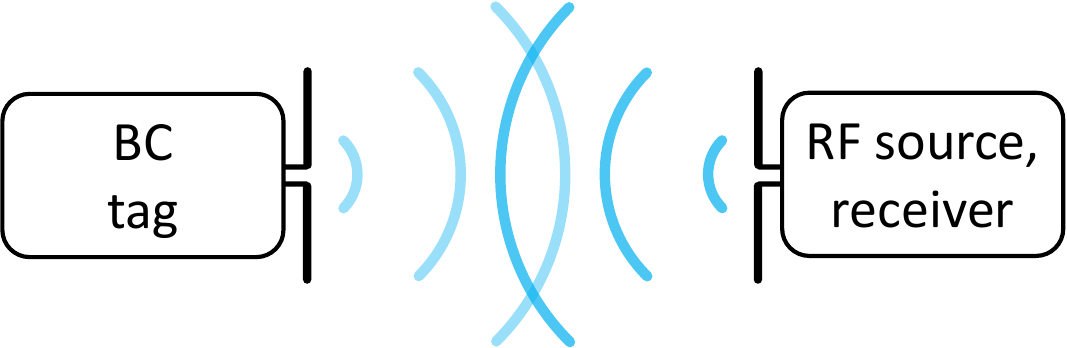}}
\else
\renewcommand\myFigWidth{42mm}
\begin{figure}[!ht]
\centering
\subfloat[bistatic BC or ambient BC]{\label{fig:ConceptBistatic}
\includegraphics[width=\myFigWidth,trim=0 0 0 5]{F01_ConceptBistatic.pdf}}
\
\subfloat[monostatic BC (e.g., RFID)]{\label{fig:ConceptMonostatic}
\includegraphics[width=\myFigWidth,trim=0 -2 0 5]{F01_ConceptMonostatic.pdf}}
\fi
\caption{Different basic setups in backscatter communication. Analogous illustrations were given in \cite{DarsenaTC2017,HuynhCST2018} among others.}
\label{fig:Concepts}
\end{figure}
The data-rate requirements of many IoT applications have however incentivized modulation schemes beyond binary \cite[Tab.~III]{HuynhCST2018} such as quadrature phase-shift keying (4-PSK) in \cite{ZhaoNAT2020,WangMercierISSCC2020} and 16-ary quadrature amplitude modulation (16-QAM) in \cite{KimionisNAT2021}.
It is unclear how well-suited these constellations are to the physical constraints of passive tags. Furthermore, the idea to increase the information rate by ever-increasing constellation sizes will inevitably stagnate: for a very large constellation, the receiver can't reliably distinguish neighboring symbols in the noise, which must be countered with increased redundancy in the employed error-correcting code to ensure reliable decoding.
This trade-off caps the information rate and is described in the most general form by Shannon's channel coding theorem and the notion of channel capacity \cite[Cpt.~7]{Cover2006}. Specific results on the information-theoretic limits of BC are however scarce. Even BC systems with unmodulated field sources, which are much simpler to analyze than AmBC systems, have received little academic attention so far.

\IEEEpubidadjcol

\subsubsection*{Related Work}
An important resource on load modulation is the book by Finkenzeller \cite{Finkenzeller2010}.
A few works address signal detection in terms of error probabilities and optimal decision thresholds in BC \cite{WangTC2016,ZhaoACCESS2018,LiuTWC2020deep} and specifically RFID \cite{FuschiniAPL2008,BarberoTC2014}.
The focus of \cite{FuschiniAPL2008} is on the effect of the propagation environment on the Euclidean symbol distances and the resulting bit error rate with PSK and ASK.
Zhao et al. \cite{ZhaoACCESS2018} calculated the channel capacity of binary load modulation with thresholding in AmBC for various cases of the ambient signal modulation.
Kim and Kim \cite{KimSPAWC2017} studied the maximization of AmBC network capacity in terms of redundancy and modulation (for 2-PSK, 4-PSK, and 16-QAM alphabets) with orthogonal frequency-division multiplexing (OFDM) in WiFi.
Hoang et al. \cite{HoangTC2017} contrasted AmBC with the harvest-then-transmit approach.

Rich work on BC information theory has been published by Darsena et al. \cite{DarsenaTC2017}.
They noted the important role of the load reflection coefficient $\Gamma$ and that $|\Gamma| \leq 1$ implies an instantaneous amplitude constraint. This was invoked in the capacity analysis, albeit based on the results of Smith \cite{Smith1971} which apply only to real-valued, but not to the complex-valued transmit variable $\Gamma \in \bbC$, as clarified in \cite{ShamaiTIT1995}.

Duan et al. \cite{DuanTVT2017} studied the achievable rate of AmBC with a multiple-input multiple-output (MIMO) legacy system and a multi-antenna tag that uses polyphase-coding modulation 
and leaves the amplitude unmodulated. Their comparison to Gaussian signaling hints that modulating also the amplitude promises significant rate gains at high SNR.

\subsubsection*{Identified Shortcomings}
Most existing studies are restricted to specific finite modulation alphabets or constant amplitude by assumption. However, there does not exist any formally correct description of the channel capacity of a given BC link with a general adaptive passive load. The related question of the capacity-achieving load modulation scheme thus remains unresolved. The answer promises valuable design criteria for high-data-rate BC systems and is also a stepping stone towards optimal ambient backscatter communication.

\subsubsection*{Contribution}
This paper presents the channel capacity solution of backscatter load modulation in the general case of a freely adaptable passive load (i.e. no specific constellation is assumed) for monostatic and bistatic links with a single tag, an unmodulated field source, and no antenna arrays.
More specifically, the presented core contributions are:
\begin{itemize}
\item
We develop a circuit-based system model that unifies load-modulation links for monostatic BC (MoBC), bistatic BC (BiBC), and ambient BC (AmBC).
\item
Based on an analysis of the tag-side physical constraints, we apply existing information theory on peak-power-limited quadrature Gaussian channels to solve the channel capacity problem for a general passive load. 
\item
We state the channel capacity also for the special cases where the load is purely reactive or purely resistive. 
\item
We characterize the capacity-achieving load distribution in terms of reflection coefficient and impedance.
\item
The rate results are discussed 
by means of bounds and asymptotes. We find that restricting the load to be purely reactive or resistive causes a capacity pre-log factor of $\f{1}{2}$.
\end{itemize}
Furthermore, we present the following complementary contributions of high technical relevance:
\begin{itemize}
\item
We characterize how the rate decreases due to an adaptive load with limited dynamic range. 
\item
We present an amplitude and phase-shift keying (APSK) constellation design 
that allows to approach capacity.
\item
We show that high rates are possible even with simple load circuits of a few switched resistors and capacitors.
\item
We show that the results describe the channel capacity of AmBC in important special cases.
\end{itemize}
This paper does not consider aspects of tag power consumption, energy harvesting, channel models, or multiple access.

\subsubsection*{Paper Structure}
\Cref{sec:model} develops the employed system model and analyzes the tag-side constraints. \Cref{sec:capacity} presents the channel capacity result, the associated distribution, a discussion guided by reference cases and asymptotes, and a result on the effect of load value-range constraints. \Cref{sec:Ambient} explains implications for the channel capacity of AmBC.
\Cref{sec:APSK} addresses the design of a near-capacity APSK scheme and suitable constellations and \Cref{sec:Switch} studies near-capacity BC with switched load circuits. \Cref{sec:summary} concludes the paper.

\subsubsection*{Notation}
For a random variable $\bm{x}$ (written boldface italic), we denote a realization as $x$ (lightface italic),
the probability density function (PDF) as $f_\r{x}(x)$,
and the expected value as $\EV{\r x}$.
We use the Rice distribution (or Rician)
$\r b \sim \mathrm{Rice}(a,s)$
with $f_\r{b}(b) = \f{b}{s^2} \exp(-\f{b^2 + a^2}{2 s^2}) I_0(\f{ab}{s^2})$, $b \geq 0$,
\cite[Eq.~(3.35)]{Goldsmith2005} and the modified Bessel function of the first kind $I_0(x)$.
We use the imaginary unit $j$ and Euler number $e$. 
Random vectors occur briefly in \Cref{sec:Ambient} and are denoted like $\vec{\r x}$. Matrices do not occur.
All information quantities are in unit bit unless explicitly stated otherwise.
We refer to modulation schemes in the style 2-PSK, 4-PSK (not BPSK, QPSK).

%% file: 02-Model.tex
For conventional (active) transmitters it is straightforward to characterize the realizable signals. This is less intuitive for BC tags, due to their passive nature. A circuit analysis can give the answer and has been conducted, e.g., in \cite{Finkenzeller2010,DarsenaTC2017,HuynhCST2018}. However, due to different assumptions and formalism, their models are not readily usable for the general capacity analysis this paper aims for. 
Therefore, in this section we develop from scratch a system model for tag-to-receiver links, based on the circuit model in \Cref{fig:Circuits}. It encapsulates both setups from \Cref{fig:ConceptBistatic,fig:ConceptMonostatic}.
\ifdefined\SingleColumnDraft
\renewcommand\myFigWidth{100mm}
\begin{figure}[b]
\else
\renewcommand\myFigWidth{82mm}
\begin{figure}[!ht]
\fi%
\vspace{4mm}
\centering
\resizebox{\myFigWidth}{!}{\hspace{2mm}\input{F02_CircuitCombo.tex}}
\caption{Unified circuit description of BC links: passive load-modulation tag (left) transmits to coupled information receiver (right). The voltage measurement $v[n]$ is the starting point for receive processing.
In monostatic BC (red), the receiver is also the power source (via source current $i\Rx$).}
\label{fig:Circuits}
\end{figure}
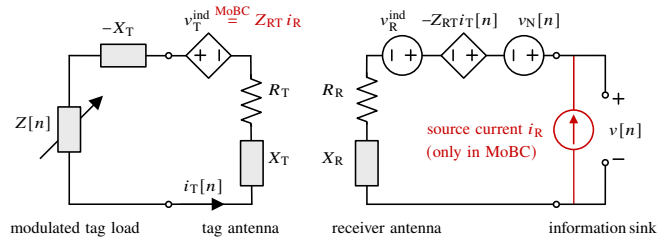
The left-hand circuit models the passive tag. Its adaptive load impedance $Z\L[n] \in \bbC$ must fulfill $\Re(Z\L[n]) \geq 0\ \forall n$ because the load is passive \cite[Sec.~4.1]{Pozar2004}. Thereby $n \in \bbZ$ is the symbol time index. The given tag-side induced voltage, represented by the phasor $v\Ind\Tx \in \bbC$, is the crucial cause for any electrical activity at the tag. The tag current phasor $i\Tx[n] \in \bbC$ is altered by $Z\L[n]$. The tag-antenna impedance $R\Tx + jX\Tx$ has its reactance $X\Tx$ purposefully canceled by the serial reactance $-X\Tx$ in order to establish a near-resonant state. The right-hand circuit is an information receiver that measures a voltage phasor $v[n] \in \bbC$. The measurement is impaired by the noise voltage $v\N[n] \in \bbC$ and the interfering induced voltage $v\Ind\Rx \in \bbC$. The tag and receiver antennas are coupled with mutual impedance $Z\RxTx \in \bbC$, which encapsulates all aspects of the propagation channel.

For the ambient and bistatic cases, the assumption is that the voltages $v\Ind\Tx , v\Ind\Rx \in \bbC$
are induced by an electromagnetic field from some external radio-frequency (RF) source (not shown in \Cref{fig:Circuits}).
These cases differ in one aspect:
AmBC assumes randomly modulated $\r{v}\Ind\Tx , \r{v}\Ind\Rx$ while BiBC assumes constant $v\Ind\Tx , v\Ind\Rx$ because of a dedicated unmodulated RF source \cite{HuynhCST2018}.

Monostatic BC does not assume any external source. Instead, the information receiver is also the system's power source (e.g., an RFID reader) and the crucial tag-side induced voltage is $v\Ind\Tx = Z\RxTx\, i\Rx$ as a consequence of the source current $i\Rx$. A prominent example of MoBC is inductive RFID, where $X\Tx, X\Rx, Z\RxTx$ are determined by inductances and where $-X\Tx$ is realized by a resonance capacitor \cite{Finkenzeller2010}.

For the tag antenna we assume the minimum-scattering property \cite{KahnTAP1965} which asserts that the electromagnetic field is left unaltered by an open-circuited tag ($Z = \infty$).
We assume that $Z\L[n]$ is piecewise constant over time and that it changes instantaneously at the symbol switching instants. We neglect any signal transients which result for $i\Tx$ and $v$. This is a meaningful assumption if the symbol duration is significantly larger than the circuits' time constants. Our previous work \cite[Appendix~E]{Dumphart2020} showed that transients do not cause significant deterioration for the receive processing of load-modulated signals (when anticipated, they can even improve the SNR).

The electrical state of the passive load is described by well-known complex quantities:
the load impedance (unit ohm)
\begin{align}
Z\L
= R\Tx\cdot z
= R\Tx\, \f{1 + \Gamma}{1 - \Gamma} \, , \ \ &&
\Re(Z\L) \geq 0
\label{eq:ZL} 
\end{align}
or equivalently by the load reflection coefficient
\begin{align}
\Gamma
= \f{Z\L - R\Tx}{Z\L + R\Tx}
= \f{z-1}{z+1} \, , &&
|\Gamma| \leq 1
\label{eq:ReflCoeff}
\end{align}
or the normalized load impedance
$z
= \f{Z\L}{R\Tx}
= \f{1 + \Gamma}{1 - \Gamma}$,
$\Re(z) \geq 0$.
The tag-antenna resistance $R\Tx$ serves as reference impedance (not the usual $50\,\Omega$).
The M\"obius transformation $\Gamma = \f{z-1}{z+1}$ in \Cref{eq:ReflCoeff} is a bijective map from the right half-plane $\Re(z) \geq 0$ to the unit disk $|\Gamma| \leq 1$. It underlies the well-known Smith chart \cite[Eq.~(2.53)]{Pozar2004}.
The load state results in a tag current
\begin{align}
\iTx[n]
= \f{v\Ind\Tx}{R\Tx + Z\L[n]}
= \f{2\cdot i\PM}{1 + z[n]}
= \big( 1 - \Gamma[n] \big) \, i\PM
\label{eq:Current}
\end{align}
where $i\PM := \iTx|_{z=1} = \f{v\Ind\Tx}{2 \, R\Tx}$ is shorthand for the current that would result from a matched load.
Now $\iTx[n] = ( 1 - \Gamma[n])\,i\PM$ together with $|\Gamma[n]| \leq 1$ characterizes the set of transmit currents that can be realized by a passive tag. Specifically, the set of realizable currents $\iTx \in \bbC$ is given by a disk
$|i\Tx[n] - i\PM| \leq |i\PM|$
with center $i\PM \in \bbC$ and radius $|i\PM|$.
Analogous observations have been made in \cite[Sec.~4.1]{Finkenzeller2010} and \cite{DarsenaTC2017}.

The achievable information rate of the link is determined by the relationship between $i\Tx[n]$ and $v[n]$. It would however be confusing to apply information theory to those quantities directly, because of the intricate constraint on $i\Tx[n]$ and the case-specific receiver differences. For this reason, we introduce two signal transformations that simplify the mathematics while leaving the mutual information unaltered.
As a preceding step, we express the receive voltage as
$v[n] = -Z\RxTx\,\iTx[n] + v\N[n] + \tilde{v}[n]$.
The term $\tilde{v}[n]$ is a case-dependent interference voltage, given
by $(R\Rx + jX\Rx) i\Rx$ in MoBC,
by $v\Ind\Rx$ in BiBC, and
by $v\Ind\Rx[n]$ in AmBC.
To unify these cases within the same system model, we use a specific receive-side signal compensation
\begin{align}
v_\mathrm{diff}[n] :=& \, v[n] - v|_{Z\RxTx\iTx \, = \, 0, \, v\N \, = \, 0}
\label{eq:TransformOne} \\
=& \, -Z\RxTx\,\iTx[n] + v\N[n]
\, . \label{eq:SignalModelVoltage}
\end{align}
The processing rule \Cref{eq:TransformOne} will 
require a high-resolution receiver that has been precisely calibrated while the tag was either absent or open-circuited (such that $Z\RxTx\iTx = 0$). MoBC faces the challenge of canceling the strong self-interference voltage at the receiver \cite[Sec.~6.2.4]{Finkenzeller2010},\cite{DuanCM2020}.
In BiBC, the direct-path interference $v\Ind\Rx$ is an a-priori unknown constant and can only be canceled after precise estimation. In AmBC, the unknown modulated $v\Ind\Rx[n]$ must be decoded in order to cancel it, requiring accurate channel estimation and high resolution.
We note that the rate of change of ambient signals can be vastly different from the backscatter symbol rate.

To declutter the mathematical model, we introduce one last receive signal transformation
\begin{align}
y[n] := 1 + \f{v_\mathrm{diff}[n]}{Z\RxTx\,i\PM}
\, . \label{eq:TransformTwo}
\end{align}
This signal has the structure
$y[n] = \Gamma[n] + v\N[n] \, / \, (Z\RxTx i\PM)$,
i.e. reflection coefficient plus unitless noise,
because of \Cref{eq:SignalModelVoltage,eq:Current}.
The transformation can be realized with a pilot sequence, phase synchronization, and automatic gain control. The signal $y[n] \in \bbC$ is henceforth considered as receive signal. We only consider BC systems whose receiver can actually realize the suggested processing steps. Related implementation details are relayed to the existing literature \cite{DuanCM2020}.

\ifdefined\SingleColumnDraft
\else
\CasesTable
\fi

For the noise voltage $v\N[n] \in \bbC$ we assume a circularly-symmetric complex Gaussian distribution
$v\N[n] \sim \calCN(0,\sigma^2)$
with variance $\sigma^2$ and statistically independent and identically distributed (i.i.d.) samples for different $n$. This is a well-established model for thermal receiver noise \cite[Sec.~2.2.4]{Tse2005}.

We have now established a simple complex-valued, discrete-time signal and noise model:
\begin{align}
y[n] &= \Gamma[n] + w[n] \, , &
w[n] &\iid \mathcal{CN}\left(0, 1 / \SNR \right)
\label{eq:SignalModel}
\end{align}
with $|\Gamma[n]| \leq 1\ \forall n$.
The value $\SNR$ defines the signal-to-noise ratio (SNR) in accordance with \cite[$s^2$ in Tab.~II]{Blachman1953},
\begin{align}
\SNR := \f{|Z\RxTx\,i\PM|^2}{\sigma^2}
= \f{|Z\RxTx|^2 \, |v\Ind\Tx|^2}{4\, R\Tx^2 \,\sigma^2}
\, . \label{eq:SNR}
\end{align}
We will often refer to the SNR in terms of its decibel (dB) value
$10 \cdot \log_{10}(\SNR)$.

%% file: F02_CircuitCombo.tex
\begin{tikzpicture}
\node (0,0) {\includegraphics[width=100mm]{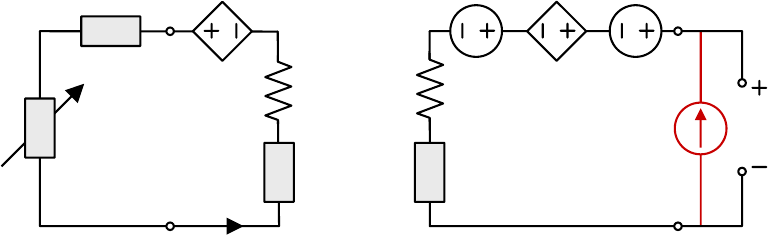}};
\put (-154, -3) {$Z\L[n]$}
\put (-112, 43) {$-X\Tx$}
\put (-72, 47){$v\Ind\Tx \color[rgb]{.784,0,0}\overset{\mathrm{MoBC}}{=} Z\RxTx\,i\Rx$}
\put (-72, -33.5) {$\iTx[n]$}
\put (-31, 13) {$R\Tx$}
\put (-5, 13) {$R\Rx$}
\put (-31, -20) {$X\Tx$}
\put (-5, -20) {$X\Rx$}
\put (23, 47) {$v\Ind\Rx$}
\put (43, 47) {$-\!Z\RxTx\iTx[n]$}
\put (90, 47) {$v\N[n]$} 
\put (134, -6.5) {$v[n]$}
\put (-156,-54) {$\mathrm{modulated\,\,tag\,\,load}$}
\put (-63,-54) {$\mathrm{tag\,\,antenna}$}
\put (0,-54) {$\mathrm{receiver\,\,antenna}$}
\put (105,-54) {$\mathrm{information\,sink}$}

\ifdefined\SingleColumnDraft
\put (38,-6) {\color[rgb]{.784,0,0}$\mathrm{source\,current}\,i\Rx$}
\put (37,-17) {\color[rgb]{.784,0,0}$\mathrm{(only\,in\,MoBC)}$}
\else 
\put (46,-6) {\color[rgb]{.784,0,0}$\mathrm{source\,\,current}\,\,i\Rx$}
\put (45,-17) {\color[rgb]{.784,0,0}$\mathrm{(only\,\,in\,\,MoBC)}$}
\fi
\end{tikzpicture}

%% file: 03-Capacity.tex
In accordance with typical conventions in information theory, we discard the time indexation $[n]$ and write the signal and noise model \Cref{eq:SignalModel} in terms of random variables,
\begin{align}
\r y &= \r\Gamma + \r w \, , &
|\r\Gamma| &\leq 1 \, , &
\r w &\sim \mathcal{CN}\left(0, 1 / \SNR \right) \, .
\label{eq:SignalModelRV}
\end{align}
Throughout the paper, we assume $\r\Gamma$ and $\r w$ to be statistically independent and independently sampled for different time steps $n$. We do not consider any symbol decision regions or other forms of thresholding or quantization. We assume that the SNR $\SNR$ from \Cref{eq:SNR} is constant, which requires that both $v\Ind\Tx$ and $Z\RxTx$ are constant for the duration of a load-modulated coding block. This is fulfilled in MoBC and BiBC with time-invariant or slow-fading propagation channels. It is \emph{not} fulfilled in almost every AmBC use case, where usually $v\Ind\Tx$ is fluctuating rapidly due to the ambient source modulation.

We are interested in the achievable information rates over the channel \Cref{eq:SignalModelRV} at a given SNR $\SNR$.
The mutual information $I(\r y;\r\Gamma)$ specifies an achievable rate \cite[Cpt.~7]{Cover2006} which is measured in bit per channel use (bpcu). From an engineering perspective, reliable communication over the channel is possible at any information rate below $I(\r y;\r\Gamma)$. A suitable error-correcting code with a very large block length allows to approach $I(\r y;\r\Gamma)$ with arbitrarily small block error rate \cite{Cover2006}.

An important aspect is that $I(\r y;\r\Gamma)$ depends on the distribution $\r\Gamma$, i.e. on the specific transmit signaling. The channel capacity is defined as the supremum of $I(\r y;\r\Gamma)$ over all eligible distributions $\r\Gamma$ (those that adhere to the technical constraints).
The channel capacity is the upper limit of achievable information rates for reliable communication \cite{Cover2006,Tse2005}.
In this paper we study the channel capacity for the different technical assumptions on the adaptive passive load.
Those are listed in \Cref{tab:cases} together with important information that provides a guideline through this section.

\ifdefined\SingleColumnDraft
\CasesTable
\fi

\subsection{Channel Capacity, General Passive Load}
\label{sec:CapacityGeneral}

The following shall provide a concise information-theoretic description of the channel capacity $\RateMaxGP$ and its computation. We note that the mutual information was not affected by the receive-side transformations \Cref{eq:TransformOne,eq:TransformTwo}
because it is invariant under smooth, uniquely invertible maps \cite[Eq.~(45)]{Kraskov2004}.
As in \cite[Eq.~(5)]{ShamaiTIT1995}, our exposition uses polar coordinates
\begin{align}
\r\Gamma &= \r a \hspace{.2mm} e^{j\r\theta} \, , &
\r y &= \r b \hspace{.2mm} e^{j\r\phi} \, .
\label{eq:PolarDef}
\end{align}
We use the probability density function (PDF) $f_\r\Gamma(\Gamma)$ of $\r\Gamma$ with argument $\Gamma \in \bbC$. The PDF is necessarily zero for $\Gamma$ outside the unit disk. Likewise, the amplitude PDFs $f_\r{a}(a)$ and $f_\r{b}(b)$ are zero unless $a \in [0,1]$ and $b \in \bbR_{\geq 0\,}$, respectively.

Consider the physical amplitude constraint $|\r\Gamma| \leq 1$ on the complex-valued reflection coefficient. Mathematically (but not physically), $|\r\Gamma|^2 \leq 1$ qualifies as a constraint on the instantaneous transmit power. A crucial consequence is that, therefore, the capacity problem of the channel in \Cref{eq:SignalModelRV} is equivalent to that of a peak-power-limited quadrature Gaussian channel. The latter has been solved by Shamai \& Bar-David \cite{ShamaiTIT1995}. In the following, we employ their results to solve the BC channel capacity problem.

Foremost, we infer from \cite{ShamaiTIT1995,ChanTIT2005} that the capacity-achieving distribution of $\r\Gamma = \r a \hspace{.2mm} e^{j\r\theta}$ has discrete amplitude $\r a$ and uniform independent phase $\r\theta$ (DAUIP). In other words, $f_\r\Gamma(\Gamma)$ is supported on a finite union of concentric circles. We note that $\r\Gamma$ has an infinite number of mass points, which is in contrary to a statement in the BC literature \cite[Sec.~IV-A]{DarsenaTC2017}.
In the following, we formalize the capacity-achieving distribution 
by concisely collecting statements from \cite{ShamaiTIT1995} in our notation:
\begin{enumerate}
\item
The radius $\r\Radius$ is chosen from a finite discrete set of radii (i.e. amplitudes).
The set always contains the unit circle $a_1 = 1$. Formally,
$\r\Radius \in \{\Radius_1 = 1 , \, \Radius_2 \, , \, \ldots , \, \Radius_K \}$, 
$\Radius_k \in [0,1] \ \forall k$,
where $K$ is the number of circles.
We define that the radii $a_k$ are in descending order.
\item
The radii are chosen with non-uniform probabilities $\CircProb_k$.
\item
The phase angle has uniform distribution
$\r\theta \sim \calU(0,2\pi)$
for any SNR $\SNR$ and is statistically independent of $\r a$.
\item
At low SNR $\SNR < 3.011$ (or rather $\SNR < 4.8\dB$), capacity is achieved by $K = 1$ and thus $\r\Radius = \Radius_1 = 1$ with $\CircProb_1 = 1$. This corresponds to uniform $\infty$-PSK modulation.
\item
The number of circles $K$ increases with the SNR $\SNR$.
\end{enumerate}
\begin{theorem}\label{prop:magnitudeDistr}
The channel capacity $\RateMaxGP$ is obtained by maximizing the mutual information with respect to the number of circles, their radii, and their probabilities:
\begin{align}
\RateMaxGP(\SNR) = &\max_{K,\,\Radius_2,\,\ldots,\,\Radius_K, \,\CircProb_1,\,\ldots,\,\CircProb_K} \!\! I(\r y;\r\Gamma)\big|_{
\,\r\Radius \in \{1, \,\Radius_2 , \,\ldots\, , \,\Radius_K \},
\, \text{UIP}\ \r\theta 
}
\nonumber \\
&\mathrm{subject\ to}\
K \in \bbN,\ \
0 \leq \Radius_K < \ldots < \Radius_2 < 1
\, , \nonumber \\
&\,0 < \CircProb_1 ,\, \ldots ,\, \CircProb_K \leq 1
\, , \ \ \ \
\CircProb_1 + \ldots + \CircProb_K = 1
\, . \label{eq:Capacity}
\end{align}
\end{theorem}
\textit{Proof Sketch.} The statement is a concise recollection of vast formal statements by Shamai \& Bar-David \cite{ShamaiTIT1995} who originally derived the channel capacity of complex-valued peak-power-limited AWGN channels. This was accomplished with rigorous application of optimization theory such as Karush–Kuhn–Tucker conditions. A full derivation of \Cref{eq:Capacity} would require an almost full recollection of \cite{ShamaiTIT1995}. Instead, we refer to this paper for the formal details. \QEDB

The calculation of the mutual information is non-trivial. 

\begin{theorem}\label{prop:MICalc}
Consider DAUIP signaling $\r\Gamma$ with fixed parameters $\Radius_k , \CircProb_k$ for $k = 1,\ldots,K$. The mutual information over a complex AWGN channel with SNR $\SNR$ is calculated as
\begin{align}
I(\r{y};\r\Gamma) = \log_2( 2\SNR/e ) - \!\int_0^\infty \!
f_\r{b}(b) \log_2\!\Big( \f{f_\r{b}(b)}{b} \Big) \, db
\label{eq:MutualInformationUP}
\end{align}
which requires numerical integration.
The receive-amplitude $\r{b}$ is a Rician mixture of the different circles, with the PDF
\begin{align}
& f_\r{b}(b) = 2\SNR b \sum_{k=1}^K \CircProb_k
e^{-\SNR(b - a_k)^2} g( 2\SNR a_k b )
\label{eq:NoisyRadiusPDFCircleSum}
\end{align}
\end{theorem}
where
$g(x) := I_0(x) e^{-x}$ for $x \in \bbR_{\geq 0}$ is an exponentially-scaled modified Bessel function of the first kind (whose direct use enables a numerically robust calculation) \cite{ScaledBesselFunction}.

\textit{Proof:} Equivalent statements are scattered across the literature, e.g., in much formal detail in \cite[Eq.~(11),(13),(46)]{ShamaiTIT1995}. \Cref{apdx:MutInfCalc} presents a concise derivation in our formalism. \QEDB

The maximization problem \Cref{eq:Capacity} is not concave; counterexamples are easily found in the parameter space. It classifies as mixed-integer programming with a variable number of parameters ($2K$).
Approximate $\SNR$-thresholds for the optimality of $K = 1,2,3$ are stated in \cite[Tab.~1]{ShamaiTIT1995} (please note that their values are $3\dB$ larger because they assumed a variance of $2$ for the complex-valued AWGN). The threshold $\SNR < 3.011$ (or rather $\SNR < 4.8\dB$) for the optimality of $K=1$ is the numerical solution of a complicated equation \cite[Eq.~(45)]{ShamaiTIT1995}.

\Cref{fig:Radii_PMF_evolution} shows a numerical evaluation of
the capacity-achieving parameters versus $\SNR$. We observe how new circles emerge at the center when $\SNR$ increases. This occurs whenever $f_{\r y}(0)$ has vanished so much that adding a circle becomes justified.
The behavior is best seen in our animation \cite{youtube2022}.

Calculating the optimal parameters is quite complicated and requires numerical methods. One approach is the vector optimization proposal in \cite[Sec.~III-B]{ShamaiTIT1995} where also formal arguments for its correctness (despite the non-concavitiy of \Cref{eq:Capacity}) are provided. 
We chose a very similar approach which we describe in \Cref{apdx:CapOptProb} for clarity and reproducibility.

\Cref{fig:CapDistr_Gamma} illustrates the PDF $f_\r\Gamma(\Gamma)$ of the capacity-achieving distribution of $\r\Gamma \in \bbC$ over the complex plane, for different SNR values. This evaluation uses specific values from \Cref{fig:Radii_PMF_evolution}.%
\footnote{We note the following distribution characteristics. Consider the lateral probability density $\lambda_k := \CircProb_k / (2\pi\Radius_k)$ over the $k$-th circumference. At high SNR, they take similar values $\forall k$, so $\r\Gamma$ closely resembles a uniform distribution over the unit disk. However, $\lambda_k$ is appreciably larger for the outmost circles. Complementary numerical evaluations implied that $\lambda_2 / \lambda_1 \rightarrow 2/3$ and $\lambda_3 / \lambda_1 \rightarrow 1/\sqrt{3}$ hold for $\rho \rightarrow \infty$ (which we can not proof formally), while for $k > 3$ any $\lambda_k / \lambda_1$ approaches a value slightly below $1/\sqrt{3}$.}

The associated distributions of the normalized load impedance $\r z = \f{1+\r\Gamma}{1-\r\Gamma}$ could be very interesting to circuit designers. They are given and discussed in \Cref{apdx:zStats}.

A tangible behavioral description of the rate $\RateMaxGP(\SNR)$ is provided by simple lower and upper bounds:
\begin{align}
\log_2(1 + \SNR/e) < \RateMaxGP < \log_2(1 + \SNR) \, .
\label{eq:RMaxBounds}
\end{align}
They characterize the low- and high-SNR behavior:
\begin{align}
\text{low SNR%
:} && \RateMaxGP &\approx \log_2(1 + \SNR) \approx \SNR \log_2(e) \, ,
\label{eq:RMaxBoundLowSNR} \\
\text{high SNR%
:} && \RateMaxGP &\approx \log_2(1 + \SNR/e) \, .
\label{eq:RMaxBoundHighSNR}
\end{align}
A derivation of the lower bound is given in \Cref{apdx:LowerBoundSNReDeriv}.
The upper bound $\log_2(1 + \SNR)$ is the well-known AWGN channel capacity with average-power constraint $\EV{|\r\Gamma|^2} \leq 1$, which is looser than the instantaneous constraint $|\r\Gamma|^2 \leq 1$ at hand \cite{ShamaiTIT1995,DarsenaTC2017}. 
This difference causes an SNR gap $e = 4.3\dB$ or rather $1.44\unit{bpcu}$ between upper and lower bound, 
The low-SNR asymptote $\SNR \log_2(e)$ is a consequence of using only the maximum amplitude ($K = 1, \Radius_1 = 1$) at small $\SNR$ \cite{Blachman1953,WynerBSTJ1966}.

\renewcommand\myFigHeight{50mm}

\ifdefined\SingleColumnDraft
\begin{figure}[!b]
\else 
\begin{figure}[t]
\fi
\centering
\subfloat[circle radii]{%
\includegraphics[height=\myFigHeight]{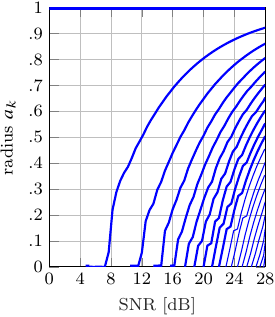}
\put(-56,132){\scriptsize{$a_1$}}%
\put(-56,106.5){\scriptsize{$a_2$}}%
\put(-56,73){\scriptsize{$a_3$}}%
\put(-46,66){\scriptsize{$...$}}%
\label{fig:Radii_evolution}%
}\!
\
\subfloat[circle probabilities]{%
\includegraphics[height=\myFigHeight]{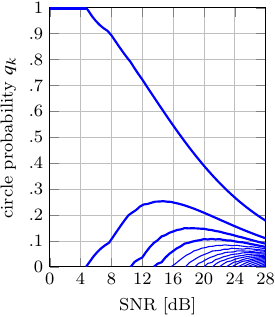}
\put(-56,108){\footnotesize{$\CircProb_1$}}%
\put(-69,52){\scriptsize{$\CircProb_2$}}%
\put(-55,40.5){\scriptsize{$\CircProb_3$}}%
\put(-43,36){\scriptsize{$...$}}%
\label{fig:PMF_evolution}%
}
\caption{Circle radii and probabilities of the capacity-achieving DAUIP distribution of load reflection coefficient $\r\Gamma$. For either plot, the number of intersection points between the graphs and a vertical line determines the optimal number of circles $K$ for a certain SNR value $\SNR$.}
\label{fig:Radii_PMF_evolution}
\end{figure}
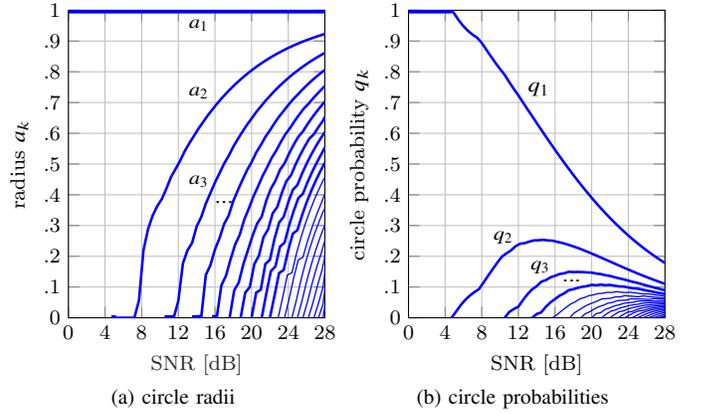

\newcommand\myLateralProbDensityLineWidth{1.7pt}

\ifdefined\SingleColumnDraft
\renewcommand\myFigHeight{31mm}
\else 
\renewcommand\myFigHeight{33mm}
\fi

\ifdefined\SingleColumnDraft
\begin{figure}[!ht]
\else 
\begin{figure}[t]
\fi
\centering
\subfloat[SNR $\SNR < 4.8\dB$]{%
\includegraphics[height=\myFigHeight]{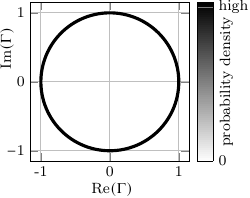}%
\label{fig:CapDistr_GammaLowSNR}} \
\subfloat[SNR $\SNR = 12\dB$]{%
\includegraphics[height=\myFigHeight]{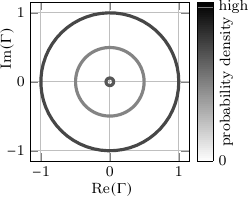}%
\label{fig:CapDistr_Gamma12dB}}
\ifdefined\SingleColumnDraft\else \\ \fi
\subfloat[SNR $\SNR = 18 \dB$]{%
\includegraphics[height=\myFigHeight]{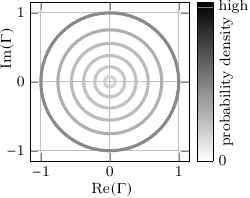}%
\label{fig:CapDistr_Gamma18dB}} \
\subfloat[SNR $\SNR = 24\dB$]{%
\includegraphics[height=\myFigHeight]{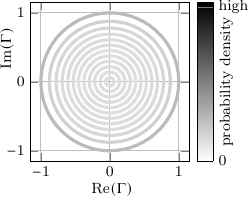}%
\label{fig:CapDistr_Gamma24dB}}
\caption{Capacity-achieving distribution of the load reflection coefficient $\r\Gamma \in \bbC$ for different SNR values. The color intensity actually describes the multiplier of a Dirac delta, because the probability density is either $0$ or $\infty$ as a result of the one-dimensional PDF support (the circles are infinitesimally thin).}
\label{fig:CapDistr_Gamma}
\end{figure}

\subsection{Uniformly Distributed Signaling over Unit Disk}
\label{sec:MaxEntropyRate}

An interesting reference case is the achievable rate $\RateME(\SNR)$ resulting from a uniformly distributed $\r\Gamma$ over the complex unit disk (henceforth abbreviated as UD signaling).
Here, no SNR-dependent adaptation is done. 
Still, $\RateME(\SNR)$ approaches the channel capacity $\RateMaxGP(\SNR)$ at high SNR. This can be seen informally at the fact that \Cref{fig:CapDistr_Gamma24dB} is similar to UD signaling. For a more formal argument, note that the distributions $\r y$ and $\r\Gamma$ are similar at high SNR. This causes similar entropies $h(\r y) \approx h(\r\Gamma)$.
Now
$I(\r y;\r\Gamma) = h(\r y) - h(\r w)$
is approximately
$h(\r\Gamma) - h(\r w)$,
which is maximized by maximizing $h(\r\Gamma)$.
This is indeed achieved with UD signaling, as shown in \Cref{apdx:MaxEntropy}.

The calculation of the achievable rate $\RateME(\SNR)$ is described in \Cref{apdx:UDRateCalc} and requires numerical double integration.
A much simpler characterization is given by the bounds
\begin{align}
\log_2(1 + \SNR / e) &\leq \RateME < \RateMaxGP
\label{eq:RateMEBounds}
\end{align}
which both become tight at high $\SNR$, i.e. UD signaling approaches channel capacity \cite{ShamaiTIT1995}.
It has however poor low-SNR behavior due to using amplitudes below $1$, which is wasteful in the power-limited regime.
In particular, $3\dB$ are being wasted, as shown in \Cref{apdx:UDRateCalc}.
In summary, the characteristics are
\begin{align}
\text{low SNR%
:} &&
\RateME &\approx \tfrac{1}{2} \,\RateMaxGP
\approx  \tfrac{1}{2} \log_2(1 + \SNR) \, ,
\label{eq:RateMEHalf} \\
\text{high SNR%
:} &&
\RateME &\approx \RateMaxGP \approx \log_2(1 + \SNR/e) \, .
\label{eq:RateMEUpper}
\end{align}

\subsection{Channel Capacity, Purely Reactive Load Modulation}
\label{sec:CapacityReact}

Consider that the load is constrained to be purely reactive for technical reasons, e.g., the load circuit should consist of capacitors and inductors. This corresponds to a purely imaginary impedance $\r z = j \r x$, a constant-amplitude reflection coefficient on the unit circle $\r\Gamma = e^{j\r\theta}$, and PSK modulation.

It is known from \cite[Eq.~(12)]{WynerBSTJ1966},\cite[Eq.~(14)]{DuanTVT2017},\cite{ShamaiTIT1995} that the mutual information under a constant-amplitude constraint is maximized by UIP signaling $\r\theta \sim \calU(0,2\pi)$. This can be regarded as $\infty$-PSK modulation. The resulting mutual information is the channel capacity for this case, which we denote $\RateReact$. It is calculated with \Cref{eq:MutualInformationUP,eq:NoisyRadiusPDFCircleSum} as the special case of a single circle $K=1$ (with $\Radius_1=1, \CircProb_1 = 1$). This still requires numerical integration but no more numerical optimization (all circle parameters have been fixed).

Following from the statements in \Cref{sec:CapacityGeneral}, at low SNR $\SNR < 4.8\dB$ the equality $\RateReact = \RateMaxGP$ holds precisely (here reactive UIP load modulation is optimal).
At higher SNR however, rate loss is to be expected becauce only the boundary of the unit disk is utilized. This is reflected by the asymptote
\begin{align}
\text{high SNR%
:} &&
\RateReact
&\approx \f{1}{2} \log_2\left( \f{4\pi\SNR}{e} \right)
\label{eq:RateReactApprox}
\end{align}
for which a derivation and details are given in \Cref{apdx:ReactAsymptote}.
It has a pre-log factor of $\f{1}{2}$ compared to the general-case asymptote $\RateMaxGP \approx \log_2(1 + \SNR / e)$ from \Cref{eq:RateMEUpper}. 
This shows that reactive load modulation can not exploit a high SNR well. The factor $\f{1}{2}$ stems from the unused dimension in the load impedance $\r z = \r r + j\r x$, where only $\r x$ is being modulated, but not $\r r$. The general-case scheme (\Cref{sec:CapacityGeneral}) instead modulates both dimensions whenever the SNR calls for that.

\subsection{Channel Capacity, Purely Resistive Load Modulation}
\label{sec:CapacityResist}

Consider that technical constraints are in place that demand a purely resistive load. The impedance becomes real-valued: $\r z = \r r \in \bbR_{\geq 0}$. Then the reflection coefficient $\r\Gamma = \f{\r r - 1}{\r r + 1}$ lies on the interval $[-1,1]$ on the real axis. Now the received imaginary part $\Im(\r y) = \Im(\r\Gamma + \r w) = \Im(\r w)$ bears no information about $\r\Gamma$. The mutual information is thus equivalently described by a real-valued channel
$\r y\reSub = \r\Gamma\reSub + \r w\reSub$
subject to
$\r\Gamma\reSub \in [-1,1]$ and with $\r w\reSub \sim \calN(0,1/(2\SNR))$.
The associated channel capacity, here termed $\RateResist$, was solved by Smith \cite{Smith1971}. He found that the optimal transmit distribution is discrete, i.e. a finite constellation with SNR-dependent symbol locations and probabilities. We give formal details in \Cref{apdx:CapCalcRes} and a rich visualization in \cite{youtube2022}.

A very interesting property%
\footnote{Another noteworthy property $\RateMaxGP < 2\RateResist$ results from the fact that the unit disk fits into the larger square
$(\Gamma\reSub, \Gamma\imSub) \in [-1,1]^2$, as explained in \cite[Fig.\,1\,\&\,Eq.\,(27)]{ShamaiTIT1995}. 
Similarly,
$\RateResist < \f{1}{2} \log_2(1 + 2\SNR)$ is from the real-valued AWGN channel capacity with just an average-power constraint.}
is
$\f{1}{2} \log_2( 1 + \f{4\,\SNR}{\pi e} ) \leq \RateResist$,
resulting from (as shown in \Cref{apdx:ResBounds}) a lower bound on the sub-capacity rate with a continuous uniform distribution
$\r\Gamma\reSub \in \mathcal{U}(-1,1)$.
This lower bound is however tight in the asymptotic sense \cite{Smith1971}
\begin{align}
\text{high SNR:} &&
\RateResist &\approx 
\f{1}{2} \log_2\left( 1 + \f{4\,\SNR}{\pi e} \right)
 \, .
\label{eq:RateResistApprox}
\end{align}
Also this asymptote has the pre-log factor $\f{1}{2}$, again due to modulating only on a one-dimensional subset of the unit disk.

A comparison between \Cref{eq:RateReactApprox,eq:RateResistApprox} reveals a significant SNR-domain gain of $\pi^2 \approx 10$ of reactive over resistive load modulation.
This is explained by an amplitude-domain gain of $\pi$, hailing from the circle's circumference-to-diameter ratio.
Loosely speaking, the interval $[-1,1] \subset \bbR$ just offers less space for modulation than the unit circle.

\subsection{Numerical Comparison}
\label{sec:results}

The numerical evaluation in \Cref{fig:Rates_Capacity} shows the evolution of the discussed rates and the most important asymptotes versus SNR.
\ifdefined\SingleColumnDraft
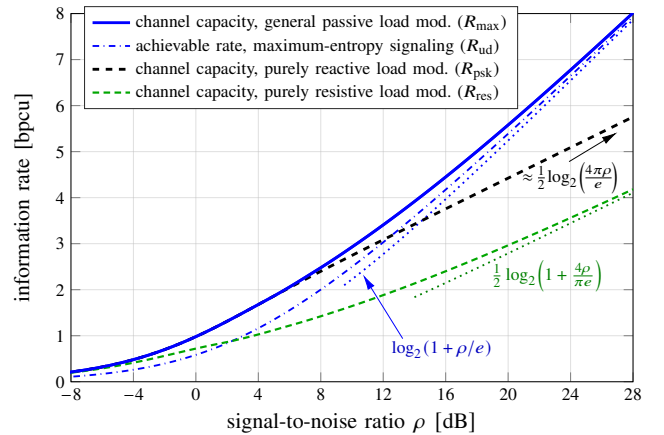
\begin{figure}[!b]
\else
\begin{figure}[!ht]
\fi
\centering
\resizebox{85mm}{!}{\input{F03e_Rates_CapacityResults}}%
\caption{Channel capacity in bit per channel use (bpcu) plotted versus SNR.}
\label{fig:Rates_Capacity}
\end{figure}
We used the computation rules stated or referenced earlier.
The graphs confirm the key high-SNR statements made earlier. Indeed, $\RateReact$ beats $\RateResist$ by an SNR difference of $\pi^2 = 9.94\dB$, i.e. reactive load modulation performs much better than resistive. And indeed both $\RateReact, \RateResist$ suffer a significant rate limitation at high SNR (recall their $\f{1}{2}$ pre-log factor). Meanwhile $\RateMaxGP, \RateME$ do not suffer this problem. The graph shows that, this way, high information rates in excess of $5\unit{bpcu}$ are possible at practically reasonably SNR values.

The gap $\RateMaxGP - \RateME$ is small but appreciable at high SNR.
The technical conclusion is that suboptimal signaling schemes can still yield near-optimal data rates, if the whole unit disk is utilized in a meaningful way.
The gap is however significant at low SNR where, as predicted, a rate loss $\RateME \approx \f{1}{2} \RateMaxGP$ occurs for the SNR-agnostic UD signaling. We conclude that adaptive modulation based on SNR estimates (e.g., resorting to PSK at low SNR) is important for BC systems to operate well over a wide range of channel conditions.

%% file: F03e_Rates_CapacityResults.tex
\begin{tikzpicture}

\begin{axis}[%
width=90mm,
height=59mm,
at={(0,0)},
scale only axis,
xmin=-8,
xmax=28,
xtick={ -8, -4, 0, 4, 8, 12, 16, 20, 24, 28},
xlabel style={font=\color{white!15!black}},
xlabel={signal-to-noise ratio $\SNR\ [\mathrm{dB}]$},
ymin=0,
ymax=8,
ytick={0,1,2,3,4,5,6,7,8,9,10},
yminorticks=true,
ylabel style={font=\color{white!15!black}},
ylabel={information rate [bpcu]},
ylabel style={at={(-0.055,0.5)}},
axis background/.style={fill=white},
xmajorgrids,
ymajorgrids,
grid style={opacity=0.5},
label style={font=\normalsize},
legend style={at={(.0234,1.02)}, anchor=north west, legend cell align=left, align=left, draw=white!15!black} 
]

\newcommand\PlotCapacity{%
\addplot [color=blue, line width=1.3pt]
  table[row sep=crcr]{%
-12	0.0864315365372339\\
-11.5	0.0968938816009866\\
-11	0.108535315924035\\
-10.5	0.121478406979223\\
-10	0.135855771055293\\
-9.5	0.151810236217422\\
-9	0.16949482352194\\
-8.5	0.189072495555821\\
-8	0.210715614271572\\
-7.5	0.234605043429964\\
-7	0.260928825391637\\
-6.5	0.289880358384887\\
-6	0.32165599985449\\
-5.5	0.356452025488289\\
-5	0.394460883742997\\
-4.5	0.435866704131751\\
-4	0.48084004635321\\
-3.5	0.529531918697995\\
-3	0.582067149935895\\
-2.5	0.63853727020936\\
-2	0.698993143127997\\
-1.5	0.763437690917999\\
-1	0.831819161673053\\
-0.5	0.904025492931994\\
0	0.979880414508299\\
0.5	1.05914198601691\\
1	1.14150425662973\\
1.5	1.22660263971832\\
2	1.31402338890148\\
2.5	1.40331722954043\\
3	1.49401674435592\\
3.5	1.58565656558761\\
4	1.67779485739644\\
4.5	1.77003408686118\\
5	1.86252447577785\\
5.5	1.9586982297103\\
6	2.05865655462784\\
6.5	2.16151423207272\\
7	2.26628092318256\\
7.5	2.37190950249353\\
8	2.47809417055567\\
8.5	2.58677238614189\\
9	2.6980725693527\\
9.5	2.81176968358008\\
10	2.9274800746115\\
10.5	3.04469941545486\\
11	3.16373469941759\\
11.5	3.28486380192205\\
12	3.40765530732148\\
12.5	3.53207680492358\\
13	3.65833650083317\\
13.5	3.78626838756433\\
14	3.91577392694257\\
14.5	4.04689071863925\\
15	4.17949655464713\\
15.5	4.31360742956774\\
16	4.44913720761297\\
16.5	4.58605498302582\\
17	4.72430591740325\\
17.5	4.86383595659652\\
18	5.00462349817165\\
18.5	5.14659618924892\\
19	5.28971625348757\\
19.5	5.43393861934742\\
20	5.57921502640108\\
20.5	5.72551423784874\\
21	5.87277857916322\\
21.5	6.02098003948519\\
22	6.17006471615952\\
22.5	6.32000983991334\\
23	6.47076539080284\\
23.5	6.6222949637335\\
24	6.77457096976575\\
24.5	6.92755334339171\\
25	7.08120458375078\\
25.5	7.23550591084472\\
26	7.39041834941432\\
26.5	7.54591308787267\\
27	7.70196209969228\\
27.5	7.85853834282628\\
28	8.01561580406267\\
};}
\PlotCapacity
\addlegendentry{channel capacity, general passive load mod. ($\RateMaxGP$)}

\addplot [color=blue, dashdotted, line width=0.85pt]
  table[row sep=crcr]{%
-12	0.0430357800270258\\
-11.5	0.0485153982601769\\
-11	0.0546231435100362\\
-10.5	0.0614300584631233\\
-10	0.0690144698516399\\
-9.5	0.0774625408774487\\
-9	0.0868688171033347\\
-8.5	0.0973367509805105\\
-8	0.108979186408522\\
-7.5	0.121918780473549\\
-7	0.136288334845428\\
-6.5	0.152231004357094\\
-6	0.169900345247842\\
-5.5	0.189460160734288\\
-5	0.21108409741799\\
-4.5	0.234954943132778\\
-4	0.261263575926185\\
-3.5	0.290207515867983\\
-3	0.321989037331367\\
-2.5	0.356812810424131\\
-2	0.394883057452775\\
-1.5	0.436400234576403\\
-1	0.481557280602153\\
-0.5	0.530535513885831\\
0	0.583500303102796\\
0.5	0.640596685316208\\
1	0.701945150566495\\
1.5	0.767637849531232\\
2	0.837735501484555\\
2.5	0.912265274981588\\
3	0.991219875545785\\
3.5	1.07455799863318\\
4	1.16220619390248\\
4.5	1.25406204850955\\
5	1.34999845239453\\
5.5	1.44986858438086\\
6	1.55351118298489\\
6.5	1.66075566133413\\
7	1.77142669592218\\
7.5	1.8853480467942\\
8	2.00234551698455\\
8.5	2.1222490914638\\
9	2.24489438158952\\
9.5	2.37012353272218\\
10	2.49778574275876\\
10.5	2.62773750902744\\
11	2.75984268815715\\
11.5	2.89397242723057\\
12	3.03000500632597\\
12.5	3.16782562061964\\
13	3.3073261223198\\
13.5	3.44840473726665\\
14	3.5909657671503\\
14.5	3.73491928545086\\
15	3.88018083307993\\
15.5	4.02667111809909\\
16	4.17431572267127\\
16.5	4.323044819469\\
17	4.47279289905068\\
17.5	4.62349850916987\\
18	4.77510400656688\\
18.5	4.92755532147541\\
19	5.08080173483939\\
19.5	5.23479566805831\\
20	5.38949248494992\\
20.5	5.54485030552707\\
21	5.70082983112254\\
21.5	5.8573941803548\\
22	6.01450873540459\\
22.5	6.17214099806219\\
23	6.33026045500549\\
23.5	6.48883845177673\\
24	6.64784807493932\\
24.5	6.80726404191326\\
25	6.96706259800815\\
25.5	7.1272214201944\\
26	7.28771952717653\\
26.5	7.4485371953555\\
27	7.60965588029071\\
27.5	7.77105814329524\\
28	7.93272758282015\\
};
\addlegendentry{achievable rate, maximum-entropy signaling ($\RateME$)}

\addplot [color=blue, dotted, line width=0.85pt, forget plot]
  table[row sep=crcr]{%
9.5	2.09718229660688\\
10	2.22613683743591\\
10.5	2.35830488636829\\
11	2.49347512494798\\
11.5	2.63143847976525\\
12	2.7719903907463\\
12.5	2.91493268063064\\
13	3.06007503578848\\
13.5	3.20723612198796\\
14	3.35624436807358\\
14.5	3.50693845616561\\
15	3.6591675595116\\
15.5	3.81279136916546\\
16	3.96767994887829\\
16.5	4.12371345454369\\
17	4.28078175074461\\
17.5	4.43878395279852\\
18	4.59762791849355\\
18.5	4.75722970966559\\
19	4.9175130400244\\
19.5	5.07840872227762\\
20	5.23985412466059\\
20.5	5.40179264446133\\
21	5.56417320401488\\
21.5	5.72694977289813\\
22	5.89008091864528\\
22.5	6.05352938718374\\
23	6.21726171331761\\
23.5	6.38124786092306\\
24	6.54546089202881\\
24.5	6.70987666360581\\
25	6.87447355065339\\
25.5	7.03923219402248\\
26	7.2041352713394\\
26.5	7.36916728936972\\
27	7.53431439617792\\
27.5	7.69956421148362\\
28	7.86490567368074\\
};

\node (destination) at (axis cs:10.45,2.465){};
\node (source) at (axis cs:13.5,.8){};
\draw[-{Latex[length=2.5mm, width=1.25mm]},color=blue](source)--(destination);
\node[anchor=west] (text) at (axis cs:12.15,.7){\textcolor{blue!75!black}{$\log_2(1 + \SNR / e )$}};

\addplot [color=black, dashed, line width=1.3pt]
  table[row sep=crcr]{%
-12	0.0864315365372339\\
-11.5	0.0968938816009866\\
-11	0.108535315924035\\
-10.5	0.121478406979223\\
-10	0.135855771055293\\
-9.5	0.151810236217422\\
-9	0.16949482352194\\
-8.5	0.189072495555821\\
-8	0.210715614271572\\
-7.5	0.234605043429964\\
-7	0.260928825391637\\
-6.5	0.289880358384887\\
-6	0.32165599985449\\
-5.5	0.356452025488289\\
-5	0.394460883742997\\
-4.5	0.435866704131751\\
-4	0.48084004635321\\
-3.5	0.529531918697995\\
-3	0.582067149935895\\
-2.5	0.63853727020936\\
-2	0.698993143127997\\
-1.5	0.763437690917999\\
-1	0.831819161673053\\
-0.5	0.904025492931994\\
0	0.979880414508299\\
0.5	1.05914198601691\\
1	1.14150425662973\\
1.5	1.22660263971832\\
2	1.31402338890148\\
2.5	1.40331722954043\\
3	1.49401674435592\\
3.5	1.58565656558761\\
4	1.67779485739644\\
4.5	1.77003408686118\\
5	1.86203881197189\\
5.5	1.95354829118792\\
6	2.0443822277425\\
6.5	2.1344388954696\\
7	2.22368611366353\\
7.5	2.31214677966389\\
8	2.39988159307618\\
8.5	2.48697192226504\\
9	2.57350535324004\\
9.5	2.65956545798701\\
10	2.7452260803892\\
10.5	2.83054939188579\\
11	2.91558641939581\\
11.5	3.00037874494083\\
12	3.08496043743093\\
12.5	3.16935973200879\\
13	3.25360032066712\\
13.5	3.33770229677027\\
14	3.42168284556869\\
14.5	3.50555676107747\\
15	3.58933684436541\\
15.5	3.67303421818009\\
16	3.75665858028019\\
16.5	3.84021841044758\\
17	3.92372114165475\\
17.5	4.00717330297593\\
18	4.09058063988499\\
18.5	4.17394821622599\\
19	4.25728050116793\\
19.5	4.34058144374078\\
20	4.42385453701641\\
20.5	4.5071028735919\\
21	4.59032919372108\\
21.5	4.67353592719623\\
22	4.75672522988997\\
22.5	4.83989901571424\\
23	4.92305898463008\\
23.5	5.00620664724228\\
24	5.08934334643103\\
24.5	5.1724702764055\\
25	5.25558849950875\\
25.5	5.33869896105639\\
26	5.42180250245275\\
26.5	5.50489987279502\\
27	5.58799173914823\\
27.5	5.67107869565019\\
28	5.75416127158544\\
};
\addlegendentry{channel capacity, purely reactive load mod. ($\RateReact$)}
\node[anchor=west] (text) at (axis cs:20.5,4.43){$\approx\! \f{1}{2}\! \log_2\!\left(\! \f{4\pi\SNR}{e} \!\right)$};

\node (destination) at (axis cs:27.4,5.5){};
\node (source) at (axis cs:23.5,4.7){};
\draw[-{Latex[length=2.5mm, width=1.25mm]}](source)--(destination);

\addplot [color=green!65!black, densely dashed, line width=1.0pt]
  table[row sep=crcr]{%
-8	0.196698596607796\\
-7.5	0.217405812207362\\
-7	0.239870927521977\\
-6.5	0.264158253000455\\
-6	0.290314802103217\\
-5.5	0.318365356624729\\
-5	0.348307052066349\\
-4.5	0.380103605609812\\
-4	0.413679378986762\\
-3.5	0.448913552399593\\
-3	0.485634781715593\\
-2.5	0.52361681476145\\
-2	0.562575645200593\\
-1.5	0.602168870814168\\
-1	0.641997977966404\\
-0.5	0.681614270700819\\
0	0.720529071618524\\
0.5	0.75822861111117\\
1	0.79419366457484\\
1.5	0.827994298285716\\
2	0.862709222960929\\
2.5	0.900183593063015\\
3	0.940332426254877\\
3.5	0.982984049651306\\
4	1.02785401406029\\
4.5	1.07454020804247\\
5	1.12252091535096\\
5.5	1.17116418030642\\
6	1.21974766993879\\
6.5	1.26788857510336\\
7	1.31762089009979\\
7.5	1.3692494296679\\
8	1.42249849126372\\
8.5	1.47692478764458\\
9	1.53192694450348\\
9.5	1.58747946412745\\
10	1.64444126743691\\
10.5	1.70265134954027\\
11	1.76174060127165\\
11.5	1.82156861637384\\
12	1.88251184285216\\
12.5	1.94440569178326\\
13	2.00707835387181\\
13.5	2.07069472179581\\
14	2.13513146815299\\
14.5	2.20038367009221\\
15	2.26643206115613\\
15.5	2.33324200218459\\
16	2.40079085949757\\
16.5	2.46905300212064\\
17	2.53799826713913\\
17.5	2.60760705700686\\
18	2.67785101582461\\
18.5	2.74870506040953\\
19	2.82014647813607\\
19.5	2.8921500690948\\
20	2.96469385464086\\
20.5	3.03775227622324\\
21	3.11130925121938\\
21.5	3.18533504415832\\
22	3.25981835716104\\
22.5	3.33473170512913\\
23	3.41005642471332\\
23.5	3.48577706095962\\
24	3.56187268794469\\
24.5	3.6383237681678\\
25	3.71511944303183\\
25.5	3.79223936444427\\
26	3.86966824100414\\
26.5	3.94739118052244\\
27	4.02539383188005\\
27.5	4.10366201662668\\
28	4.18218325159116\\
};
\addlegendentry{channel capacity, purely resistive load mod. ($\RateResist$)}

\addplot [color=green!50!black, dotted, line width=0.85pt, forget plot]
  table[row sep=crcr]{%
14	1.83709710541944\\
14.5	1.91397390563754\\
15	1.99147698572113\\
15.5	2.06954690952106\\
16	2.14812907418407\\
16.5	2.2271734782031\\
17	2.30663446539012\\
17.5	2.386470454655\\
18	2.46664366362063\\
18.5	2.54711983244506\\
19	2.62786795277147\\
19.5	2.70886000548825\\
20	2.7900707099427\\
20.5	2.87147728639759\\
21	2.95305923282961\\
21.5	3.03479811662156\\
22	3.116677381275\\
22.5	3.19868216794785\\
23	3.28079915138357\\
23.5	3.36301638962993\\
24	3.4453231868321\\
24.5	3.52770996831514\\
25	3.6101681671354\\
25.5	3.69269012127055\\
26	3.77526898062788\\
26.5	3.85789862307388\\
27	3.94057357872213\\
27.5	4.0232889617562\\
28	4.10604040910854\\
};
\node[anchor=west] (text) at (axis cs:18.5,2.3){\textcolor{green!50!black}{$\f{1}{2} \log_2\!\left( 1 + \f{4\SNR}{\pi e} \right)$}};

\PlotCapacity 

\end{axis}

\end{tikzpicture}%

%% file: 03f-ValueRange.tex
In practice it may not be possible to realize any desired reflection coefficient $\Gamma$ on the unit disk. In the following we analyze the BC performance loss when $\Gamma$ is restricted by technical constraints.
\begin{theorem}
Consider that $\r\Gamma$ is restricted to a set $\mathcal{G} \subset \bbC$ with non-empty interior. Then the mutual information is lower-bounded by
\begin{align}
\log_2\!\left( 1 + \f{\mathrm{area}(\mathcal{G})}{\pi} \cdot \f{\SNR}{e} \right)
\leq I(\r y;\r\Gamma) \, .
\label{eq:AreaInequality}
\end{align}
\end{theorem}

\textit{Proof:} In \Cref{apdx:LowerBoundSNReDeriv} we derive
$\log_2(1 + \f{\SNR}{\pi e} 2^{h(\r\Gamma)}) \leq I(\r y;\r\Gamma)$ for complex AWGN channels, 
which holds for any distribution $\r\Gamma$. We consider specifically a uniform distribution
$\r\Gamma \sim \mathcal{U}(\mathcal{G})$
and deduct $h(\r\Gamma) = \log_2(\mathrm{area}(\mathcal{G}))$ from \Cref{apdx:MaxEntropy}. \QEDB

\definecolor{darkRedLoss}{rgb}{.65,0,0}%

\ifdefined\SingleColumnDraft

\renewcommand\myFigWidth{39mm}
\begin{figure}[!b]\centering
\newcommand\myPut[1]{\put(-94,62){#1}}
\subfloat[$\Delta =  5\%$]{\includegraphics[width=\myFigWidth]{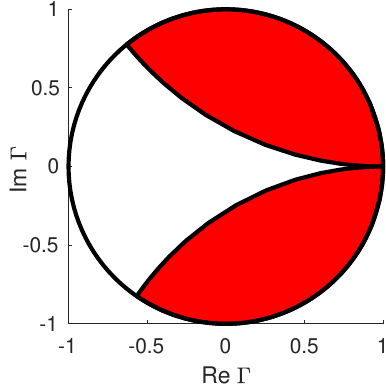}
\myPut{\scriptsize\textcolor{darkRedLoss}{\footnotesize\begin{tabular}{l}%
$-61.2\%$\,area,\\[0mm]$-1.36\unit{bpcu}$%
\end{tabular}}}}\
\renewcommand\myPut[1]{\put(-82,62){#1}}
\
\subfloat[$\Delta = 25\%$]{\includegraphics[width=\myFigWidth]{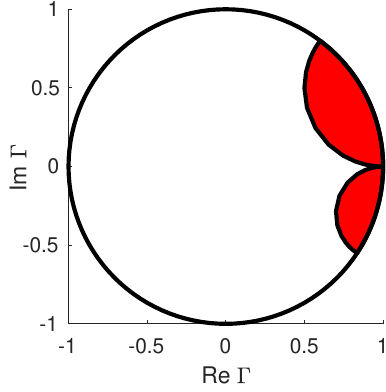}
\myPut{\scriptsize\textcolor{darkRedLoss}{\footnotesize\begin{tabular}{l}%
$-11.0\%$\,area,\\[0mm]$-0.168\unit{bpcu}$%
\end{tabular}}}}\
\renewcommand\myPut[1]{\put(-76,62){#1}}
\subfloat[$\Delta = 50\%$]{\includegraphics[width=\myFigWidth]{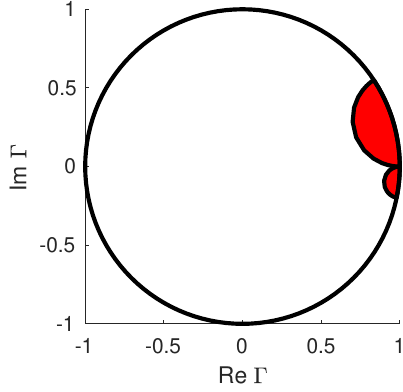}
\myPut{\scriptsize\textcolor{darkRedLoss}{\footnotesize\begin{tabular}{l}%
$-3.85\%$\,area,\\[0mm]$-0.0566\unit{bpcu}$%
\end{tabular}}}}\
\caption{Non-realizable values (red) of reflection coefficient $\Gamma$ due to value-range constraints on the load reactance. This experiment assumes an inductive RFID tag.
Each bpcu value is a high-SNR rate loss $\log_2( \pi / \mathrm{area}(\mathcal{G}) )$.}
\label{fig:DiskConstraintOutage}
\end{figure}

\fi

Analogous to the arguments in \Cref{sec:capacity}, this bound becomes tight at high SNR, in which case it constitutes an accurate approximation of the achievable rate with uniform signaling and of the channel capacity.
The expression can therefore be used to assess the rate loss due to a constraint $\Gamma \in \mathcal{G}$ with $\mathrm{area}(\mathcal{G}) < \pi$.
In particular, the absolute rate loss at high SNR can be quantified as
$\log_2( \pi / \mathrm{area}(\mathcal{G}) )$ bpcu.

The plots in \Cref{fig:DiskConstraintOutage} show the regions of unrealizable reflection coefficients for an exemplary inductive RFID tag. Thereby, the tag antenna coil with $Z\Tx = R\Tx + j\omega L\Tx$ is loaded with an impedance $R+\f{1}{j\omega C}$ with adaptive resistance $R \in \bbR_{\geq 0}$ and adaptive capacitance $C$ with the restricted value range
$(1-\Delta)C_\text{res} \leq C \leq (1+\Delta)C_\text{res}$
about the resonance value
$C_\text{res} = 1 / (\omega^2 L\Tx)$.
An equivalent description in our formalism from \Cref{sec:model} is
$z = r + jx$
whereby the reactance
$x = x\Tx (1 - \f{C_\text{res}}{C})$
is restricted to
$\f{-\Delta}{1-\Delta} x\Tx \leq x \leq \f{\Delta}{1+\Delta} x\Tx$.
Thereby $x\Tx 
= \omega L\Tx / R\Tx$ is the coil Q-factor.
From the loss numbers in \Cref{fig:DiskConstraintOutage} we infer that load value-range constraints can have a significant effect, but mild constraints do not prohibit high data rates in BC.

In the low-SNR regime, the rate will be approximately proportional to $d_\mathrm{max}^{\,2}$ where
$d_\mathrm{max} := \max \{ |\Gamma_l - \Gamma_m| \ \big| \ \Gamma_l , \Gamma_m \in \mathcal{G} \}$
is the maximum pairwise Euclidean distance within $\mathcal{G}$ \cite{Blachman1953}.
Since $d_\mathrm{max} \leq 2$, the channel capacity at low SNR can be approximated as
$(d_\mathrm{max}/2)^2 \SNR \cdot \log_2(e)$ analogous to \Cref{eq:RMaxBoundLowSNR}. In the examples in \Cref{fig:DiskConstraintOutage}, the low-SNR rate loss is minimal because the entire real axis is attainable and thus $d_\mathrm{max} = 2$.

\ifdefined\SingleColumnDraft
\else

\renewcommand\myFigWidth{28mm}
\begin{figure}[t]\centering
\newcommand\myPut[1]{\put(-70,43.5){#1}}
\subfloat[$\Delta =  5\%$]{\includegraphics[width=\myFigWidth]{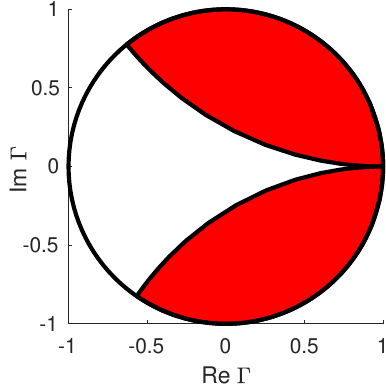}
\myPut{\scriptsize\textcolor{darkRedLoss}{\scriptsize\begin{tabular}{l}%
$-61.2\%$\,area,\\[0mm]$-1.36\unit{bpcu}$%
\end{tabular}}}}
\renewcommand\myPut[1]{\put(-60,45){#1}}
\subfloat[$\Delta = 25\%$]{\includegraphics[width=\myFigWidth]{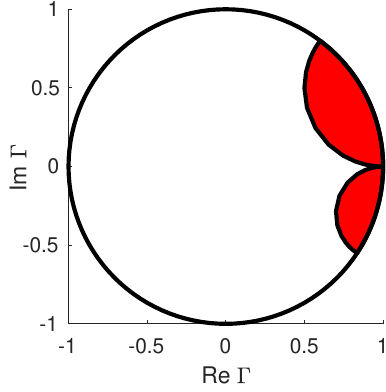}
\myPut{\scriptsize\textcolor{darkRedLoss}{\scriptsize\begin{tabular}{l}%
$-11.0\%$\,area,\\[0mm]$-0.168\unit{bpcu}$%
\end{tabular}}}}
\renewcommand\myPut[1]{\put(-57,45){#1}}
\subfloat[$\Delta = 50\%$]{\includegraphics[width=\myFigWidth]{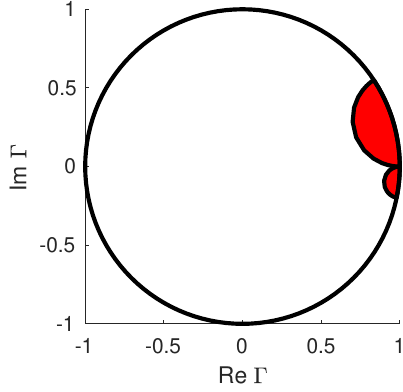}
\myPut{\scriptsize\textcolor{darkRedLoss}{\scriptsize\begin{tabular}{l}%
$-3.85\%$\,area,\\[0mm]$-0.0566\unit{bpcu}$%
\end{tabular}}}}
\caption{Non-realizable values (red) of reflection coefficient $\Gamma$ due to value-range constraints on the load reactance. This experiment assumes an inductive RFID tag.
Each bpcu value is a high-SNR rate loss $\log_2( \pi / \mathrm{area}(\mathcal{G}) )$.}
\label{fig:DiskConstraintOutage}
\end{figure}

\fi

%% file: 04a-APSK.tex
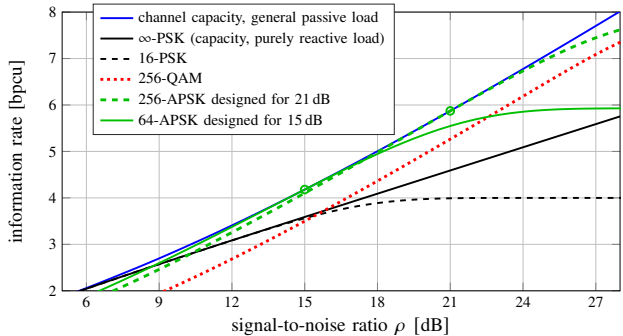
\begin{figure}[t]
\centering
\resizebox{82mm}{!}{\input{F04a_Rates_APSK_PSK}}
\caption{Achievable information rate versus SNR for different signaling strategies and symbol constellations.}
\label{fig:Constellation_rate}
\end{figure}

A simple observation is that the capacity-achieving distribution of a purely-reactive load (a uniform distribution over the unit circle, cf. \Cref{sec:CapacityReact}) is resembled by $M$-PSK modulation $\Gamma_m = \exp(j2\pi\f{m - 1/2}{M})$ with $\SymbProb_m = \f{1}{M}$ and large $M$.
The evaluation in \Cref{fig:Constellation_rate} shows that the rate of 16-PSK (black dashed graph) is indeed very close to channel capacity (black solid graph) below an SNR-threshold of about $\rho < 15\dB$. At higher SNR, the rate is bottlenecked by the small $M$. This problem could be remedied entirely by increasing $M$. This evaluation uses the calculations from \Cref{apdx:RateFinte}.

In the case of a general passive load and high SNR, the capacity-achieving $\r\Gamma$ distribution over the concentric circles in \Cref{fig:CapDistr_Gamma} is naturally resembled by amplitude-and-phase-shift keying (APSK). This long-established scheme \cite{Blachman1953} has been used by optical \cite{ZouOPT2017} and satellite \cite{DeTWC2006} communication systems.
APSK constellation design comes with various degrees of freedom and there are different proposals for settling the details \cite{Kayhan2016,MericCL2015,JovanovicACCESS2020,HuleihelISIT2018,YodaTB2018}.
These nuances are however secondary, since any decent discretization of the optimal continuous distribution (for a certain SNR) will allow for a near-capacity rate.
Nevertheless, we want to provide the interested reader with the following specific near-capacity APSK design for a target design-SNR $\SNR$, tailored to the formalism at hand:
First, choose the optimal number of circles $K$ (cf. \Cref{fig:Radii_PMF_evolution}) and use the constellation size $M = 4K^2$, i.e. $K = \sqrt{M/4}$ circles.
As in \cite{Kayhan2016}, put $M_k = 8(K-k) + 4$ symbols on the $k$-th circle but enforce a constant angular spacing of $2\pi/M_k$. For circles with even $k$, introduce a rotation angle $\pi/M_k$ to increase Euclidean distances between some symbols on neighboring circles.
Set the symbol probabilities to $\SymbProb_m = \CircProb_{k(m)} / M_{k(m)}$ based on the odds $\CircProb_k$ of choosing the circle $k$ that holds symbol $\Gamma_m$. Set the $\CircProb_k$ and the radii $\Radius_k$ according to the capacity-achieving parameter set for the target design-SNR (cf. \Cref{eq:Capacity} and \Cref{fig:Radii_PMF_evolution}).

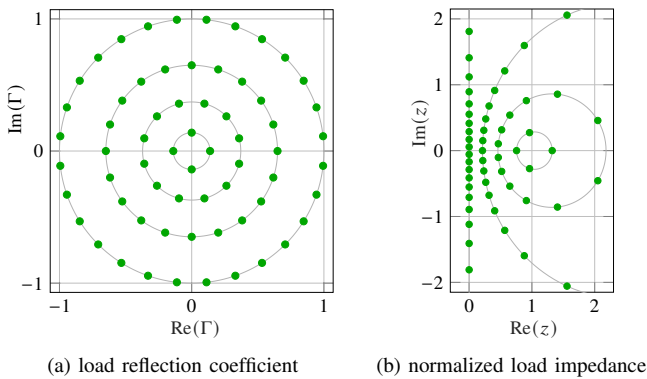
\begin{figure}[t]
\centering
\vspace{-5mm}
\subfloat[load reflection coefficient]{%
\resizebox{!}{45mm}{\input{F04a_64APSK_Gamma.tex}\vspace{-1mm}}
\label{fig:Constellation_Gamma}} \ \
\subfloat[normalized load impedance\!]{
\resizebox{!}{45mm}{\ \ \ \ \input{F04a_64APSK_z.tex}\ \ \ \ \vspace{-1mm}}
\label{fig:Constellation_z}}
\caption{64-APSK is a suitable constellation for near-capacity backscatter information rates in the mid-SNR range.}
\label{fig:Constellation}
\end{figure}

We evaluate 
a 256-APSK designed for $21\dB$ SNR and
a 64-APSK designed for $15\dB$ SNR. 
Indeed, the associated information rates (green graphs in \Cref{fig:Constellation_rate}) are very close to channel capacity near the respective target SNR-values.
They perform significantly better than the 256-QAM benchmark, which provides worse coverage of the complex unit disk than APSK. In detail, the disadvantage of a square-shaped QAM constellation is equivalent to an SNR loss of
$\f{2}{\pi} = -1.96 \dB$ by the property in \Cref{eq:AreaInequality}, whereby $2$ is the area of a square confined in the unit disk (which itself has area $\pi$).

The 64-APSK constellation design is depicted in \Cref{fig:Constellation_Gamma} and the associated normalized load-impedance constellation $z_m = \f{1 + \Gamma_m}{1 - \Gamma_m}$ in \Cref{fig:Constellation_z}.
An accurate realization with a low-cost passive load circuit is left as an interesting circuit-design challenge for future work.

%% file: F04a_Rates_APSK_PSK.tex
\begin{tikzpicture}

\begin{axis}[%
width=102mm,
height=51mm,
at={(0,0)},
scale only axis,
xmin=5,
xmax=28,
xtick={-9, -6, -3, 0, 3, 6, 9, 12, 15, 18, 21, 24, 27, 30},
xlabel style={font=\color{white!15!black}},
xlabel={signal-to-noise ratio $\SNR$ $[\mathrm{dB}]$},
ymin=2,
ymax=8,
ytick={0,1,2,3,4,5,6,7,8,9,10},
yticklabels={0,1,2,3,4,5,6,7,8,9,10},
yminorticks=true,
ylabel style={font=\color{white!15!black}},
ylabel={information rate [bpcu]},
ylabel style={at={(-0.055,0.5)}},
axis background/.style={fill=white},
xmajorgrids,
ymajorgrids,
yminorgrids,
label style={font=\normalsize},
legend style={at={(0.055,1.03)}, anchor=north west, legend cell align=left, align=left, draw=white!15!black} 
]

\addplot [color=blue, line width=1pt]
  table[row sep=crcr]{%
-12	0.0864315365372339\\
-11.5	0.0968938816009866\\
-11	0.108535315924035\\
-10.5	0.121478406979223\\
-10	0.135855771055293\\
-9.5	0.151810236217422\\
-9	0.16949482352194\\
-8.5	0.189072495555821\\
-8	0.210715614271572\\
-7.5	0.234605043429964\\
-7	0.260928825391637\\
-6.5	0.289880358384887\\
-6	0.32165599985449\\
-5.5	0.356452025488289\\
-5	0.394460883742997\\
-4.5	0.435866704131751\\
-4	0.48084004635321\\
-3.5	0.529531918697995\\
-3	0.582067149935895\\
-2.5	0.63853727020936\\
-2	0.698993143127997\\
-1.5	0.763437690917999\\
-1	0.831819161673053\\
-0.5	0.904025492931994\\
0	0.979880414508299\\
0.5	1.05914198601691\\
1	1.14150425662973\\
1.5	1.22660263971832\\
2	1.31402338890148\\
2.5	1.40331722954043\\
3	1.49401674435592\\
3.5	1.58565656558761\\
4	1.67779485739644\\
4.5	1.77003408686118\\
5	1.86252447577785\\
5.5	1.9586982297103\\
6	2.05865655462784\\
6.5	2.16151423207272\\
7	2.26628092318256\\
7.5	2.37190950249353\\
8	2.47809417055567\\
8.5	2.58677238614189\\
9	2.6980725693527\\
9.5	2.81176968358008\\
10	2.9274800746115\\
10.5	3.04469941545486\\
11	3.16373469941759\\
11.5	3.28486380192205\\
12	3.40765530732148\\
12.5	3.53207680492358\\
13	3.65833650083317\\
13.5	3.78626838756433\\
14	3.91577392694257\\
14.5	4.04689071863925\\
15	4.17949655464713\\
15.5	4.31360742956774\\
16	4.44913720761297\\
16.5	4.58605498302582\\
17	4.72430591740325\\
17.5	4.86383595659652\\
18	5.00462349817165\\
18.5	5.14659618924892\\
19	5.28971625348757\\
19.5	5.43393861934742\\
20	5.57921502640108\\
20.5	5.72551423784874\\
21	5.87277857916322\\
21.5	6.02098003948519\\
22	6.17006471615952\\
22.5	6.32000983991334\\
23	6.47076539080284\\
23.5	6.6222949637335\\
24	6.77457096976575\\
24.5	6.92755334339171\\
25	7.08120458375078\\
25.5	7.23550591084472\\
26	7.39041834941432\\
26.5	7.54591308787267\\
27	7.70196209969228\\
27.5	7.85853834282628\\
28	8.01561580406267\\
};
\addlegendentry{channel capacity, general passive load}

\addplot [color=black, solid, line width=1pt]
  table[row sep=crcr]{%
-12	0.0864315365372339\\
-11.5	0.0968938816009866\\
-11	0.108535315924035\\
-10.5	0.121478406979223\\
-10	0.135855771055293\\
-9.5	0.151810236217422\\
-9	0.16949482352194\\
-8.5	0.189072495555821\\
-8	0.210715614271572\\
-7.5	0.234605043429964\\
-7	0.260928825391637\\
-6.5	0.289880358384887\\
-6	0.32165599985449\\
-5.5	0.356452025488289\\
-5	0.394460883742997\\
-4.5	0.435866704131751\\
-4	0.48084004635321\\
-3.5	0.529531918697995\\
-3	0.582067149935895\\
-2.5	0.63853727020936\\
-2	0.698993143127997\\
-1.5	0.763437690917999\\
-1	0.831819161673053\\
-0.5	0.904025492931994\\
0	0.979880414508299\\
0.5	1.05914198601691\\
1	1.14150425662973\\
1.5	1.22660263971832\\
2	1.31402338890148\\
2.5	1.40331722954043\\
3	1.49401674435592\\
3.5	1.58565656558761\\
4	1.67779485739644\\
4.5	1.77003408686118\\
5	1.86203881197189\\
5.5	1.95354829118792\\
6	2.0443822277425\\
6.5	2.1344388954696\\
7	2.22368611366353\\
7.5	2.31214677966389\\
8	2.39988159307618\\
8.5	2.48697192226504\\
9	2.57350535324004\\
9.5	2.65956545798701\\
10	2.7452260803892\\
10.5	2.83054939188579\\
11	2.91558641939581\\
11.5	3.00037874494083\\
12	3.08496043743093\\
12.5	3.16935973200879\\
13	3.25360032066712\\
13.5	3.33770229677027\\
14	3.42168284556869\\
14.5	3.50555676107747\\
15	3.58933684436541\\
15.5	3.67303421818009\\
16	3.75665858028019\\
16.5	3.84021841044758\\
17	3.92372114165475\\
17.5	4.00717330297593\\
18	4.09058063988499\\
18.5	4.17394821622599\\
19	4.25728050116793\\
19.5	4.34058144374078\\
20	4.42385453701641\\
20.5	4.5071028735919\\
21	4.59032919372108\\
21.5	4.67353592719623\\
22	4.75672522988997\\
22.5	4.83989901571424\\
23	4.92305898463008\\
23.5	5.00620664724228\\
24	5.08934334643103\\
24.5	5.1724702764055\\
25	5.25558849950875\\
25.5	5.33869896105639\\
26	5.42180250245275\\
26.5	5.50489987279502\\
27	5.58799173914823\\
27.5	5.67107869565019\\
28	5.75416127158544\\
};
\addlegendentry{$\infty$-PSK (capacity, purely reactive load)}

\addplot [color=black, dashed, line width=1pt]
  table[row sep=crcr]{%
-10	0.135845722114546\\
-9.5	0.151802008096499\\
-9	0.169488483576718\\
-8.5	0.18905923309727\\
-8	0.210713279406907\\
-7.5	0.234604839475176\\
-7	0.26093084734744\\
-6.5	0.289884708058856\\
-6	0.321662785365794\\
-5.5	0.35646136051411\\
-5	0.394472886497251\\
-4.5	0.435881496071827\\
-4	0.480857750615182\\
-3.5	0.529552658265581\\
-3	0.582091045554569\\
-2.5	0.638564438064716\\
-2	0.699023692319533\\
-1.5	0.763471720774281\\
-1	0.831856758974726\\
-0.5	0.904066729132218\\
0	0.97992534306572\\
0.5	1.05919063998102\\
1	1.14155664663861\\
1.5	1.22665875258012\\
2	1.31407658688953\\
2.5	1.40338065078657\\
3	1.49408370386177\\
3.5	1.58572695768212\\
4	1.6778685590099\\
4.5	1.77011096170197\\
5	1.86211871527636\\
5.5	1.9536310745194\\
6	2.04446774277742\\
6.5	2.13452699422969\\
7	2.22377663847292\\
7.5	2.31223952473787\\
8	2.39996940090029\\
8.5	2.48706753473879\\
9	2.57359990047238\\
9.5	2.65965378118065\\
10	2.74528852132688\\
10.5	2.83057258068499\\
11	2.91550269466717\\
11.5	3.00006554518213\\
12	3.08418317779411\\
12.5	3.16769584889137\\
13	3.25033377080114\\
13.5	3.33169132999113\\
14	3.41121248021376\\
14.5	3.48818516827619\\
15	3.56182432289567\\
15.5	3.63121867531651\\
16	3.69550757799288\\
16.5	3.75390476220108\\
17	3.80578243819862\\
17.5	3.85073255303752\\
18	3.88860934585104\\
18.5	3.91954829651083\\
19	3.94394007711519\\
19.5	3.9624850160949\\
20	3.97595164508903\\
20.5	3.9852786251636\\
21	3.99139922475132\\
21.5	3.99518081442782\\
22	3.99736517359146\\
22.5	3.99853560812626\\
23	3.999086028985\\
23.5	3.99934424216306\\
24	3.99947603627656\\
24.5	3.99948582537725\\
25	3.99952520270651\\
25.5	3.99952755155828\\
26	3.99952671649184\\
26.5	3.99952447576636\\
27	3.99952123218142\\
27.5	3.99951702016356\\
28	3.99951180123193\\
28.5	3.99950553847216\\
29	3.99949821594524\\
29.5	3.99948984576612\\
30	3.99948047039416\\
30.5	3.99947016039835\\
31	3.99945900758263\\
31.5	3.99944226571668\\
32	3.99943444174723\\
};
\addlegendentry{16-PSK}

\addplot [color=red, dotted, line width=1.5pt]
  table[row sep=crcr]{%
-10	0.0523893797386918\\
-9.5	0.0588558524921403\\
-9	0.0660663877593919\\
-8.5	0.0740980129008246\\
-8	0.0830583057224974\\
-7.5	0.0930299047948653\\
-7	0.104127039943767\\
-6.5	0.116467971105785\\
-6	0.130180964289316\\
-5.5	0.145404555902071\\
-5	0.162287672290283\\
-4.5	0.180989563638398\\
-4	0.201679506683174\\
-3.5	0.224536226924372\\
-3	0.249746988840512\\
-2.5	0.277506302886468\\
-2	0.308014201772923\\
-1.5	0.341474046840725\\
-1	0.378089839442044\\
-0.5	0.418063033235306\\
0	0.46158887203099\\
0.5	0.508852314566949\\
1	0.560023651713588\\
1.5	0.61525397116015\\
2	0.674669286341361\\
2.5	0.738373310467188\\
3	0.806429983546598\\
3.5	0.878874692091409\\
4	0.955705833215154\\
4.5	1.0368861362753\\
5	1.12234414726832\\
5.5	1.21197725848936\\
6	1.30565611045523\\
6.5	1.40323002549568\\
7	1.50453299644549\\
7.5	1.60938968412206\\
8	1.71762010420804\\
8.5	1.82904815486022\\
9	1.94349707976411\\
9.5	2.06079965331628\\
10	2.18079474144366\\
10.5	2.30333208146032\\
11	2.42826527285893\\
11.5	2.55545852423312\\
12	2.68478277475999\\
12.5	2.81611598845862\\
13	2.94934275103061\\
13.5	3.08435388229005\\
14	3.22104606670461\\
14.5	3.35932095790479\\
15	3.49908756060386\\
15.5	3.64025647381893\\
16	3.78274501788201\\
16.5	3.92647421920332\\
17	4.07136906524633\\
17.5	4.21735822250805\\
18	4.36437375822655\\
18.5	4.51235086230597\\
19	4.66122727062761\\
19.5	4.81094444344249\\
20	4.9614443095996\\
20.5	5.11267185856036\\
21	5.26457323312365\\
21.5	5.41709534675565\\
22	5.57018447362994\\
22.5	5.72378280260564\\
23	5.87782015838527\\
23.5	6.03219673537801\\
24	6.18675100206489\\
24.5	6.34121098826739\\
25	6.49513197359972\\
25.5	6.64783226447937\\
26	6.79834639982816\\
26.5	6.94541424188541\\
27	7.08751966162427\\
27.5	7.22297976196588\\
28	7.35006321160876\\
28.5	7.46711256617669\\
29	7.57268691311134\\
29.5	7.66573401954179\\
30	7.74569750175579\\
30.5	7.8124851680342\\
31	7.86642900127781\\
31.5	7.90836517343358\\
32	7.93968455617391\\
};
\addlegendentry{256-QAM}

\addplot [color=green!75!black, dashed, line width=1.5pt]
  table[row sep=crcr]{%
-10	0.0848574278125458\\
-9.5	0.0950907695928676\\
-9	0.106477900089102\\
-8.5	0.119132564948686\\
-8	0.133205951988807\\
-7.5	0.148818512749449\\
-7	0.166128456437634\\
-6.5	0.185297428058457\\
-6	0.206496991934916\\
-5.5	0.229907981498903\\
-5	0.255719505221599\\
-4.5	0.284127558370587\\
-4	0.315333195167296\\
-3.5	0.349540225717353\\
-3	0.386952417977585\\
-2.5	0.427770208057026\\
-2	0.472186953017522\\
-1.5	0.520384799065881\\
-1	0.572530283585498\\
-0.5	0.628769839334459\\
0	0.689225418965627\\
0.5	0.753990501325144\\
1	0.823126769300881\\
1.5	0.896661752576589\\
2	0.974584542531037\\
2.5	1.05686185827255\\
3	1.1434082764323\\
3.5	1.23412099422443\\
4	1.32886845168219\\
4.5	1.42749869657455\\
5	1.52984492600327\\
5.5	1.63573086470961\\
6	1.74497555782302\\
6.5	1.85739730316936\\
7	1.97281662752825\\
7.5	2.09105836965499\\
8	2.21195031806442\\
8.5	2.33533758042084\\
9	2.46105588328957\\
9.5	2.58895886637215\\
10	2.71890175250829\\
10.5	2.85075770387309\\
11	2.98439003399186\\
11.5	3.11967947326498\\
12	3.25651004352385\\
12.5	3.39477142921372\\
13	3.53435855948972\\
13.5	3.67517115984956\\
14	3.81711327216077\\
14.5	3.96008955713149\\
15	4.10402063473617\\
15.5	4.24881064952799\\
16	4.39437836256923\\
16.5	4.54064040206779\\
17	4.68751340613708\\
17.5	4.83491269247515\\
18	4.9827504795122\\
18.5	5.13093342269217\\
19	5.2793550245957\\
19.5	5.42791130038586\\
20	5.57645299624274\\
20.5	5.72481810828333\\
21	5.87280106604538\\
21.5	6.02014564186318\\
22	6.16653376825364\\
22.5	6.31157510184695\\
23	6.45479201527438\\
23.5	6.59563182819\\
24	6.73343176795364\\
24.5	6.86740498908686\\
25	6.9966982269355\\
25.5	7.12032506453036\\
26	7.23726086488317\\
26.5	7.34646713347254\\
27	7.4469703407575\\
27.5	7.53793211215102\\
28	7.61873281014813\\
28.5	7.68903116545939\\
29	7.7487925256234\\
29.5	7.79831150448237\\
30	7.83819905035287\\
30.5	7.86933494241906\\
31	7.89280680190553\\
31.5	7.90982562158474\\
32	7.92165545486135\\
};
\addlegendentry{256-APSK designed for $21\dB\!\!$}
\addplot [color=green!75!black, line width=1pt,mark=o,forget plot]
  table[row sep=crcr]{%
21	5.87280106604538\\
};

\addplot [color=green!75!black, line width=1pt]
  table[row sep=crcr]{%
-10	0.0979836519136912\\
-9.5	0.109712666557631\\
-9	0.122751611429623\\
-8.5	0.137226585322979\\
-8	0.153303548556964\\
-7.5	0.171114519782486\\
-7	0.190831210457365\\
-6.5	0.212627660384219\\
-6	0.236686854483617\\
-5.5	0.263199589506007\\
-5	0.292362920953627\\
-4.5	0.324378143604787\\
-4	0.359448270004937\\
-3.5	0.397774988917599\\
-3	0.439555111206975\\
-2.5	0.48497654477078\\
-2	0.53421388307292\\
-1.5	0.587423742496343\\
-1	0.644740039438338\\
-0.5	0.706269453915189\\
0	0.772087374991744\\
0.5	0.842234654600862\\
1	0.916715498358251\\
1.5	0.995496782660305\\
2	1.07850475045885\\
2.5	1.16564886925552\\
3	1.25678353743518\\
3.5	1.35175594092835\\
4	1.45039090426861\\
4.5	1.55250129370439\\
5	1.65789363041921\\
5.5	1.7663726901753\\
6	1.87774486257497\\
6.5	1.99182030765846\\
7	2.10841414012418\\
7.5	2.22734693586808\\
8	2.34844067719631\\
8.5	2.47153917360543\\
9	2.59646633688402\\
9.5	2.72306678611894\\
10	2.8511800040712\\
10.5	2.98066474889366\\
11	3.11135453908259\\
11.5	3.24309853768041\\
12	3.37573726319014\\
12.5	3.50910299484191\\
13	3.64301438373101\\
13.5	3.77726920184794\\
14	3.91163521477081\\
14.5	4.04583484135714\\
15	4.17956020577363\\
15.5	4.31241510868479\\
16	4.4439608425769\\
16.5	4.57369015903625\\
17	4.70103314209761\\
17.5	4.82535801919001\\
18	4.94597177690581\\
18.5	5.06212338250764\\
19	5.17300924456445\\
19.5	5.27782355432529\\
20	5.3757412059266\\
20.5	5.4660197879889\\
21	5.54802443096749\\
21.5	5.62128175580628\\
22	5.68551924750556\\
22.5	5.74069189279654\\
23	5.78699086292538\\
23.5	5.82485793882206\\
24	5.85494509167533\\
24.5	5.87807967517405\\
25	5.8952452433295\\
25.5	5.9074716973508\\
26	5.91579909100944\\
26.5	5.9211950873403\\
27	5.92450391175172\\
27.5	5.9264128976602\\
28	5.92744263131906\\
28.5	5.92795838593969\\
29	5.92819641117371\\
29.5	5.92829674930115\\
30	5.92833498136117\\
30.5	5.92834796345013\\
31	5.92835179422533\\
31.5	5.92835272035735\\
32	5.92835294982383\\
};
\addlegendentry{64-APSK designed for $15\dB$}
\addplot [color=green!75!black, line width=1pt,mark=o,forget plot]
  table[row sep=crcr]{%
15	4.17956020577363\\
};

\end{axis}

\end{tikzpicture}%

%% file: F04a_64APSK_Gamma.tex
\begin{tikzpicture}[>=latex']

\begin{axis}[%
width=1.65in,
height=1.65in,
at={(0,0)},
scale only axis,
xmin=-1.07,
xmax=1.07,
xlabel style={font=\color{white!15!black},at={(0.515,-0.077)},anchor=north},
xlabel={$\Re(\Gamma)$},
ymin=-1.07,
ymax=1.07,
ylabel style={at={(-0.065,0.65)}},
ylabel={$\Im(\Gamma)$},
xtick={-1,0,1},
ytick={-1,0,1},
axis background/.style={fill=white},
xmajorgrids,
ymajorgrids,
legend style={legend cell align=left, align=left, draw=white!15!black}
]
\addplot [color=white!70!black, forget plot]
  table[row sep=crcr]{%
1	0\\
0.999390827019096	0.034899496702501\\
0.997564050259824	0.0697564737441253\\
0.994521895368273	0.104528463267653\\
0.99026806874157	0.139173100960065\\
0.984807753012208	0.17364817766693\\
0.978147600733806	0.207911690817759\\
0.970295726275996	0.241921895599668\\
0.961261695938319	0.275637355816999\\
0.951056516295154	0.309016994374947\\
0.939692620785908	0.342020143325669\\
0.927183854566787	0.374606593415912\\
0.913545457642601	0.4067366430758\\
0.898794046299167	0.438371146789077\\
0.882947592858927	0.469471562785891\\
0.866025403784439	0.5\\
0.848048096156426	0.529919264233205\\
0.829037572555042	0.559192903470747\\
0.809016994374947	0.587785252292473\\
0.788010753606722	0.615661475325658\\
0.766044443118978	0.642787609686539\\
0.743144825477394	0.669130606358858\\
0.719339800338651	0.694658370458997\\
0.694658370458997	0.719339800338651\\
0.669130606358858	0.743144825477394\\
0.642787609686539	0.766044443118978\\
0.615661475325658	0.788010753606722\\
0.587785252292473	0.809016994374947\\
0.559192903470747	0.829037572555042\\
0.529919264233205	0.848048096156426\\
0.5	0.866025403784439\\
0.469471562785891	0.882947592858927\\
0.438371146789077	0.898794046299167\\
0.4067366430758	0.913545457642601\\
0.374606593415912	0.927183854566787\\
0.342020143325669	0.939692620785908\\
0.309016994374947	0.951056516295154\\
0.275637355816999	0.961261695938319\\
0.241921895599668	0.970295726275996\\
0.207911690817759	0.978147600733806\\
0.17364817766693	0.984807753012208\\
0.139173100960066	0.99026806874157\\
0.104528463267653	0.994521895368273\\
0.0697564737441255	0.997564050259824\\
0.0348994967025011	0.999390827019096\\
6.12323399573677e-17	1\\
-0.0348994967025007	0.999390827019096\\
-0.0697564737441253	0.997564050259824\\
-0.104528463267653	0.994521895368273\\
-0.139173100960065	0.99026806874157\\
-0.17364817766693	0.984807753012208\\
-0.207911690817759	0.978147600733806\\
-0.241921895599668	0.970295726275996\\
-0.275637355816999	0.961261695938319\\
-0.309016994374947	0.951056516295154\\
-0.342020143325669	0.939692620785908\\
-0.374606593415912	0.927183854566787\\
-0.4067366430758	0.913545457642601\\
-0.438371146789078	0.898794046299167\\
-0.469471562785891	0.882947592858927\\
-0.5	0.866025403784439\\
-0.529919264233205	0.848048096156426\\
-0.559192903470747	0.829037572555042\\
-0.587785252292473	0.809016994374947\\
-0.615661475325658	0.788010753606722\\
-0.642787609686539	0.766044443118978\\
-0.669130606358858	0.743144825477394\\
-0.694658370458997	0.719339800338651\\
-0.719339800338651	0.694658370458997\\
-0.743144825477394	0.669130606358858\\
-0.766044443118978	0.642787609686539\\
-0.788010753606722	0.615661475325658\\
-0.809016994374947	0.587785252292473\\
-0.829037572555042	0.559192903470747\\
-0.848048096156426	0.529919264233205\\
-0.866025403784439	0.5\\
-0.882947592858927	0.469471562785891\\
-0.898794046299167	0.438371146789077\\
-0.913545457642601	0.4067366430758\\
-0.927183854566787	0.374606593415912\\
-0.939692620785908	0.342020143325669\\
-0.951056516295154	0.309016994374948\\
-0.961261695938319	0.275637355817\\
-0.970295726275996	0.241921895599668\\
-0.978147600733806	0.207911690817759\\
-0.984807753012208	0.17364817766693\\
-0.99026806874157	0.139173100960066\\
-0.994521895368273	0.104528463267654\\
-0.997564050259824	0.0697564737441255\\
-0.999390827019096	0.0348994967025007\\
-1	1.22464679914735e-16\\
-0.999390827019096	-0.0348994967025009\\
-0.997564050259824	-0.0697564737441248\\
-0.994521895368273	-0.104528463267653\\
-0.99026806874157	-0.139173100960066\\
-0.984807753012208	-0.17364817766693\\
-0.978147600733806	-0.207911690817759\\
-0.970295726275997	-0.241921895599668\\
-0.961261695938319	-0.275637355816999\\
-0.951056516295154	-0.309016994374948\\
-0.939692620785908	-0.342020143325669\\
-0.927183854566787	-0.374606593415912\\
-0.913545457642601	-0.4067366430758\\
-0.898794046299167	-0.438371146789077\\
-0.882947592858927	-0.469471562785891\\
-0.866025403784439	-0.5\\
-0.848048096156426	-0.529919264233205\\
-0.829037572555042	-0.559192903470747\\
-0.809016994374947	-0.587785252292473\\
-0.788010753606722	-0.615661475325658\\
-0.766044443118978	-0.642787609686539\\
-0.743144825477394	-0.669130606358858\\
-0.719339800338651	-0.694658370458997\\
-0.694658370458997	-0.719339800338651\\
-0.669130606358858	-0.743144825477394\\
-0.642787609686539	-0.766044443118978\\
-0.615661475325658	-0.788010753606722\\
-0.587785252292473	-0.809016994374947\\
-0.559192903470747	-0.829037572555041\\
-0.529919264233205	-0.848048096156426\\
-0.5	-0.866025403784438\\
-0.469471562785891	-0.882947592858927\\
-0.438371146789078	-0.898794046299167\\
-0.4067366430758	-0.913545457642601\\
-0.374606593415912	-0.927183854566787\\
-0.342020143325669	-0.939692620785908\\
-0.309016994374948	-0.951056516295154\\
-0.275637355816999	-0.961261695938319\\
-0.241921895599668	-0.970295726275996\\
-0.20791169081776	-0.978147600733806\\
-0.17364817766693	-0.984807753012208\\
-0.139173100960065	-0.99026806874157\\
-0.104528463267653	-0.994521895368273\\
-0.0697564737441256	-0.997564050259824\\
-0.0348994967025016	-0.999390827019096\\
-1.83697019872103e-16	-1\\
0.0348994967025013	-0.999390827019096\\
0.0697564737441252	-0.997564050259824\\
0.104528463267653	-0.994521895368273\\
0.139173100960065	-0.99026806874157\\
0.17364817766693	-0.984807753012208\\
0.207911690817759	-0.978147600733806\\
0.241921895599667	-0.970295726275997\\
0.275637355816999	-0.961261695938319\\
0.309016994374947	-0.951056516295154\\
0.342020143325668	-0.939692620785909\\
0.374606593415912	-0.927183854566787\\
0.406736643075801	-0.913545457642601\\
0.438371146789077	-0.898794046299167\\
0.46947156278589	-0.882947592858927\\
0.5	-0.866025403784439\\
0.529919264233205	-0.848048096156426\\
0.559192903470746	-0.829037572555042\\
0.587785252292473	-0.809016994374948\\
0.615661475325659	-0.788010753606722\\
0.642787609686539	-0.766044443118978\\
0.669130606358858	-0.743144825477395\\
0.694658370458997	-0.719339800338652\\
0.719339800338651	-0.694658370458998\\
0.743144825477394	-0.669130606358858\\
0.766044443118978	-0.64278760968654\\
0.788010753606722	-0.615661475325659\\
0.809016994374947	-0.587785252292473\\
0.829037572555041	-0.559192903470747\\
0.848048096156425	-0.529919264233206\\
0.866025403784438	-0.5\\
0.882947592858927	-0.469471562785891\\
0.898794046299167	-0.438371146789077\\
0.913545457642601	-0.4067366430758\\
0.927183854566787	-0.374606593415912\\
0.939692620785908	-0.342020143325669\\
0.951056516295154	-0.309016994374948\\
0.961261695938319	-0.275637355817\\
0.970295726275996	-0.241921895599668\\
0.978147600733806	-0.20791169081776\\
0.984807753012208	-0.173648177666931\\
0.99026806874157	-0.139173100960066\\
0.994521895368273	-0.104528463267653\\
0.997564050259824	-0.0697564737441248\\
0.999390827019096	-0.0348994967025008\\
1	-2.44929359829471e-16\\
};
\addplot [color=white!70!black, forget plot]
  table[row sep=crcr]{%
0.649246512705834	0\\
0.648851009272347	0.0226583765292875\\
0.647664980831898	0.0452891473170294\\
0.645689872377447	0.0678647402550238\\
0.642928090274405	0.0903576504607794\\
0.639382999328844	0.112740473787978\\
0.635058918688001	0.134985940214204\\
0.629961116578065	0.157066947065269\\
0.624095803885649	0.178956592035644\\
0.617470126590787	0.200628205964773\\
0.610092157060656	0.22205538532934\\
0.60197088421464	0.243212024411893\\
0.593116202572713	0.264072347106641\\
0.5835389002005	0.284610938323666\\
0.573250645565669	0.304802774953297\\
0.562263973321708	0.324623256352917\\
0.550592269036381	0.344048234319049\\
0.53824975288347	0.363054042508232\\
0.52525146231769	0.381617525270807\\
0.51161323375386	0.399716065862512\\
0.497351683272679	0.417327613999504\\
0.482484186376584	0.434430712723229\\
0.46702885682038	0.451004524542421\\
0.451004524542421	0.46702885682038\\
0.434430712723229	0.482484186376584\\
0.417327613999504	0.497351683272679\\
0.399716065862512	0.51161323375386\\
0.381617525270807	0.52525146231769\\
0.363054042508232	0.53824975288347\\
0.344048234319049	0.550592269036381\\
0.324623256352917	0.562263973321708\\
0.304802774953297	0.573250645565668\\
0.284610938323666	0.5835389002005\\
0.264072347106641	0.593116202572713\\
0.243212024411893	0.60197088421464\\
0.22205538532934	0.610092157060656\\
0.200628205964773	0.617470126590787\\
0.178956592035644	0.624095803885649\\
0.157066947065269	0.629961116578065\\
0.134985940214204	0.635058918688001\\
0.112740473787978	0.639382999328844\\
0.0903576504607795	0.642928090274405\\
0.0678647402550238	0.645689872377447\\
0.0452891473170295	0.647664980831898\\
0.0226583765292876	0.648851009272347\\
3.9754883182139e-17	0.649246512705834\\
-0.0226583765292873	0.648851009272347\\
-0.0452891473170294	0.647664980831898\\
-0.0678647402550237	0.645689872377447\\
-0.0903576504607793	0.642928090274405\\
-0.112740473787978	0.639382999328844\\
-0.134985940214204	0.635058918688001\\
-0.157066947065269	0.629961116578065\\
-0.178956592035644	0.624095803885649\\
-0.200628205964773	0.617470126590787\\
-0.22205538532934	0.610092157060656\\
-0.243212024411893	0.60197088421464\\
-0.264072347106641	0.593116202572713\\
-0.284610938323666	0.5835389002005\\
-0.304802774953297	0.573250645565669\\
-0.324623256352917	0.562263973321708\\
-0.344048234319049	0.550592269036381\\
-0.363054042508232	0.53824975288347\\
-0.381617525270807	0.52525146231769\\
-0.399716065862512	0.51161323375386\\
-0.417327613999504	0.497351683272679\\
-0.434430712723229	0.482484186376584\\
-0.451004524542421	0.46702885682038\\
-0.46702885682038	0.451004524542421\\
-0.482484186376583	0.434430712723229\\
-0.497351683272679	0.417327613999504\\
-0.51161323375386	0.399716065862512\\
-0.525251462317689	0.381617525270807\\
-0.53824975288347	0.363054042508232\\
-0.550592269036381	0.344048234319049\\
-0.562263973321708	0.324623256352917\\
-0.573250645565668	0.304802774953298\\
-0.5835389002005	0.284610938323666\\
-0.593116202572713	0.264072347106641\\
-0.601970884214639	0.243212024411893\\
-0.610092157060656	0.22205538532934\\
-0.617470126590787	0.200628205964773\\
-0.624095803885649	0.178956592035644\\
-0.629961116578065	0.157066947065269\\
-0.635058918688001	0.134985940214204\\
-0.639382999328844	0.112740473787977\\
-0.642928090274405	0.0903576504607796\\
-0.645689872377447	0.067864740255024\\
-0.647664980831898	0.0452891473170295\\
-0.648851009272347	0.0226583765292873\\
-0.649246512705834	7.9509766364278e-17\\
-0.648851009272347	-0.0226583765292874\\
-0.647664980831898	-0.0452891473170291\\
-0.645689872377447	-0.0678647402550236\\
-0.642928090274405	-0.0903576504607794\\
-0.639382999328844	-0.112740473787978\\
-0.635058918688001	-0.134985940214204\\
-0.629961116578065	-0.157066947065269\\
-0.624095803885649	-0.178956592035644\\
-0.617470126590787	-0.200628205964773\\
-0.610092157060656	-0.22205538532934\\
-0.60197088421464	-0.243212024411893\\
-0.593116202572714	-0.26407234710664\\
-0.5835389002005	-0.284610938323665\\
-0.573250645565668	-0.304802774953297\\
-0.562263973321708	-0.324623256352917\\
-0.550592269036381	-0.344048234319049\\
-0.53824975288347	-0.363054042508232\\
-0.52525146231769	-0.381617525270807\\
-0.51161323375386	-0.399716065862512\\
-0.497351683272679	-0.417327613999504\\
-0.482484186376584	-0.434430712723229\\
-0.46702885682038	-0.451004524542421\\
-0.451004524542421	-0.46702885682038\\
-0.434430712723229	-0.482484186376583\\
-0.417327613999504	-0.497351683272679\\
-0.399716065862512	-0.51161323375386\\
-0.381617525270807	-0.525251462317689\\
-0.363054042508232	-0.53824975288347\\
-0.34404823431905	-0.550592269036381\\
-0.324623256352917	-0.562263973321708\\
-0.304802774953297	-0.573250645565669\\
-0.284610938323666	-0.5835389002005\\
-0.264072347106641	-0.593116202572713\\
-0.243212024411893	-0.601970884214639\\
-0.22205538532934	-0.610092157060656\\
-0.200628205964773	-0.617470126590787\\
-0.178956592035644	-0.624095803885649\\
-0.157066947065269	-0.629961116578065\\
-0.134985940214204	-0.635058918688001\\
-0.112740473787978	-0.639382999328844\\
-0.0903576504607791	-0.642928090274405\\
-0.0678647402550238	-0.645689872377447\\
-0.0452891473170296	-0.647664980831898\\
-0.0226583765292879	-0.648851009272347\\
-1.19264649546417e-16	-0.649246512705834\\
0.0226583765292877	-0.648851009272347\\
0.0452891473170293	-0.647664980831898\\
0.0678647402550235	-0.645689872377447\\
0.0903576504607794	-0.642928090274405\\
0.112740473787977	-0.639382999328844\\
0.134985940214203	-0.635058918688001\\
0.157066947065269	-0.629961116578065\\
0.178956592035644	-0.624095803885649\\
0.200628205964773	-0.617470126590787\\
0.222055385329339	-0.610092157060656\\
0.243212024411893	-0.60197088421464\\
0.264072347106641	-0.593116202572713\\
0.284610938323666	-0.5835389002005\\
0.304802774953297	-0.573250645565669\\
0.324623256352917	-0.562263973321708\\
0.344048234319049	-0.550592269036381\\
0.363054042508232	-0.538249752883471\\
0.381617525270807	-0.52525146231769\\
0.399716065862512	-0.51161323375386\\
0.417327613999504	-0.497351683272679\\
0.434430712723228	-0.482484186376584\\
0.451004524542421	-0.46702885682038\\
0.46702885682038	-0.451004524542421\\
0.482484186376584	-0.434430712723228\\
0.497351683272679	-0.417327613999504\\
0.51161323375386	-0.399716065862513\\
0.525251462317689	-0.381617525270807\\
0.53824975288347	-0.363054042508233\\
0.550592269036381	-0.34404823431905\\
0.562263973321708	-0.324623256352917\\
0.573250645565669	-0.304802774953297\\
0.5835389002005	-0.284610938323665\\
0.593116202572713	-0.264072347106641\\
0.601970884214639	-0.243212024411893\\
0.610092157060656	-0.22205538532934\\
0.617470126590787	-0.200628205964773\\
0.624095803885649	-0.178956592035644\\
0.629961116578065	-0.157066947065269\\
0.635058918688001	-0.134985940214204\\
0.639382999328844	-0.112740473787978\\
0.642928090274405	-0.0903576504607797\\
0.645689872377447	-0.0678647402550238\\
0.647664980831898	-0.045289147317029\\
0.648851009272347	-0.0226583765292874\\
0.649246512705834	-1.59019532728556e-16\\
};
\addplot [color=white!70!black, forget plot]
  table[row sep=crcr]{%
0.371321962980076	0\\
0.371095763673012	0.0129589496225893\\
0.370417441340833	0.0259021107612367\\
0.369287822414813	0.0388137141678358\\
0.367708283161608	0.0516780290425158\\
0.365680748006491	0.0644793821991975\\
0.363207687188728	0.077202177160957\\
0.360292113751981	0.0898309131619296\\
0.356937579873373	0.102350204032606\\
0.353148172535709	0.114744796945508\\
0.348928508548115	0.126999590998414\\
0.344283728921173	0.139099655612476\\
0.339219492603382	0.151030248722833\\
0.333741969586612	0.162776834739547\\
0.327857833388909	0.174325102256981\\
0.32157425292385	0.185660981490038\\
0.31489888376632	0.196770661416031\\
0.307839858825375	0.207640606601286\\
0.300405778435546	0.21825757369198\\
0.292605699878657	0.228608627549133\\
0.284449126348918	0.238681157008076\\
0.275945995374752	0.24846289024322\\
0.267106666711444	0.257941909719375\\
0.257941909719376	0.267106666711444\\
0.24846289024322	0.275945995374752\\
0.238681157008077	0.284449126348918\\
0.228608627549133	0.292605699878657\\
0.21825757369198	0.300405778435546\\
0.207640606601286	0.307839858825375\\
0.196770661416031	0.31489888376632\\
0.185660981490038	0.32157425292385\\
0.174325102256981	0.327857833388909\\
0.162776834739547	0.333741969586612\\
0.151030248722833	0.339219492603382\\
0.139099655612476	0.344283728921173\\
0.126999590998414	0.348928508548115\\
0.114744796945509	0.353148172535709\\
0.102350204032606	0.356937579873373\\
0.0898309131619296	0.360292113751981\\
0.077202177160957	0.363207687188728\\
0.0644793821991975	0.365680748006491\\
0.0516780290425158	0.367708283161608\\
0.0388137141678358	0.369287822414813\\
0.0259021107612368	0.370417441340833\\
0.0129589496225894	0.371095763673012\\
2.27369126708331e-17	0.371321962980076\\
-0.0129589496225893	0.371095763673012\\
-0.0259021107612367	0.370417441340833\\
-0.0388137141678358	0.369287822414813\\
-0.0516780290425157	0.367708283161608\\
-0.0644793821991975	0.365680748006491\\
-0.0772021771609569	0.363207687188728\\
-0.0898309131619296	0.360292113751981\\
-0.102350204032606	0.356937579873373\\
-0.114744796945508	0.353148172535709\\
-0.126999590998414	0.348928508548115\\
-0.139099655612476	0.344283728921173\\
-0.151030248722832	0.339219492603382\\
-0.162776834739547	0.333741969586612\\
-0.174325102256981	0.327857833388909\\
-0.185660981490038	0.32157425292385\\
-0.196770661416031	0.31489888376632\\
-0.207640606601286	0.307839858825375\\
-0.21825757369198	0.300405778435546\\
-0.228608627549133	0.292605699878657\\
-0.238681157008077	0.284449126348918\\
-0.24846289024322	0.275945995374752\\
-0.257941909719375	0.267106666711444\\
-0.267106666711444	0.257941909719375\\
-0.275945995374752	0.24846289024322\\
-0.284449126348918	0.238681157008077\\
-0.292605699878657	0.228608627549133\\
-0.300405778435546	0.21825757369198\\
-0.307839858825375	0.207640606601286\\
-0.31489888376632	0.196770661416031\\
-0.32157425292385	0.185660981490038\\
-0.327857833388909	0.174325102256981\\
-0.333741969586612	0.162776834739547\\
-0.339219492603382	0.151030248722833\\
-0.344283728921172	0.139099655612476\\
-0.348928508548115	0.126999590998414\\
-0.353148172535709	0.114744796945509\\
-0.356937579873373	0.102350204032606\\
-0.360292113751981	0.0898309131619296\\
-0.363207687188728	0.077202177160957\\
-0.365680748006491	0.0644793821991975\\
-0.367708283161608	0.0516780290425159\\
-0.369287822414813	0.0388137141678359\\
-0.370417441340833	0.0259021107612368\\
-0.371095763673012	0.0129589496225892\\
-0.371321962980076	4.54738253416662e-17\\
-0.371095763673012	-0.0129589496225893\\
-0.370417441340833	-0.0259021107612365\\
-0.369287822414813	-0.0388137141678357\\
-0.367708283161608	-0.0516780290425158\\
-0.365680748006491	-0.0644793821991976\\
-0.363207687188728	-0.0772021771609569\\
-0.360292113751981	-0.0898309131619295\\
-0.356937579873373	-0.102350204032606\\
-0.353148172535709	-0.114744796945509\\
-0.348928508548115	-0.126999590998414\\
-0.344283728921173	-0.139099655612476\\
-0.339219492603382	-0.151030248722832\\
-0.333741969586612	-0.162776834739547\\
-0.327857833388909	-0.174325102256981\\
-0.32157425292385	-0.185660981490038\\
-0.31489888376632	-0.196770661416031\\
-0.307839858825375	-0.207640606601286\\
-0.300405778435546	-0.21825757369198\\
-0.292605699878657	-0.228608627549133\\
-0.284449126348918	-0.238681157008076\\
-0.275945995374752	-0.24846289024322\\
-0.267106666711444	-0.257941909719376\\
-0.257941909719376	-0.267106666711444\\
-0.24846289024322	-0.275945995374752\\
-0.238681157008077	-0.284449126348918\\
-0.228608627549133	-0.292605699878657\\
-0.21825757369198	-0.300405778435546\\
-0.207640606601286	-0.307839858825375\\
-0.196770661416031	-0.31489888376632\\
-0.185660981490038	-0.32157425292385\\
-0.174325102256981	-0.327857833388909\\
-0.162776834739547	-0.333741969586612\\
-0.151030248722832	-0.339219492603382\\
-0.139099655612476	-0.344283728921172\\
-0.126999590998414	-0.348928508548115\\
-0.114744796945509	-0.353148172535709\\
-0.102350204032606	-0.356937579873373\\
-0.0898309131619296	-0.360292113751981\\
-0.0772021771609571	-0.363207687188728\\
-0.0644793821991975	-0.365680748006491\\
-0.0516780290425156	-0.367708283161608\\
-0.0388137141678358	-0.369287822414813\\
-0.0259021107612368	-0.370417441340833\\
-0.0129589496225896	-0.371095763673012\\
-6.82107380124992e-17	-0.371321962980076\\
0.0129589496225895	-0.371095763673012\\
0.0259021107612367	-0.370417441340833\\
0.0388137141678356	-0.369287822414813\\
0.0516780290425158	-0.367708283161608\\
0.0644793821991974	-0.365680748006491\\
0.0772021771609567	-0.363207687188728\\
0.0898309131619295	-0.360292113751981\\
0.102350204032606	-0.356937579873373\\
0.114744796945508	-0.353148172535709\\
0.126999590998414	-0.348928508548115\\
0.139099655612476	-0.344283728921173\\
0.151030248722833	-0.339219492603382\\
0.162776834739547	-0.333741969586612\\
0.174325102256981	-0.327857833388909\\
0.185660981490038	-0.32157425292385\\
0.196770661416031	-0.31489888376632\\
0.207640606601285	-0.307839858825375\\
0.21825757369198	-0.300405778435546\\
0.228608627549133	-0.292605699878657\\
0.238681157008076	-0.284449126348918\\
0.248462890243219	-0.275945995374752\\
0.257941909719375	-0.267106666711444\\
0.267106666711444	-0.257941909719376\\
0.275945995374752	-0.24846289024322\\
0.284449126348918	-0.238681157008077\\
0.292605699878657	-0.228608627549133\\
0.300405778435546	-0.21825757369198\\
0.307839858825375	-0.207640606601286\\
0.31489888376632	-0.196770661416031\\
0.32157425292385	-0.185660981490038\\
0.327857833388909	-0.174325102256981\\
0.333741969586612	-0.162776834739547\\
0.339219492603382	-0.151030248722833\\
0.344283728921172	-0.139099655612476\\
0.348928508548115	-0.126999590998414\\
0.353148172535709	-0.114744796945509\\
0.356937579873373	-0.102350204032606\\
0.360292113751981	-0.0898309131619296\\
0.363207687188728	-0.0772021771609572\\
0.365680748006491	-0.0644793821991979\\
0.367708283161608	-0.0516780290425159\\
0.369287822414813	-0.0388137141678358\\
0.370417441340833	-0.0259021107612365\\
0.371095763673012	-0.0129589496225893\\
0.371321962980076	-9.09476506833323e-17\\
};
\addplot [color=white!70!black, forget plot]
  table[row sep=crcr]{%
0.13916925794608	0\\
0.139084479794367	0.00485693705877873\\
0.138830248628346	0.00970795668790513\\
0.138406874189532	0.0145471486672034\\
0.137814872294462	0.0193686171866672\\
0.137054964206255	0.0241664880295958\\
0.136128075755862	0.0289349157294224\\
0.135035336214083	0.0336680906915148\\
0.133778076915726	0.0383602462712714\\
0.13235782963758	0.0430056657998894\\
0.130776324732182	0.0475986895492452\\
0.129035489019646	0.0521337216274014\\
0.127137443440133	0.0566052367963387\\
0.12508450046981	0.061007787203608\\
0.122879161303454	0.0653360090196989\\
0.120524112807135	0.06958462897304\\
0.118022224244676	0.0737484707746678\\
0.115376543781905	0.0778224614247378\\
0.112590294772929	0.0818016373931929\\
0.109666871832979	0.0856811506670607\\
0.106609836702586	0.0894562746570102\\
0.103422913908158	0.0931224099559729\\
0.100109986224211	0.0966750899428118\\
0.0966750899428118	0.100109986224211\\
0.0931224099559729	0.103422913908158\\
0.0894562746570102	0.106609836702586\\
0.0856811506670607	0.109666871832979\\
0.0818016373931929	0.112590294772929\\
0.0778224614247378	0.115376543781905\\
0.0737484707746678	0.118022224244676\\
0.06958462897304	0.120524112807135\\
0.0653360090196989	0.122879161303454\\
0.061007787203608	0.12508450046981\\
0.0566052367963387	0.127137443440133\\
0.0521337216274014	0.129035489019646\\
0.0475986895492453	0.130776324732182\\
0.0430056657998894	0.13235782963758\\
0.0383602462712714	0.133778076915726\\
0.0336680906915148	0.135035336214083\\
0.0289349157294224	0.136128075755862\\
0.0241664880295958	0.137054964206255\\
0.0193686171866672	0.137814872294462\\
0.0145471486672034	0.138406874189532\\
0.00970795668790515	0.138830248628346\\
0.00485693705877874	0.139084479794367\\
8.52165931416896e-18	0.13916925794608\\
-0.00485693705877869	0.139084479794367\\
-0.00970795668790514	0.138830248628346\\
-0.0145471486672034	0.138406874189532\\
-0.0193686171866672	0.137814872294462\\
-0.0241664880295958	0.137054964206255\\
-0.0289349157294224	0.136128075755862\\
-0.0336680906915148	0.135035336214083\\
-0.0383602462712714	0.133778076915726\\
-0.0430056657998894	0.13235782963758\\
-0.0475986895492452	0.130776324732182\\
-0.0521337216274014	0.129035489019646\\
-0.0566052367963387	0.127137443440133\\
-0.061007787203608	0.12508450046981\\
-0.0653360090196989	0.122879161303454\\
-0.06958462897304	0.120524112807135\\
-0.0737484707746678	0.118022224244676\\
-0.0778224614247378	0.115376543781905\\
-0.0818016373931929	0.112590294772929\\
-0.0856811506670607	0.109666871832979\\
-0.0894562746570102	0.106609836702586\\
-0.0931224099559729	0.103422913908158\\
-0.0966750899428118	0.100109986224211\\
-0.100109986224211	0.0966750899428118\\
-0.103422913908158	0.0931224099559729\\
-0.106609836702586	0.0894562746570102\\
-0.109666871832979	0.0856811506670607\\
-0.112590294772929	0.0818016373931929\\
-0.115376543781905	0.0778224614247378\\
-0.118022224244676	0.0737484707746678\\
-0.120524112807135	0.06958462897304\\
-0.122879161303454	0.065336009019699\\
-0.12508450046981	0.061007787203608\\
-0.127137443440133	0.0566052367963387\\
-0.129035489019646	0.0521337216274014\\
-0.130776324732182	0.0475986895492453\\
-0.13235782963758	0.0430056657998894\\
-0.133778076915726	0.0383602462712715\\
-0.135035336214083	0.0336680906915148\\
-0.136128075755862	0.0289349157294224\\
-0.137054964206255	0.0241664880295957\\
-0.137814872294462	0.0193686171866672\\
-0.138406874189532	0.0145471486672035\\
-0.138830248628346	0.00970795668790516\\
-0.139084479794367	0.00485693705877869\\
-0.13916925794608	1.70433186283379e-17\\
-0.139084479794367	-0.00485693705877872\\
-0.138830248628346	-0.00970795668790507\\
-0.138406874189532	-0.0145471486672034\\
-0.137814872294462	-0.0193686171866672\\
-0.137054964206255	-0.0241664880295958\\
-0.136128075755862	-0.0289349157294223\\
-0.135035336214083	-0.0336680906915148\\
-0.133778076915726	-0.0383602462712714\\
-0.13235782963758	-0.0430056657998895\\
-0.130776324732182	-0.0475986895492452\\
-0.129035489019646	-0.0521337216274014\\
-0.127137443440133	-0.0566052367963387\\
-0.12508450046981	-0.061007787203608\\
-0.122879161303454	-0.0653360090196989\\
-0.120524112807135	-0.06958462897304\\
-0.118022224244676	-0.0737484707746678\\
-0.115376543781905	-0.0778224614247378\\
-0.112590294772929	-0.0818016373931929\\
-0.109666871832979	-0.0856811506670607\\
-0.106609836702586	-0.0894562746570102\\
-0.103422913908158	-0.0931224099559729\\
-0.100109986224211	-0.0966750899428118\\
-0.0966750899428118	-0.100109986224211\\
-0.0931224099559729	-0.103422913908158\\
-0.0894562746570102	-0.106609836702586\\
-0.0856811506670607	-0.109666871832979\\
-0.0818016373931929	-0.112590294772929\\
-0.0778224614247378	-0.115376543781905\\
-0.0737484707746678	-0.118022224244676\\
-0.0695846289730401	-0.120524112807135\\
-0.0653360090196989	-0.122879161303454\\
-0.0610077872036081	-0.12508450046981\\
-0.0566052367963387	-0.127137443440133\\
-0.0521337216274014	-0.129035489019646\\
-0.0475986895492453	-0.130776324732182\\
-0.0430056657998894	-0.13235782963758\\
-0.0383602462712714	-0.133778076915726\\
-0.0336680906915148	-0.135035336214083\\
-0.0289349157294224	-0.136128075755862\\
-0.0241664880295958	-0.137054964206255\\
-0.0193686171866671	-0.137814872294462\\
-0.0145471486672034	-0.138406874189532\\
-0.00970795668790517	-0.138830248628346\\
-0.00485693705877882	-0.139084479794367\\
-2.55649779425069e-17	-0.13916925794608\\
0.00485693705877877	-0.139084479794367\\
0.00970795668790512	-0.138830248628346\\
0.0145471486672033	-0.138406874189532\\
0.0193686171866672	-0.137814872294462\\
0.0241664880295957	-0.137054964206255\\
0.0289349157294223	-0.136128075755862\\
0.0336680906915148	-0.135035336214083\\
0.0383602462712714	-0.133778076915726\\
0.0430056657998894	-0.13235782963758\\
0.0475986895492452	-0.130776324732182\\
0.0521337216274014	-0.129035489019646\\
0.0566052367963388	-0.127137443440133\\
0.061007787203608	-0.12508450046981\\
0.0653360090196989	-0.122879161303454\\
0.06958462897304	-0.120524112807135\\
0.0737484707746678	-0.118022224244676\\
0.0778224614247377	-0.115376543781905\\
0.0818016373931929	-0.112590294772929\\
0.0856811506670607	-0.109666871832979\\
0.0894562746570102	-0.106609836702586\\
0.0931224099559728	-0.103422913908158\\
0.0966750899428117	-0.100109986224212\\
0.100109986224211	-0.0966750899428118\\
0.103422913908158	-0.0931224099559728\\
0.106609836702586	-0.0894562746570102\\
0.109666871832979	-0.0856811506670608\\
0.112590294772929	-0.0818016373931929\\
0.115376543781905	-0.0778224614247379\\
0.118022224244676	-0.073748470774668\\
0.120524112807135	-0.0695846289730401\\
0.122879161303454	-0.0653360090196989\\
0.12508450046981	-0.061007787203608\\
0.127137443440133	-0.0566052367963387\\
0.129035489019646	-0.0521337216274014\\
0.130776324732182	-0.0475986895492452\\
0.13235782963758	-0.0430056657998894\\
0.133778076915726	-0.0383602462712715\\
0.135035336214083	-0.0336680906915148\\
0.136128075755862	-0.0289349157294225\\
0.137054964206255	-0.0241664880295959\\
0.137814872294462	-0.0193686171866672\\
0.138406874189532	-0.0145471486672034\\
0.138830248628346	-0.00970795668790506\\
0.139084479794367	-0.00485693705877871\\
0.13916925794608	-3.40866372566758e-17\\
};
\addplot [color=green!65!black, draw=none, mark size=1.5pt, mark=*, mark options={solid, green!65!black}, forget plot]
  table[row sep=crcr]{%
0.993712209893243	0.111964476103308\\
0.943883330308368	0.330279061955167\\
0.846724199228284	0.532032076515337\\
0.707106781186548	0.707106781186547\\
0.532032076515337	0.846724199228284\\
0.330279061955167	0.943883330308367\\
0.111964476103308	0.993712209893243\\
-0.111964476103308	0.993712209893243\\
-0.330279061955167	0.943883330308367\\
0.649246512705834	0\\
0.617470126590787	0.200628205964773\\
0.52525146231769	0.381617525270807\\
-0.532032076515336	0.846724199228284\\
0.381617525270807	0.52525146231769\\
0.200628205964773	0.617470126590787\\
3.9754883182139e-17	0.649246512705834\\
-0.707106781186547	0.707106781186548\\
-0.200628205964773	0.617470126590787\\
0.13916925794608	0\\
0.358669473910808	0.0961051958840967\\
-0.846724199228284	0.532032076515337\\
-0.381617525270807	0.52525146231769\\
8.52165931416896e-18	0.13916925794608\\
0.262564278026712	0.262564278026712\\
-0.943883330308368	0.330279061955167\\
-0.525251462317689	0.381617525270807\\
0.0961051958840967	0.358669473910808\\
-0.0961051958840967	0.358669473910808\\
-0.993712209893243	0.111964476103308\\
-0.617470126590787	0.200628205964773\\
-0.262564278026712	0.262564278026712\\
-0.358669473910808	0.0961051958840969\\
-0.993712209893243	-0.111964476103308\\
-0.943883330308368	-0.330279061955167\\
-0.846724199228284	-0.532032076515336\\
-0.707106781186548	-0.707106781186547\\
-0.532032076515337	-0.846724199228284\\
-0.330279061955167	-0.943883330308367\\
-0.111964476103308	-0.993712209893243\\
0.111964476103308	-0.993712209893243\\
0.330279061955167	-0.943883330308368\\
-0.649246512705834	7.9509766364278e-17\\
-0.617470126590787	-0.200628205964772\\
-0.52525146231769	-0.381617525270807\\
0.532032076515336	-0.846724199228284\\
-0.381617525270807	-0.525251462317689\\
-0.200628205964773	-0.617470126590787\\
-1.19264649546417e-16	-0.649246512705834\\
0.707106781186547	-0.707106781186548\\
0.200628205964773	-0.617470126590787\\
-0.13916925794608	1.70433186283379e-17\\
-0.358669473910808	-0.0961051958840968\\
0.846724199228284	-0.532032076515337\\
0.381617525270807	-0.52525146231769\\
-2.55649779425069e-17	-0.13916925794608\\
-0.262564278026712	-0.262564278026712\\
0.943883330308367	-0.330279061955167\\
0.525251462317689	-0.381617525270807\\
-0.0961051958840967	-0.358669473910808\\
0.0961051958840966	-0.358669473910809\\
0.993712209893243	-0.111964476103308\\
0.617470126590787	-0.200628205964773\\
0.262564278026712	-0.262564278026712\\
0.358669473910808	-0.096105195884097\\
};

\end{axis}
\end{tikzpicture}%

%% file: F04a_64APSK_z.tex
%
%
\begin{tikzpicture}

\begin{axis}[%
width=0.95in,
height=1.65in,
at={(0,0)},
scale only axis,
unbounded coords=jump,
xmin=-0.3,
xmax=2.3,
ymin=-2.15,
ymax=2.15,
xlabel style={font=\color{white!15!black},at={(0.515,-0.077)},anchor=north},
xlabel={$\Re(z)$},
xtick={0, 1, 2},
ytick={-2, -1,  0,  1, 2},
ylabel style={font=\color{white!15!black},at={(-0.09,0.6)}},
ylabel={$\Im(z)$},
axis background/.style={fill=white},
xmajorgrids,
ymajorgrids,
legend style={legend cell align=left, align=left, draw=white!15!black}
]
\addplot [color=white!70!black, forget plot]
  table[row sep=crcr]{%
nan	0\\
-5.24738234479477e-14	57.2899616307594\\
2.04665041035633e-14	28.6362532829156\\
8.87093286382056e-15	19.0811366877282\\
-4.76307868370449e-15	14.3006662567119\\
2.85522912881897e-15	11.4300523027613\\
-1.9805749653287e-15	9.51436445422259\\
-1.13025511560763e-16	8.14434642797459\\
-5.92473547088374e-16	7.11536972238421\\
1.05145084550313e-15	6.31375151467504\\
-4.72229074145823e-16	5.67128181961771\\
-2.85580115997903e-16	5.14455401597031\\
5.22318866988775e-16	4.70463010947845\\
-3.60667818467899e-16	4.33147587428416\\
-4.45286133913907e-16	4.01078093353584\\
-0	3.73205080756888\\
1.93590390598126e-16	3.48741444384091\\
1.81568794359829e-16	3.27085261848414\\
-1.70845756274088e-16	3.07768353717525\\
1.61216089238127e-16	2.90421087767582\\
3.05031269071623e-16	2.74747741945462\\
-1.44611493041114e-16	2.6050890646938\\
-1.3739492063049e-16	2.4750868534163\\
-1.30776076957759e-16	2.35585236582375\\
-0	2.24603677390422\\
-0	2.14450692050956\\
-2.27629453239874e-16	2.0503038415793\\
-0	1.96261050550515\\
-0	1.88072646534633\\
-2.00289535542566e-16	1.80404775527142\\
-1.92296268638356e-16	1.73205080756888\\
-0	1.66427948235052\\
-8.88364120622015e-17	1.60033452904105\\
8.54796768820267e-17	1.53986496381458\\
8.22986661326708e-17	1.48256096851274\\
7.92781399828809e-17	1.42814800674211\\
7.64045449392444e-17	1.37638192047117\\
-0	1.32704482162041\\
-1.42102067023706e-16	1.27994163219308\\
6.85505628114675e-17	1.23489715653505\\
6.6155613908892e-17	1.19175359259421\\
6.38582746249067e-17	1.15036840722101\\
1.23302758540022e-16	1.11061251482919\\
5.9528421637849e-17	1.07236871002468\\
-0	1.03553031379057\\
-5.55111512312578e-17	1\\
-5.55111512312578e-17	0.965688774807074\\
1.11022302462516e-16	0.932515086137662\\
-5.55111512312578e-17	0.90040404429784\\
-5.55111512312578e-17	0.869286737816227\\
0	0.83909963117728\\
0	0.809784033195007\\
0	0.781285626506717\\
0	0.753554050102794\\
0	0.726542528005361\\
0	0.70020753820971\\
-5.55111512312578e-17	0.674508516842427\\
0	0.649407593197511\\
0	0.624869351909327\\
0	0.600860619027561\\
5.55111512312578e-17	0.577350269189626\\
-2.77555756156289e-17	0.554309051452769\\
0	0.531709431661479\\
0	0.509525449494429\\
-5.55111512312578e-17	0.487732588565861\\
0	0.466307658154999\\
0	0.445228685308536\\
-2.77555756156289e-17	0.424474816209605\\
0	0.404026225835157\\
-2.77555756156289e-17	0.383864035035416\\
1.38777878078145e-17	0.363970234266202\\
-1.38777878078145e-17	0.344327613289665\\
0	0.324919696232906\\
1.38777878078145e-17	0.30573068145866\\
1.38777878078145e-17	0.286745385758808\\
0	0.267949192431123\\
6.93889390390723e-18	0.249328002843181\\
6.93889390390723e-18	0.230868191125563\\
2.08166817117217e-17	0.212556561670022\\
6.93889390390723e-18	0.194380309137719\\
6.93889390390723e-18	0.176326980708465\\
6.93889390390723e-18	0.158384440324536\\
2.42861286636753e-17	0.140540834702392\\
-1.73472347597681e-18	0.122784560902905\\
-2.25514051876985e-17	0.105104235265676\\
2.60208521396521e-17	0.087488663525924\\
1.12757025938492e-17	0.0699268119435106\\
9.97465998686664e-18	0.0524077792830413\\
1.7130394325271e-17	0.0349207694917478\\
-1.1221492485225e-17	0.0174550649282175\\
-3.74939945665464e-33	6.12323399573677e-17\\
-1.47451495458029e-17	-0.0174550649282176\\
-1.40946282423116e-17	-0.0349207694917475\\
-9.97465998686664e-18	-0.052407779283041\\
2.60208521396521e-17	-0.0699268119435105\\
9.54097911787244e-18	-0.0874886635259241\\
3.46944695195361e-18	-0.105104235265676\\
-2.77555756156289e-17	-0.122784560902904\\
1.04083408558608e-17	-0.140540834702391\\
-2.77555756156289e-17	-0.158384440324536\\
-6.93889390390723e-18	-0.176326980708465\\
-6.93889390390723e-18	-0.194380309137718\\
-6.93889390390723e-18	-0.212556561670022\\
6.93889390390723e-18	-0.230868191125563\\
6.93889390390723e-18	-0.249328002843181\\
0	-0.267949192431123\\
1.38777878078145e-17	-0.286745385758808\\
-1.38777878078145e-17	-0.30573068145866\\
1.38777878078145e-17	-0.324919696232906\\
2.77555756156289e-17	-0.344327613289665\\
0	-0.363970234266202\\
-2.77555756156289e-17	-0.383864035035416\\
0	-0.404026225835157\\
2.77555756156289e-17	-0.424474816209605\\
0	-0.445228685308536\\
0	-0.466307658154998\\
2.77555756156289e-17	-0.487732588565862\\
0	-0.509525449494429\\
0	-0.531709431661478\\
0	-0.554309051452769\\
-5.55111512312578e-17	-0.577350269189626\\
-5.55111512312578e-17	-0.60086061902756\\
0	-0.624869351909327\\
0	-0.649407593197511\\
0	-0.674508516842426\\
-5.55111512312578e-17	-0.700207538209709\\
0	-0.726542528005361\\
0	-0.753554050102794\\
0	-0.781285626506717\\
-5.55111512312578e-17	-0.809784033195007\\
0	-0.83909963117728\\
0	-0.869286737816227\\
-5.55111512312578e-17	-0.90040404429784\\
5.55111512312578e-17	-0.932515086137661\\
0	-0.965688774807073\\
0	-1\\
0	-1.03553031379057\\
-5.9528421637849e-17	-1.07236871002468\\
6.16513792700109e-17	-1.11061251482919\\
-6.38582746249066e-17	-1.15036840722101\\
6.6155613908892e-17	-1.19175359259421\\
6.85505628114674e-17	-1.23489715653505\\
-1.42102067023706e-16	-1.27994163219308\\
7.36657857836282e-17	-1.32704482162041\\
0	-1.37638192047117\\
7.92781399828809e-17	-1.42814800674211\\
8.22986661326708e-17	-1.48256096851274\\
0	-1.53986496381458\\
-8.88364120622015e-17	-1.60033452904105\\
0	-1.66427948235052\\
-1.92296268638356e-16	-1.73205080756888\\
2.00289535542566e-16	-1.80404775527142\\
0	-1.88072646534633\\
1.08946768579152e-16	-1.96261050550515\\
0	-2.0503038415793\\
0	-2.14450692050956\\
1.24680087027163e-16	-2.24603677390421\\
1.30776076957759e-16	-2.35585236582375\\
-1.3739492063049e-16	-2.47508685341629\\
1.44611493041114e-16	-2.6050890646938\\
1.52515634535811e-16	-2.74747741945462\\
-1.61216089238126e-16	-2.90421087767582\\
-3.41691512548176e-16	-3.07768353717525\\
-1.81568794359829e-16	-3.27085261848414\\
1.93590390598125e-16	-3.4874144438409\\
2.07170436781694e-16	-3.73205080756887\\
-4.45286133913907e-16	-4.01078093353584\\
4.80890424623865e-16	-4.33147587428416\\
-5.22318866988775e-16	-4.70463010947846\\
-4.28370173996854e-16	-5.14455401597031\\
1.57409691381941e-16	-5.67128181961771\\
-3.50483615167709e-16	-6.31375151467504\\
5.92473547088373e-16	-7.11536972238419\\
-1.24328062716839e-15	-8.14434642797459\\
-2.24465162737252e-15	-9.51436445422256\\
-7.9311920244971e-16	-11.4300523027613\\
1.9846161182102e-16	-14.3006662567119\\
9.66534476207315e-15	-19.0811366877282\\
-9.53778832010726e-15	-28.6362532829158\\
-4.41257151721381e-14	-57.2899616307597\\
nan	0\\
};
\addplot [color=white!70!black, forget plot]
  table[row sep=crcr]{%
4.70201030766277	0\\
4.67197192930518	0.365991869637532\\
4.58415127025054	0.717786866931589\\
4.44500717593989	1.04293941627194\\
4.26402970195866	1.33207161602342\\
4.05224910838217	1.57949557872703\\
3.82078435358214	1.78313196803917\\
3.57972312780359	1.94391228142398\\
3.33745387298521	2.06493029672143\\
3.10042096722724	2.15057740464389\\
2.87319156356013	2.20581109273682\\
2.65870446583615	2.23561765853505\\
2.45859295417093	2.24466731010007\\
2.27350893542847	2.23712718948239\\
2.10340886496734	2.2165883182426\\
1.94778595818631	2.18606607283233\\
1.80584782759174	2.14804268894303\\
1.6766460002214	2.10452979031895\\
1.55916638765564	2.05713691795909\\
1.45238979897535	2.00713792870003\\
1.35533041271651	1.95553111113407\\
1.26705857727167	1.9030913598828\\
1.18671280744496	1.85041419719072\\
1.11350456244722	1.79795219416276\\
1.04671837151349	1.74604468530562\\
0.985709100048292	1.69494176680968\\
0.929897581167914	1.64482353460575\\
0.878765430293869	1.59581542158991\\
0.831849574224937	1.54800037436301\\
0.788736828372058	1.50142848997428\\
0.749058721469456	1.45612462311894\\
0.712486677353512	1.41209437832488\\
0.678727604705366	1.36932882072663\\
0.64751990824899	1.32780817217801\\
0.618629911921954	1.28750470505128\\
0.591848671209438	1.24838500224524\\
0.566989144853011	1.21041171687394\\
0.543883693258096	1.17354493721443\\
0.522381870599768	1.13774324037035\\
0.502348478841893	1.10296450060263\\
0.483661853962423	1.06916650444574\\
0.466212357177438	1.03630741380316\\
0.449901046598026	1.00434610959137\\
0.434638507364784	0.973242441689999\\
0.420343820783651	0.942957405575831\\
0.406943655279904	0.913453261764076\\
0.394371464070694	0.884693610815908\\
0.382566776325566	0.856643434006816\\
0.371474570245675	0.829269107639382\\
0.361044717958862	0.802538397309878\\
0.351231493415829	0.776420437109421\\
0.341993135599768	0.750885697685265\\
0.333291460345465	0.725905946249321\\
0.325091514920723	0.701454200954583\\
0.317361270268286	0.677504681530217\\
0.3100713464543	0.654032757644829\\
0.303194767432257	0.631014896132614\\
0.296706741720335	0.608428607951123\\
0.29058446601498	0.586252395528397\\
0.284806949132852	0.564465700989965\\
0.279354853994468	0.543048855623966\\
0.274210355642403	0.521983030838361\\
0.269357013530352	0.501250190782135\\
0.264779656531617	0.480833046738231\\
0.260464279300807	0.460715013345898\\
0.256397948784344	0.44088016667151\\
0.252568719816899	0.421313204117251\\
0.248965558864808	0.401999406134592\\
0.245578275086186	0.382924599692721\\
0.242397457972859	0.364075123439848\\
0.239414420923129	0.345437794486543\\
0.236621150168216	0.32699987673441\\
0.234010258540393	0.308749050669579\\
0.231574943628332	0.290673384538477\\
0.229308949916115	0.27276130682251\\
0.22720653454746	0.255001579928476\\
0.225262436396851	0.23738327501243\\
0.223471848164885	0.219895747856089\\
0.221830391247056	0.202528615716661\\
0.220334093153642	0.185271735072922\\
0.218979367283949	0.168115180192481\\
0.21776299488117	0.151049222447315\\
0.216682109014931	0.134064310306772\\
0.21573418045745	0.117151049939297\\
0.214917005336441	0.100300186356072\\
0.214228694463682	0.0835025850315858\\
0.213667664252696	0.0667492139377686\\
0.213232629152484	0.0500311259298247\\
0.212922595536923	0.033339441423153\\
0.212736857001306	0.0166653313018298\\
0.212674991028905	5.84627613092968e-17\\
0.212736857001306	-0.0166653313018299\\
0.212922595536923	-0.0333394414231526\\
0.213232629152484	-0.0500311259298244\\
0.213667664252696	-0.0667492139377685\\
0.214228694463682	-0.0835025850315859\\
0.214917005336441	-0.100300186356072\\
0.21573418045745	-0.117151049939296\\
0.216682109014931	-0.134064310306772\\
0.21776299488117	-0.151049222447315\\
0.218979367283949	-0.168115180192481\\
0.220334093153642	-0.185271735072922\\
0.221830391247056	-0.202528615716661\\
0.223471848164885	-0.219895747856089\\
0.225262436396851	-0.23738327501243\\
0.22720653454746	-0.255001579928476\\
0.229308949916115	-0.27276130682251\\
0.231574943628332	-0.290673384538477\\
0.234010258540393	-0.308749050669578\\
0.236621150168216	-0.326999876734409\\
0.239414420923129	-0.345437794486543\\
0.242397457972859	-0.364075123439848\\
0.245578275086186	-0.382924599692721\\
0.248965558864808	-0.401999406134591\\
0.252568719816899	-0.421313204117251\\
0.256397948784344	-0.440880166671509\\
0.260464279300807	-0.460715013345898\\
0.264779656531617	-0.480833046738231\\
0.269357013530352	-0.501250190782135\\
0.274210355642403	-0.521983030838361\\
0.279354853994468	-0.543048855623966\\
0.284806949132852	-0.564465700989965\\
0.29058446601498	-0.586252395528397\\
0.296706741720335	-0.608428607951123\\
0.303194767432256	-0.631014896132613\\
0.3100713464543	-0.654032757644829\\
0.317361270268286	-0.677504681530217\\
0.325091514920723	-0.701454200954583\\
0.333291460345465	-0.725905946249321\\
0.341993135599768	-0.750885697685265\\
0.351231493415829	-0.776420437109421\\
0.361044717958862	-0.802538397309878\\
0.371474570245675	-0.829269107639382\\
0.382566776325566	-0.856643434006815\\
0.394371464070694	-0.884693610815907\\
0.406943655279904	-0.913453261764076\\
0.420343820783651	-0.942957405575831\\
0.434638507364784	-0.973242441689999\\
0.449901046598026	-1.00434610959137\\
0.466212357177438	-1.03630741380316\\
0.483661853962423	-1.06916650444573\\
0.502348478841892	-1.10296450060263\\
0.522381870599767	-1.13774324037035\\
0.543883693258096	-1.17354493721443\\
0.566989144853011	-1.21041171687394\\
0.591848671209438	-1.24838500224524\\
0.618629911921954	-1.28750470505128\\
0.64751990824899	-1.32780817217801\\
0.678727604705366	-1.36932882072663\\
0.712486677353512	-1.41209437832488\\
0.749058721469456	-1.45612462311894\\
0.788736828372058	-1.50142848997428\\
0.831849574224937	-1.54800037436301\\
0.878765430293869	-1.59581542158991\\
0.929897581167914	-1.64482353460575\\
0.985709100048292	-1.69494176680968\\
1.04671837151349	-1.74604468530562\\
1.11350456244722	-1.79795219416276\\
1.18671280744496	-1.85041419719072\\
1.26705857727168	-1.9030913598828\\
1.35533041271651	-1.95553111113407\\
1.45238979897535	-2.00713792870003\\
1.55916638765564	-2.05713691795909\\
1.6766460002214	-2.10452979031895\\
1.80584782759174	-2.14804268894303\\
1.94778595818631	-2.18606607283233\\
2.10340886496734	-2.2165883182426\\
2.27350893542847	-2.23712718948239\\
2.45859295417093	-2.24466731010007\\
2.65870446583615	-2.23561765853505\\
2.87319156356013	-2.20581109273682\\
3.10042096722724	-2.15057740464389\\
3.3374538729852	-2.06493029672143\\
3.57972312780359	-1.94391228142398\\
3.82078435358213	-1.78313196803917\\
4.05224910838216	-1.57949557872703\\
4.26402970195865	-1.33207161602342\\
4.44500717593989	-1.04293941627194\\
4.58415127025054	-0.717786866931584\\
4.67197192930518	-0.36599186963753\\
4.70201030766277	-2.58509479615664e-15\\
};
\addplot [color=white!70!black, forget plot]
  table[row sep=crcr]{%
2.18127862312552	0\\
2.17878472326781	0.0655007689731416\\
2.17134013689902	0.13047439506974\\
2.15905483682386	0.194406665725809\\
2.14210764819703	0.256808567903064\\
2.12073987025765	0.317227292418165\\
2.09524694401921	0.375255476741131\\
2.0659686875247	0.430538332959436\\
2.033278674141	0.482778469817829\\
1.99757333156882	0.531738380655843\\
1.95926129527241	0.57724071640573\\
1.91875347013776	0.619166582287041\\
1.87645415122004	0.657452181224573\\
1.83275344173291	0.692084174290164\\
1.78802109584736	0.72309414104863\\
1.74260181465167	0.75055250620881\\
1.69681194183321	0.774562260862194\\
1.65093744418178	0.795252754713443\\
1.60523302110024	0.812773777370572\\
1.55992216516646	0.827290088018696\\
1.51519798940195	0.838976498137266\\
1.47122464268467	0.848013564236132\\
1.42813914908317	0.854583908368418\\
1.38605352651112	0.858869153788823\\
1.34505706227427	0.861047441116101\\
1.30521864569712	0.861291475769684\\
1.26658907957522	0.859767049065187\\
1.22920331173959	0.856631971884776\\
1.19308254501952	0.852035360059756\\
1.15823619816071	0.846117213426245\\
1.12466370184438	0.839008235028677\\
1.09235612305874	0.830829842422513\\
1.06129761798153	0.821694328913289\\
1.03146671856365	0.811705138463575\\
1.00283746148251	0.800957223629385\\
0.975380370364441	0.789537461081803\\
0.949063303431193	0.77752510393409\\
0.923852179240612	0.764992254192193\\
0.899711593163562	0.75200434217979\\
0.876605336827073	0.738620602786003\\
0.854496832085235	0.724894540888668\\
0.833349490252784	0.710874380370424\\
0.813127006426119	0.696603492823351\\
0.793793597777532	0.68212080338412\\
0.775314193779647	0.667461172205845\\
0.757654585425199	0.652655750901016\\
0.740781539669399	0.637732313922676\\
0.724662884548187	0.622715565324069\\
0.709267569720524	0.607627421680698\\
0.694565706547207	0.592487272198991\\
0.680528591251149	0.577312217194278\\
0.667128714201079	0.562117286215265\\
0.654339757917851	0.546915637137558\\
0.642136586015084	0.531718737556845\\
0.630495224948431	0.516536529792733\\
0.619392840155325	0.501377580774729\\
0.608807707914574	0.486249218028372\\
0.598719184038071	0.471157652916971\\
0.589107670320822	0.456108092226363\\
0.579954579516609	0.441104839109405\\
0.571242299471275	0.42615138433572\\
0.562954156930874	0.411250488721973\\
0.55507438144475	0.396404257549913\\
0.547588069701777	0.381614207714231\\
0.54048115056914	0.366881328280544\\
0.533740351045469	0.352206135075745\\
0.527353163292079	0.337588719878811\\
0.521307812866193	0.32302879472981\\
0.515593228247094	0.308525731828413\\
0.510199011719082	0.294078599450418\\
0.505115411653059	0.279686194271623\\
0.500333296210692	0.265347070452491\\
0.495844128480833	0.251059565804446\\
0.491639943046548	0.236821825328882\\
0.487713323972304	0.222631822393042\\
0.484057384194224	0.208487377782559\\
0.480665746291318	0.19438617684838\\
0.477532524612211	0.180325784945983\\
0.474652308729595	0.166303661346937\\
0.472020148193448	0.152317171786835\\
0.469631538553626	0.138363599799346\\
0.467482408622789	0.124440156973337\\
0.46556910895144	0.110543992258702\\
0.463888401488216	0.0966722004365162\\
0.462437450400268	0.082821829860316\\
0.461213814030571	0.0689898895676433\\
0.460215437971288	0.0551733558543447\\
0.45944064923473	0.0413691783984604\\
0.458888151506107	0.0275742860157958\\
0.458557021464958	0.0137855921253841\\
0.45844670616499	4.83629319566573e-17\\
0.458557021464958	-0.0137855921253842\\
0.458888151506107	-0.0275742860157955\\
0.45944064923473	-0.0413691783984601\\
0.460215437971288	-0.0551733558543446\\
0.461213814030571	-0.0689898895676434\\
0.462437450400268	-0.0828218298603159\\
0.463888401488216	-0.0966722004365161\\
0.46556910895144	-0.110543992258702\\
0.467482408622789	-0.124440156973337\\
0.469631538553626	-0.138363599799346\\
0.472020148193448	-0.152317171786835\\
0.474652308729595	-0.166303661346937\\
0.477532524612211	-0.180325784945983\\
0.480665746291318	-0.19438617684838\\
0.484057384194225	-0.208487377782559\\
0.487713323972304	-0.222631822393042\\
0.491639943046548	-0.236821825328882\\
0.495844128480833	-0.251059565804446\\
0.500333296210692	-0.26534707045249\\
0.505115411653059	-0.279686194271623\\
0.510199011719082	-0.294078599450418\\
0.515593228247094	-0.308525731828413\\
0.521307812866193	-0.32302879472981\\
0.527353163292079	-0.337588719878811\\
0.533740351045469	-0.352206135075745\\
0.54048115056914	-0.366881328280544\\
0.547588069701777	-0.381614207714231\\
0.55507438144475	-0.396404257549913\\
0.562954156930874	-0.411250488721973\\
0.571242299471275	-0.42615138433572\\
0.579954579516609	-0.441104839109405\\
0.589107670320822	-0.456108092226362\\
0.598719184038071	-0.471157652916971\\
0.608807707914574	-0.486249218028372\\
0.619392840155325	-0.501377580774729\\
0.630495224948431	-0.516536529792733\\
0.642136586015084	-0.531718737556845\\
0.654339757917851	-0.546915637137558\\
0.667128714201079	-0.562117286215265\\
0.680528591251149	-0.577312217194278\\
0.694565706547207	-0.592487272198991\\
0.709267569720524	-0.607627421680698\\
0.724662884548187	-0.622715565324069\\
0.740781539669398	-0.637732313922675\\
0.757654585425199	-0.652655750901016\\
0.775314193779646	-0.667461172205845\\
0.793793597777532	-0.68212080338412\\
0.813127006426119	-0.696603492823351\\
0.833349490252784	-0.710874380370424\\
0.854496832085235	-0.724894540888668\\
0.876605336827072	-0.738620602786003\\
0.899711593163562	-0.75200434217979\\
0.923852179240612	-0.764992254192193\\
0.949063303431193	-0.77752510393409\\
0.97538037036444	-0.789537461081803\\
1.00283746148251	-0.800957223629385\\
1.03146671856365	-0.811705138463575\\
1.06129761798153	-0.821694328913289\\
1.09235612305874	-0.830829842422513\\
1.12466370184438	-0.839008235028677\\
1.15823619816071	-0.846117213426245\\
1.19308254501952	-0.852035360059756\\
1.22920331173959	-0.856631971884776\\
1.26658907957522	-0.859767049065187\\
1.30521864569712	-0.861291475769684\\
1.34505706227427	-0.861047441116101\\
1.38605352651112	-0.858869153788823\\
1.42813914908317	-0.854583908368418\\
1.47122464268467	-0.848013564236132\\
1.51519798940195	-0.838976498137266\\
1.55992216516646	-0.827290088018696\\
1.60523302110024	-0.812773777370572\\
1.65093744418177	-0.795252754713443\\
1.69681194183321	-0.774562260862194\\
1.74260181465167	-0.750552506208811\\
1.78802109584736	-0.72309414104863\\
1.83275344173291	-0.692084174290164\\
1.87645415122004	-0.657452181224573\\
1.91875347013776	-0.619166582287041\\
1.95926129527241	-0.577240716405729\\
1.99757333156881	-0.531738380655843\\
2.033278674141	-0.48277846981783\\
2.0659686875247	-0.430538332959436\\
2.09524694401921	-0.375255476741132\\
2.12073987025765	-0.317227292418167\\
2.14210764819703	-0.256808567903064\\
2.15905483682386	-0.194406665725809\\
2.17134013689902	-0.130474395069739\\
2.17878472326781	-0.0655007689731413\\
2.18127862312552	-4.60219380835795e-16\\
};
\addplot [color=white!70!black, forget plot]
  table[row sep=crcr]{%
1.32333710019237	0\\
1.32303437432559	0.0131056200947381\\
1.32212739540714	0.0261773153812059\\
1.32061974830724	0.0391813716560451\\
1.3185173707533	0.0520844931694248\\
1.31582849842316	0.0648540050227677\\
1.31256358941145	0.0774580475422887\\
1.30873522936044	0.0898657602318436\\
1.30435801885401	0.102047453134962\\
1.29944844493801	0.113974763704457\\
1.29402473884605	0.125620797580068\\
1.28810672217289	0.136960252000714\\
1.28171564384567	0.147969520918424\\
1.27487401029578	0.158626781225921\\
1.26760541123368	0.168912059849873\\
1.25993434337697	0.178807281788308\\
1.25188603438473	0.188296299475891\\
1.24348626911358	0.19736490413835\\
1.23476122013874	0.206000820042172\\
1.22573728428564	0.214193682754212\\
1.21644092669916	0.221935002695829\\
1.20689853374757	0.229218115406782\\
1.1971362758218	0.236038120025866\\
1.18717998085569	0.242391807549592\\
1.17705501916303	0.248277580449739\\
1.16678619996935	0.253695365218331\\
1.15639767981249	0.258646519368448\\
1.14591288280048	0.263133734355171\\
1.1353544325494	0.267160935797194\\
1.12474409547955	0.270733182280376\\
1.11410273502551	0.273856563913864\\
1.10345027621501	0.276538101691207\\
1.09280567999184	0.278785648586548\\
1.08218692659844	0.280607793192768\\
1.07161100729265	0.282013766586821\\
1.06109392364947	0.28301335298997\\
1.05065069368908	0.283616804678732\\
1.04029536407713	0.28383476149771\\
1.03004102765804	0.283678175229\\
1.01989984560704	0.283158238985178\\
1.00988307351858	0.282286321714479\\
1.00000109078642	0.281073907837582\\
0.990263432672919	0.279532541975424\\
0.980678824509782	0.27767377867635\\
0.97125521751891	0.275509137008116\\
0.96199982578918	0.273050059845418\\
0.952919163991514	0.270307877655942\\
0.944019085460306	0.267293776566846\\
0.935304820313267	0.264018770478411\\
0.926781013323663	0.260493676981591\\
0.918451761298381	0.256729096830768\\
0.910320649752158	0.252735396721472\\
0.902390788702284	0.248522695124632\\
0.89466484743934	0.244100850933429\\
0.887145088157801	0.239479454685637\\
0.879833398355909	0.234667822132893\\
0.87273132193695	0.229674989938322\\
0.865840088964285	0.224509713294894\\
0.859160644040208	0.219180465268604\\
0.852693673294094	0.21369543768269\\
0.846439629978587	0.208062543371416\\
0.840398758683877	0.202289419644346\\
0.834571118189623	0.196383432814183\\
0.828956602981943	0.190351683653246\\
0.823554963469295	0.184201013655166\\
0.818365824936198	0.177938011989552\\
0.813388705277614	0.171569023047937\\
0.808623031559769	0.165100154489405\\
0.804068155455146	0.158537285703761\\
0.799723367600619	0.151886076619031\\
0.795587910928219	0.145151976788339\\
0.791660993017988	0.138340234698991\\
0.787941797521819	0.131455907253679\\
0.784429494706222	0.12450386938036\\
0.781123251160658	0.117488823733397\\
0.778022238716453	0.110415310454094\\
0.775125642619489	0.103287716963823\\
0.772432668997835	0.0961102877675315\\
0.769942551663279	0.0888871342496017\\
0.767654558283446	0.0816222444477588\\
0.765567995958781	0.0743194927941467\\
0.763682216236213	0.0669826498156848\\
0.76199661958883	0.0596153917885284\\
0.760510659388333	0.052221310343844\\
0.7592238453945	0.044803922024209\\
0.758135746783314	0.0373666777917837\\
0.757245994732845	0.029912972490991\\
0.756554284583411	0.0224461542697904\\
0.756060377585983	0.0149695339647714\\
0.755764102250232	0.00748639445621443\\
0.755665355301104	2.62668289601501e-17\\
0.755764102250232	-0.00748639445621447\\
0.756060377585983	-0.0149695339647713\\
0.756554284583411	-0.0224461542697902\\
0.757245994732845	-0.029912972490991\\
0.758135746783314	-0.0373666777917838\\
0.7592238453945	-0.0448039220242089\\
0.760510659388333	-0.0522213103438439\\
0.76199661958883	-0.0596153917885283\\
0.763682216236213	-0.0669826498156848\\
0.765567995958781	-0.0743194927941467\\
0.767654558283446	-0.0816222444477587\\
0.769942551663279	-0.0888871342496016\\
0.772432668997835	-0.0961102877675315\\
0.775125642619489	-0.103287716963823\\
0.778022238716453	-0.110415310454094\\
0.781123251160658	-0.117488823733397\\
0.784429494706222	-0.12450386938036\\
0.787941797521819	-0.131455907253679\\
0.791660993017988	-0.138340234698991\\
0.795587910928219	-0.145151976788339\\
0.799723367600619	-0.151886076619031\\
0.804068155455146	-0.158537285703761\\
0.808623031559769	-0.165100154489405\\
0.813388705277614	-0.171569023047937\\
0.818365824936198	-0.177938011989552\\
0.823554963469295	-0.184201013655166\\
0.828956602981943	-0.190351683653246\\
0.834571118189623	-0.196383432814183\\
0.840398758683877	-0.202289419644346\\
0.846439629978587	-0.208062543371416\\
0.852693673294094	-0.21369543768269\\
0.859160644040208	-0.219180465268604\\
0.865840088964285	-0.224509713294894\\
0.87273132193695	-0.229674989938322\\
0.879833398355909	-0.234667822132893\\
0.887145088157801	-0.239479454685637\\
0.89466484743934	-0.244100850933429\\
0.902390788702284	-0.248522695124632\\
0.910320649752157	-0.252735396721472\\
0.918451761298381	-0.256729096830768\\
0.926781013323663	-0.260493676981591\\
0.935304820313267	-0.264018770478411\\
0.944019085460306	-0.267293776566846\\
0.952919163991514	-0.270307877655942\\
0.96199982578918	-0.273050059845418\\
0.97125521751891	-0.275509137008116\\
0.980678824509782	-0.27767377867635\\
0.990263432672919	-0.279532541975424\\
1.00000109078642	-0.281073907837582\\
1.00988307351858	-0.282286321714479\\
1.01989984560704	-0.283158238985178\\
1.03004102765804	-0.283678175229\\
1.04029536407713	-0.28383476149771\\
1.05065069368908	-0.283616804678732\\
1.06109392364947	-0.28301335298997\\
1.07161100729265	-0.282013766586821\\
1.08218692659844	-0.280607793192768\\
1.09280567999184	-0.278785648586548\\
1.10345027621501	-0.276538101691207\\
1.11410273502551	-0.273856563913864\\
1.12474409547955	-0.270733182280376\\
1.1353544325494	-0.267160935797194\\
1.14591288280048	-0.263133734355171\\
1.15639767981249	-0.258646519368448\\
1.16678619996935	-0.253695365218331\\
1.17705501916303	-0.248277580449739\\
1.18717998085569	-0.242391807549592\\
1.1971362758218	-0.236038120025866\\
1.20689853374757	-0.229218115406782\\
1.21644092669916	-0.221935002695829\\
1.22573728428564	-0.214193682754212\\
1.23476122013874	-0.206000820042172\\
1.24348626911358	-0.19736490413835\\
1.25188603438473	-0.188296299475892\\
1.25993434337697	-0.178807281788308\\
1.26760541123368	-0.168912059849873\\
1.27487401029578	-0.158626781225921\\
1.28171564384567	-0.147969520918424\\
1.28810672217289	-0.136960252000714\\
1.29402473884605	-0.125620797580068\\
1.29944844493801	-0.113974763704458\\
1.30435801885401	-0.102047453134962\\
1.30873522936044	-0.0898657602318437\\
1.31256358941145	-0.0774580475422889\\
1.31582849842316	-0.064854005022768\\
1.3185173707533	-0.052084493169425\\
1.32061974830724	-0.039181371656045\\
1.32212739540714	-0.0261773153812057\\
1.32303437432559	-0.013105620094738\\
1.32333710019237	-9.19980491987052e-17\\
};
\addplot [color=green!65!black, draw=none, mark size=1.35pt, mark=*, mark options={solid, green!65!black}, forget plot]
  table[row sep=crcr]{%
-3.21251978754258e-15	17.8066497453503\\
-0	5.88557845235807\\
-5.78050408034917e-16	3.47107680296989\\
-0	2.41421356237309\\
-0	1.80936375494128\\
-0	1.40936810645926\\
-6.21170068981929e-17	1.11900051648029\\
0	0.893654636680067\\
5.55111512312578e-17	0.709537838565317\\
4.70201030766277	0\\
3.10042096722724	2.15057740464389\\
1.55916638765564	2.05713691795909\\
0	0.552680464206852\\
0.878765430293869	1.59581542158991\\
0.566989144853011	1.21041171687394\\
0.406943655279904	0.913453261764076\\
0	0.414213562373095\\
0.317361270268286	0.677504681530217\\
1.32333710019237	0\\
2.05002578214554	0.457055002561778\\
-1.38777878078145e-17	0.288095037005344\\
0.264779656531617	0.480833046738231\\
0.96199982578918	0.273050059845418\\
1.40696526798316	0.857000927663578\\
-2.77555756156289e-17	0.16990683381331\\
0.234010258540393	0.308749050669579\\
0.91165031859925	0.758551339106386\\
0.648166474263825	0.539315938343701\\
-1.69135538907739e-17	0.0561587954107495\\
0.21776299488117	0.151049222447315\\
0.518409840148601	0.315770207003146\\
0.464699867760614	0.103605233211993\\
-3.03576608295941e-18	-0.0561587954107494\\
2.08166817117217e-17	-0.16990683381331\\
-1.38777878078145e-17	-0.288095037005344\\
-5.55111512312578e-17	-0.414213562373095\\
0	-0.552680464206852\\
5.55111512312578e-17	-0.709537838565317\\
-5.55111512312578e-17	-0.893654636680067\\
0	-1.11900051648029\\
7.82356460981716e-17	-1.40936810645926\\
0.212674991028905	5.84627613092968e-17\\
0.217762994881169	-0.151049222447314\\
0.234010258540393	-0.308749050669578\\
1.00439865032902e-16	-1.80936375494128\\
0.264779656531617	-0.480833046738231\\
0.317361270268286	-0.677504681530217\\
0.406943655279904	-0.913453261764076\\
1.34015774165447e-16	-2.41421356237309\\
0.566989144853011	-1.21041171687394\\
0.755665355301105	2.62668289601501e-17\\
0.464699867760614	-0.103605233211992\\
1.92683469344972e-16	-3.47107680296988\\
0.878765430293869	-1.59581542158991\\
0.96199982578918	-0.273050059845418\\
0.518409840148601	-0.315770207003146\\
1.6335761777614e-16	-5.88557845235807\\
1.55916638765564	-2.05713691795909\\
0.648166474263825	-0.539315938343701\\
0.91165031859925	-0.758551339106385\\
-7.41350720202131e-15	-17.8066497453502\\
3.10042096722724	-2.15057740464389\\
1.40696526798316	-0.857000927663579\\
2.05002578214553	-0.45705500256178\\
};

\end{axis}
\end{tikzpicture}%

%% file: 04b-Switch.tex
Running a BC tag will always require some amount of power, despite the passive transmission nature. For example, the tag of the recent work \cite{KimionisNAT2021} uses a high-electron mobility transistor as adaptive tag load for 16-QAM modulation, with a power requirement of $0.17 \f{\unit{pJ}}{\unit{bit}} \cdot 2 \f{\unit{Gbit}}{\unit{s}} = 0.34\unit{mW}$. Such power could drain a small battery fairly quickly and may not be sustainable with energy harvesting. A tag design that minimizes this power requirement is thus crucial for ensuring long and reliable uptimes. Therefore, it is worthwhile to explore the simplest imaginable switched load circuits as alternative to complicated analog electronics, which may be the key to minimal power consumption. A detailed technological comparison (which must also account for the power consumption of the switching logic and other required circuits) is out of scope. We focus on the achievable information rates with such circuits. This shall help clarify whether it is technologically worthwhile to follow this avenue.

We consider a simple low-cost circuit for the adaptive tag load, composed of $L+1$ lumped elements (resistors, capacitors, inductors) in some topology.
Their impedances are denoted $Z_\ell$, $\ell \in \{0,\ldots,L\}$.
The elements $\ell \geq 1$ are combined with individual on-off switches that allow to detach the effect of $Z_\ell$ on the compound load impedance. 
The element $\ell = 0$ is reactive and not switched; it establishes a near-resonant state by compensating the tag-antenna reactance $X\Tx$ (like the $-X\Tx$ element in \Cref{fig:Circuits}).
An example circuit is shown in \Cref{fig:5C3R_circuit}.
It is clear that such a circuit allows for $M = 2^L$ different load states, giving rise to an $M$-ary symbol constellation.

\renewcommand\myFigHeight{46mm}

\ifdefined\SingleColumnDraft
\begin{figure}[t]
\else
\begin{figure}[!ht]
\fi
\centering
\subfloat[termination load circuit for BC tag]{
\ifdefined\SingleColumnDraft
\includegraphics[width=66mm,trim=0 -15mm 0 0]{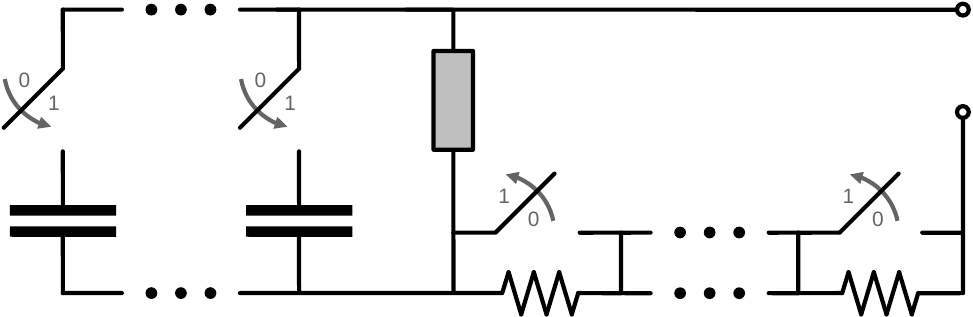}
\put(-94,58){\footnotesize{$X_0$}}
\put(1,74){\footnotesize{$+$}}
\put(1,54){\footnotesize{$-$}}
\ \ \
\else
\includegraphics[width=70mm,trim=0 0 0 0]{F04b_CircuitMixedRC.pdf}
\put(-100,43){\footnotesize{$X_0$}}
\put(1,61){\footnotesize{$+$}}
\put(1,40){\footnotesize{$-$}}
\fi
\label{fig:5C3R_circuit}}
\ \
\subfloat[symbols $\Gamma_m$]{
\includegraphics[height=\myFigHeight,trim=0 0 0 2mm]{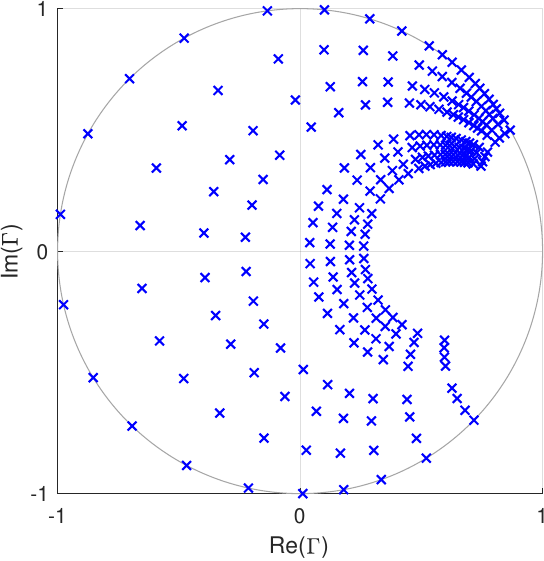}
\label{fig:5C3R_20dB_optPMF_Gamma}}
\subfloat[symbols $z_m$]{
\ \
\includegraphics[height=\myFigHeight,trim=0 0 0 2mm]{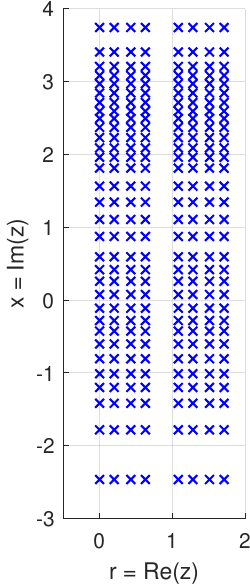}
\ \
\label{fig:5C3R_20dB_optPMF_z}}\\
\caption{8-bit symbol alphabet realized by 5 switched capacitors and 3 switched resistors. The component values and symbol probabilities are optimized for maximum achievable rate at $20\dB$ SNR.}
\label{fig:5C3R_20dB_optPMF}
\end{figure}

\ifdefined\SingleColumnDraft
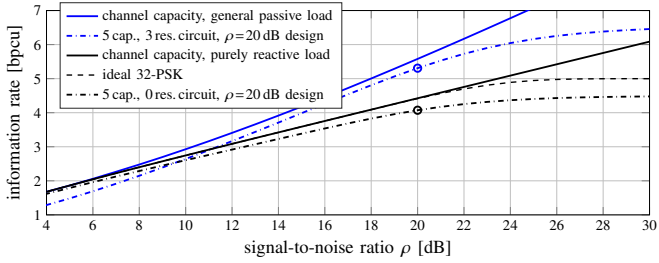
\begin{figure}[t]
\centering
\resizebox{110mm}{!}{\input{F04b_RatesVsSNR}}
\vspace{-5mm}
\else
\begin{figure}[t]
\centering
\resizebox{.99\columnwidth}{!}{\input{F04b_RatesVsSNR}}
\vspace{-6mm}
\fi
\caption{Achievable rates with the adaptive tag load of \Cref{fig:5C3R_circuit}, compared to benchmark cases.}
\label{fig:SwitchedRatesVsSNR}
\end{figure}

Technologically, it would be delightful if such a simple circuit could establish or approximate a near-capacity-achieving $M$-APSK constellation of the type discussed in \Cref{sec:APSK}. Unfortunately, this is not the case: there seems to be no topology that establishes constellation points on concentric circles. Nevertheless, we shall investigate how close we can get. To do so, we maximize $I(\r y;\r\Gamma)$ with respect to all component values of the topology in \Cref{fig:5C3R_circuit} (in omitted experiments we also evaluated other topologies, which yielded no discernible benefit) with the interior-point algorithm for gradient-based numerical optimization.
In particular, we conduct a joint optimization of all component values and all symbol probabilities.
The optimization is done for a design-SNR of $\SNR = 20\dB$. The symbol constellation resulting from the optimized circuit is shown in \Cref{fig:5C3R_20dB_optPMF_Gamma} in terms of reflection coefficients $\Gamma_m$ and in \Cref{fig:5C3R_20dB_optPMF_z} in terms of impedances $z_m$. Clearly, the constellation is far off the desired APSK constellation. The approach struggles with covering all regions of the unit disk with symbols $\Gamma_m$ in a somewhat uniform fashion. We observe the following trade-off. The resistance and reactance spread is kept small to enable a decent coverage of the left, upper, and lower regions of the disk. This however prevents coverage of the entire region around $\Gamma = 1$, which is reached either via $r \rightarrow \infty$, $x \rightarrow \infty$, or $x \rightarrow -\infty$. There is no apparent way to mitigate this trade-off.

Nevertheless, we find that the non-optimal constellations from such simple low-cost designs can still yield a high information rate. This is demonstrated by \Cref{fig:SwitchedRatesVsSNR}, which evaluates a general-passive circuit with $L = 8$ switched components (5 capacitors, 3 resistors) and a purely-reactive circuit with $L = 5$ switched capacitors (and 0 resistors), both optimized at a $20\dB$ target-SNR. They perform only slightly worse than the respective benchmark cases over a wide SNR range. Thereby, the optimization of the symbol probabilities is crucial, because it mitigates the non-uniform symbol spacing to a large extend. A small associated disadvantage is that the high-SNR rate limit $H(\r\Gamma)$ drops below $\log_2(M) = L$, which is noticeable in \Cref{fig:SwitchedRatesVsSNR}.

\begin{figure}[t]
\centering
\ifdefined\SingleColumnDraft
\includegraphics[width=105mm,trim=0 12 0 0]{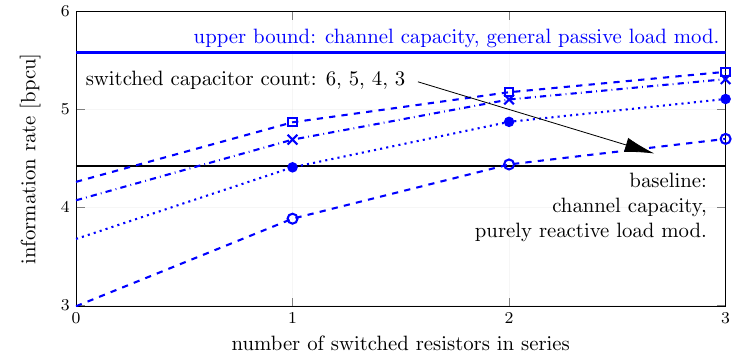}
\else
\includegraphics[width=.95\columnwidth]{F04b_RateVsNoResistors}
\fi
\caption{Achievable rates with the adaptive tag load of \Cref{fig:5C3R_circuit} at $20\dB$ SNR, for various configurations, compared to channel capacity.}
\label{fig:RateVsNoResistors}
\end{figure}

The evaluation in \Cref{fig:RateVsNoResistors} investigates whether the residual performance gap can be closed by adding more components to the circuit. In particular, we evaluate the achievable information rate as a function of the number of switched capacitors and resistors. Every data point is the result of a numerical optimization. The results indicate that the actual channel capacity can be approached with a reasonable number of components. Furthermore, they confirm the importance of modulating both resistance and reactance.


%% file: F04b_RatesVsSNR.tex
\begin{tikzpicture}

\begin{axis}[%
width=115mm,
height=39mm,
at={(0,0)},
scale only axis,
xmin=4,
xmax=30,
xlabel={\normalsize{signal-to-noise ratio $\SNR$ [dB]}},
ymin=1,
ymax=7,
ytick={0,1,2,3,4,5,6,7,8,9,10},
yminorticks=true,
ylabel={\normalsize{information rate [bpcu]}},
axis background/.style={fill=white},
xmajorgrids,
ymajorgrids,
yminorgrids,
legend style={at={(0.022,1.04)}, anchor=north west, legend cell align=left, align=left, draw=white!15!black}
]
\addplot [color=blue, line width=1pt]
  table[row sep=crcr]{%
0	0.979880414508299\\
1	1.14150425662973\\
2	1.31402338890148\\
3	1.49401674435592\\
4	1.67779485739644\\
5	1.86252447577785\\
6	2.05865655462784\\
7	2.26628092318256\\
8	2.47809417055567\\
9	2.6980725693527\\
10	2.9274800746115\\
11	3.16373469941759\\
12	3.40765530732148\\
13	3.65833650083317\\
14	3.91577392694257\\
15	4.17949655464713\\
16	4.44913720761297\\
17	4.72430591740325\\
18	5.00462349817165\\
19	5.28971625348757\\
20	5.57921502640108\\
21	5.87277857916322\\
22	6.17006471615952\\
23	6.47076539080284\\
24	6.77457096976575\\
25	7.08120458375078\\
26	7.39041834941432\\
27	7.70196209969228\\
28	8.01561580406267\\
};
\addlegendentry{channel capacity, general passive load}
\addplot [color=blue, dashdotted, line width=1pt]
  table[row sep=crcr]{%
-10	0.0821816844994565\\
-9	0.103130394068327\\
-8	0.129025503839823\\
-7	0.16091937184741\\
-6	0.200023914290171\\
-5	0.247702716133764\\
-4	0.305445151340146\\
-3	0.374816634468869\\
-2	0.457380643402549\\
-1	0.554592307289731\\
0	0.667670824914947\\
1	0.797468233087062\\
2	0.94435876684162\\
3	1.10820391758387\\
4	1.28834487279822\\
5	1.4837378929476\\
6	1.69308314140216\\
7	1.91497161046748\\
8	2.14798990113639\\
9	2.39079977803412\\
10	2.64212964799778\\
11	2.90082035281762\\
12	3.16575194772309\\
13	3.43582800109396\\
14	3.70985636867082\\
15	3.98636647222447\\
16	4.26334754894296\\
17	4.53799870627981\\
18	4.80662122731335\\
19	5.06473689580929\\
20	5.30746822665171\\
21	5.53006506646327\\
22	5.7285857478596\\
23	5.90044254926041\\
24	6.04474619023712\\
25	6.16231779455816\\
26	6.25548392855168\\
27	6.32764680447641\\
28	6.3827286615892\\
29	6.42458388282653\\
30	6.45650029701071\\
};
\addlegendentry{5\,cap., 3\,res.\,circuit, $\rho \!=\! 20\dB$ design}
\addplot [color=blue, solid, line width=1pt, mark=o, forget plot]
  table[row sep=crcr]{%
20	5.30746822665171\\
};

\addplot [color=black, line width=1pt]
  table[row sep=crcr]{%
0	0.979880414508299\\
1	1.14150425662973\\
2	1.31402338890148\\
3	1.49401674435592\\
4	1.67779485739644\\
5	1.86203881197189\\
6	2.0443822277425\\
7	2.22368611366353\\
8	2.39988159307618\\
9	2.57350535324004\\
10	2.7452260803892\\
11	2.91558641939581\\
12	3.08496043743093\\
13	3.25360032066712\\
14	3.42168284556869\\
15	3.58933684436541\\
16	3.75665858028019\\
17	3.92372114165475\\
18	4.09058063988499\\
19	4.25728050116793\\
20	4.42385453701641\\
21	4.59032919372108\\
22	4.75672522988997\\
23	4.92305898463008\\
24	5.08934334643103\\
25	5.25558849950875\\
26	5.42180250245275\\
27	5.58799173914823\\
28	5.75416127158544\\
29	5.92043394691552\\
30	6.08657137245093\\
};
\addlegendentry{channel capacity, purely reactive load}

\addplot [color=black, dashed, line width=0.75pt]
  table[row sep=crcr]{%
-10	0.135845722114547\\
-9	0.169488483576718\\
-8	0.210713279406908\\
-7	0.260930847347437\\
-6	0.321662785365795\\
-5	0.394472886497239\\
-4	0.480857750615142\\
-3	0.582091045554382\\
-2	0.699023692318542\\
-1	0.831856758969879\\
0	0.979925343043067\\
1	1.14155664653966\\
2	1.31407658659306\\
3	1.49408370236157\\
4	1.67786855390502\\
5	1.8621186997632\\
6	2.04446770361315\\
7	2.22377658832689\\
8	2.39996981962465\\
9	2.57360433121034\\
10	2.74532166763331\\
11	2.91569242015949\\
12	3.08506953456217\\
13	3.25371228408124\\
14	3.42179744890732\\
15	3.58945337571384\\
16	3.75676958758425\\
17	3.92376184839247\\
18	4.09016871399929\\
19	4.2548487832904\\
20	4.41489755907908\\
21	4.56497559755714\\
22	4.69816023801974\\
23	4.80784836033666\\
24	4.89008392202034\\
25	4.94487291980742\\
26	4.97643237311737\\
27	4.99160481580761\\
28	4.99743512808884\\
29	4.99913552416871\\
30	4.99949502846644\\
};
\addlegendentry{ideal 32-PSK}

\addplot [color=black, dashdotted, line width=1pt]
  table[row sep=crcr]{%
-10	0.133067171382758\\
-9	0.165982326122979\\
-8	0.206288781585417\\
-7	0.255348210276816\\
-6	0.31462165377861\\
-5	0.385599074963798\\
-4	0.469688529788468\\
-3	0.56805979837936\\
-2	0.681445413289989\\
-1	0.809916706365157\\
0	0.952674043282806\\
1	1.10791511353084\\
2	1.27285524774969\\
3	1.44401982953562\\
4	1.61767688888907\\
5	1.79055909745067\\
6	1.96041850190647\\
7	2.12630964984826\\
8	2.28842107140936\\
9	2.44763663165361\\
10	2.60495331772867\\
11	2.76128507919889\\
12	2.91723075499731\\
13	3.07312177031276\\
14	3.22891789417071\\
15	3.38405001908389\\
16	3.53715381364441\\
17	3.68587427597708\\
18	3.82699332309899\\
19	3.9569230697527\\
20	4.07248204760764\\
21	4.17138605587235\\
22	4.2528191515421\\
23	4.31737959862634\\
24	4.36689410152222\\
25	4.40364004338231\\
26	4.43013208483838\\
27	4.44875247428112\\
28	4.46170756870665\\
29	4.47094627601132\\
30	4.47798020432834\\
};
\addlegendentry{5\,cap., 0\,res.\,circuit, $\rho \!=\! 20\dB$ design}

\addplot [color=black, solid, line width=1pt, mark=o, forget plot]
  table[row sep=crcr]{%
20	4.07248204760764\\
};

\end{axis}

\end{tikzpicture}%

%% file: 05-Ambient.tex
\newcommand\Amb{_\mathrm{A}}
\newcommand\TxAmb{_\mathrm{TA}}
\newcommand\AmbRate{_\ell}
\newcommand\meanSNR{\SNR}
\newcommand\sqrtSNR{{\r s}}
\newcommand\fadingRV{\r\psi}
\newcommand\fadingRVCombined{\r a}

So far we have considered monostatic and bistatic BC setups, which operate under the assumption that the induced voltage phasor $v\Ind\Tx$ at the tag antenna would be constant due to a dedicated RF source. Now we consider ambient BC where $\r v\Ind\Tx$ is a random variable due to a modulated ambient source. In particular, we model $\r v\Ind\Tx = \r Z\TxAmb\,\r i\Amb$ via the mutual impedance $\r Z\TxAmb$ from the random ambient-source antenna feed current $\r i\Amb$ to the BC tag.
Other than that, we employ the same assumptions as in \Cref{sec:model}. Consequently, we find that the random instantaneous SNR of the BC link is given by
$|\r Z\RxTx \r Z\TxAmb \r i\Amb / (2 R\Tx \sigma)|^2$.
We decompose this into a convenient product $|\fadingRV|^2 \SNR$ that is composed of the mean SNR
\begin{align}
\SNR := \f{\EV{|\r Z\RxTx \r Z\TxAmb \r i\Amb|^2}}{4 R\Tx^2\,\sigma^2}
\label{eq:RedefSNR}
\end{align}
and a random variable $\fadingRV$ that encompasses the ambient signal modulation in $\r i\Amb$ but also potential time-varying channel fading in $\r Z\RxTx$ and $\r Z\TxAmb$. In particular,
\begin{align}
\fadingRV &:= \f{\r Z\RxTx\, \r Z\TxAmb\, \r i\Amb}{\sqrt{\EV{|\r Z\RxTx\, \r Z\TxAmb\, \r i\Amb|^2}}}
\in \bbC \, , &
\EV{|\fadingRV|^2} &= 1 \, .
\label{eq:fadingRV}
\end{align}
We refrain from assumptions on the distribution of $\fadingRV$ or the correlation between $\r Z\RxTx$, $\r Z\TxAmb$, $\r i\Amb$.

We define the ratio $N\Amb := T / T\Amb$ of the BC load-modulation symbol duration $T$ to the coherence time $T\Amb$ of $\fadingRV$. Usually, $T\Amb$ will be determined by the ambient-modulation symbol duration.
Most BC link designs will exhibit $N\Amb \gg 1$ because a large $T$ may be necessary for sufficient noise averaging at the BC receiver and to avoid significant distortions from transients. In \cite{DarsenaTC2017} for example, $T$ is matched to the duration of an entire ambient OFDM symbol
(e.g., $N\Amb = 64$ would be typical for an ambient WiFi signal with $64$ OFDM subcarriers).
The circumstance that $\fadingRV$ decorrelates $N\Amb$ times per change of $\r\Gamma[n]$ is captured by the following notation, which is analogous to \cite[Eq.~(38)]{DarsenaTC2017}. For simplicity we assume $N\Amb \in \bbN$ and stack the discretized temporal evolution of the random variables into the random vector $\vec{\fadingRV}[n] \in \bbC^{N\Amb}$. As before, $n \in \bbZ$ is the time index of the BC transmit signal $\r\Gamma[n] \in \bbC$.
We obtain a vector-form signal and noise model
$\vec{\r y}[n] = \vec\fadingRV[n] \, \r\Gamma[n] + \vec{\r w}[n]$
with
$\vec{\r y}, \vec{\fadingRV}, \vec{\r w} \in \bbC^{N\Amb}$.
Regarding the AWGN vector $\vec{\r w}[n]$, all elements are i.i.d. $\mathcal{CN}(0, N\Amb / \meanSNR)$ whereby the factor $N\Amb$ is due to the shorter time window for noise-averaging.
We assume that the fluctuations in $\vec\fadingRV[n]$ are caused by a digitally modulated ambient source and that the BC receiver is able to obtain full knowledge of $\vec\fadingRV[n]$ by decoding the ambient signal and through estimating all relevant channels (cf. \Cref{sec:model}). As noted in \cite[Eq.~(40)]{DarsenaTC2017}, the SNR-optimal strategy for the decoding of $\r\Gamma[n]$ involves maximum-ratio combining, which is implemented with a projection $\r y[n] := \vec{\r u}[n]\H \vec{\r y}[n]$ onto
$\vec{\r u}[n] := \f{1}{\sqrt{N\Amb}}\,\vec{\fadingRV}[n] \, / \, \|\vec{\fadingRV}[n]\|$.
This yields the scalar model
\begin{align}
{\r y}[n] &= \fadingRVCombined[n] \, \r\Gamma[n] + {\r w}[n] \, , &
\fadingRVCombined[n] &:= \f{1}{\sqrt{N\Amb}} \big\| \vec{\fadingRV}[n] \big\| \, ,
\label{eq:SignalModelAfterMRC}
\end{align}
with
$|\r\Gamma[n]| \leq 1\ \forall n$,
$\r w[n] \iid \mathcal{CN}(0, 1 / \meanSNR)$,
and
$\EV{ |\fadingRVCombined[n]|^2 } = 1$.
Because of the ever-changing effect of the ambient data in $\vec{\fadingRV}[n]$, it is reasonable to assume statistical independence between $\fadingRVCombined[n]$ for different $n$.

\begin{theorem}
Under the employed assumptions, the AmBC channel capacity is given by
\begin{align}
\RateMaxAmbient(\SNR) = \EVVarLR{\,\fadingRVCombined\!}{\RateMaxGP\big( |\fadingRVCombined|^2 \SNR \big)} \, .
\label{eq:ErgodicCapacity}
\end{align}
\end{theorem}

\textit{Proof Sketch:} From the formal perspective of information theory, the channel \Cref{eq:SignalModelAfterMRC} classifies as a fast-fading AWGN channel whose coherence time equals one symbol duration (even without interleaving techniques). The statement \Cref{eq:ErgodicCapacity} follows from the ergodic capacity arguments in \cite[Sec.~5.4.5 \& Apdx.~B.7.1]{Tse2005}. We refer to this source for the mathematical background (and details on the special error-correcting codes that are required to approach capacity over a fast-fading channel). The function $\RateMaxGP(\,.\,)$ in \Cref{eq:ErgodicCapacity} is the complicated expression from \Cref{eq:Capacity} that describes the non-ambient-case channel capacity for a constant SNR $\SNR$. \QEDB

A similar observation has been made in \cite[Eq.~(41)]{DarsenaTC2017}.
A more detailed analytical description of $\RateMaxAmbient$ seems to be infeasible at this point. However, by Jensen's inequality, one can discern that
$\RateMaxAmbient(\SNR) \leq \RateMaxGP(\SNR)$, because of $\EV{|\fadingRVCombined[n]|^2} = 1$ and the apparent concavity of $\RateMaxGP(\SNR)$.
At low SNR, the function is approximately linear, cf. \Cref{eq:RMaxBoundLowSNR}, and thus
\begin{align}
&\SNR \ll 1: &
\RateMaxAmbient(\SNR) &\approx \RateMaxGP(\SNR) \approx \SNR \cdot \log_2(e) \, .
\end{align}

Consider now the special case that the ambient rate of change is much faster than the  BC symbol rate ($N\Amb \gg 1$). Then
$|\fadingRVCombined[n]|^2 = \f{1}{N\Amb} \| \vec\fadingRV[n] \|^2 \approx \EV{|\fadingRV|^2} = 1$ by the law of large numbers.
Employing this in \Cref{eq:ErgodicCapacity} yields
$\RateMaxAmbient(\SNR) \approx \RateMaxGP(\SNR)$
because of
$|\fadingRVCombined|^2 \approx 1$. The effect of the ambient fluctuations is remedied entirely.

Consider now the special case of a PSK-modulated ambient source (i.e. $\r i\Amb$ has constant envelope) and that both propagation channels $\r Z\RxTx, \r Z\TxAmb$ are time-invariant for the duration of a BC coding block. In consequence,
$|\fadingRV|^2 = 1$,
so
$|\fadingRVCombined|^2 = 1$
and
$\RateMaxAmbient(\SNR) = \RateMaxGP(\SNR)$. The effect of the ambient PSK modulation is remedied entirely.

Finally, consider that the coherence period of $\fadingRVCombined[n]$ is longer than the coding block length of the BC load modulation. While this is unlikely in AmBC, such conditions can certainly occur in BiBC or MoBC with slow-fading propagation channels.
In this regime, the ergodic-capacity perks of the fast-fading channel are unavailable; the channel capacity is actually zero. Still, the communications performance can be meaningfully described with the outage capacity \cite[Sec.~5.4.1]{Tse2005}, given by
$\RateMaxGP(\SNR \cdot F_{|\fadingRVCombined|^2}^{-1}(\epsilon))$. It is the information rate that 
can be decoded with outage probability $\epsilon$. Thereby $F_{|\fadingRVCombined|^2}^{-1}$ is the inverse cumulative distribution function (CDF) of
$|\fadingRVCombined|^2$.

%% file: 99-Summary.tex
For the first time this paper stated the channel capacity of load modulation with a freely adaptable passive impedance. The obtained insights on the capacity-achieving transmit distribution and its approximation with finite symbol alphabets and simple switched load circuits have important implications for practical high-data-rate backscatter communication systems. This applies even to the ambient backscatter case, under certain identified conditions.

Future work should incorporate the presented insights in practical BC systems in order to realize near-capacity data rates. It should also investigate the use of microwave components such as transmission lines, 
waveguides, ferrite phase shifters \cite[Sec.~9.5]{Pozar2004} and metamaterial structures \cite{YangNAT2016} for capacity-approaching load modulation.

%% file: AP-Impedance.tex
We characterize the distributions of the normalized load impedance $\r z = \f{1 + \r\Gamma}{1 - \r\Gamma}$ associated with the 
various different distributions of the reflection coefficient $\r\Gamma$ described in \Cref{sec:capacity}.

\subsubsection{Capacity-Achieving Distribution, General Impedance}
\label{sec:zStatsHighSNR}

For values $\Gamma = \Radius \, e^{j\theta}$ on a circle with a fixed radius $\Radius < 1$, the impedance values
are $z
= \f{1 + \Radius\,e^{j\theta}}{1 - \Radius\,e^{j\theta}}$.
Due to the circle preservation property of the M\"obius transformation, this is another circle
$z = \f{1 + \Radius^2}{1 - \Radius^2} + \f{2 \Radius}{1 - \Radius^2} \, e^{j\beta}$
with center
$\f{1 + \Radius^2}{1 - \Radius^2} \in \bbR$
and radius
$\f{2 \Radius}{1 - \Radius^2} \in \bbR$.
The $z$-domain angle $\beta \in [0,2\pi)$ is a rather intricate function
$\beta(\theta) = 2 \arctan\left( \f{\sin(\theta)}{\cos(\theta) - \Radius} \right) - \theta$
of the $\Gamma$-domain angle $\theta$,
which behaves as follows.
For a small radius $\Radius \ll 1$, the approximate identity $\beta \approx \theta$ holds, so a random $\r\beta$ has similar distribution as $\r\theta$ (uniform).
For a large radius $\Radius = 1 - \varepsilon$, the angle $\r\beta$ is pushed towards the value $\pi$, causing a concentration of probability mass near $z = \f{1-a}{1+a} \approx 0$.
These properties can be observed in \Cref{fig:CapDistr_z} and especially in \Cref{fig:CapDistr_z24dB}.
The conditional distribution $\r\beta|\r a = a_k$ is determined by
$\r\theta \sim \calU(0,2\pi)$ and $\beta(\theta)$, a monotonously increasing bijective map from and to $[0,2\pi)$.
A change of variables yields the conditional PDF
$
f_{\r\beta|\r a}(\beta|a)
= f_{\r\theta}(\theta) \cdot |\fp{\theta}{\beta}|
= \f{1}{2\pi} / \fp{\beta}{\theta}
= \f{1 - 2 \Radius \cos(\theta) + \Radius^2}{2\pi(1 - \Radius^2)}
$
after some rearrangements. This is an implicit formulation in terms of $\theta$;
an explicit one is prohibited by the unavailability of the inverse map $\theta(\beta)$ in closed form.
The maximum radius $a = 1$, associated with $z = jx$, will be covered in \Cref{sec:zStatsReactive}.

\ifdefined\SingleColumnDraft
\renewcommand\myFigHeight{31mm}
\begin{figure}[!ht]
\else 
\renewcommand\myFigHeight{33mm}
\begin{figure}[t]
\fi
\centering
\subfloat[SNR $\SNR < 4.8\dB$]{
\includegraphics[height=\myFigHeight]{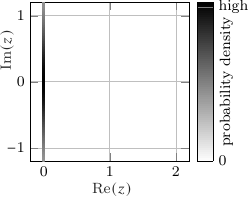}
\label{fig:CapDistr_zLowSNR}} \
\subfloat[SNR $\SNR = 12\dB$]{
\includegraphics[height=\myFigHeight]{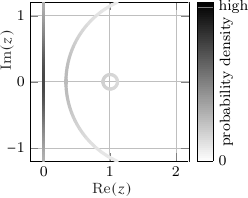}
\label{fig:CapDistr_z12dB}}
\ifdefined\SingleColumnDraft\else \\ \fi
\subfloat[SNR $\SNR = 18 \dB$]{
\includegraphics[height=\myFigHeight]{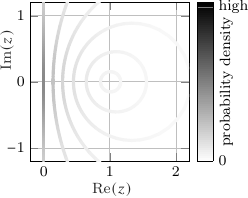}
\label{fig:CapDistr_z18dB}} \
\subfloat[SNR $\SNR = 24\dB$]{
\includegraphics[height=\myFigHeight]{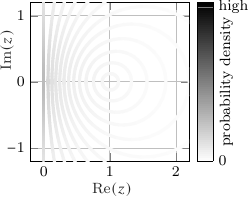}
\label{fig:CapDistr_z24dB}}
\caption{Capacity-achieving distribution of the normalized load impedance $\r z = \r Z\L / R\Tx$, $\r z \in \bbC$, for different SNR values.}
\label{fig:CapDistr_z}
\end{figure}

\subsubsection{Maximum-Entropy Transmit Signaling}
\label{sec:zStatsME}

In \Cref{sec:MaxEntropyRate} we noted that a uniform distribution over the unit disk $|\r\Gamma| \leq 1$ yields a near-capacity rate at high SNR. We shall describe the associated impedance distribution. For $\r\Gamma = \r\Radius \, e^{j\r\theta}$ we consider the joint PDF
$f_{\r a,\r\theta}(a,\theta) = f_{\r a}(a) \, f_\r\theta(\theta) = 2a \, \f{1}{2\pi}$. We write
$z = r + jx = \f{1 + \Radius e^{j\theta}}{1 - \Radius e^{j\theta}}$
in vector form
$[r,x] = [1-\Radius^2, 2\Radius\sin(\theta)] \, / \, (1 + \Radius^2 + 2\Radius\cos(\theta))$.
A two-dimensional change of variables yields the joint PDF
$f_\r{z}(z) = f_{\r{r},\r{x}}(r,x) =
\f{a}{\pi} \cdot |\det( \fp{[r,x]}{[a,\theta]} )|^{-1}$.
A detailed expansion of the $2 \times 2$ Jacobian matrix 
$\fp{[r,x]}{[a,\theta]}$
is omitted.
To evaluate the expression use $a = |\f{z-1}{z+1}|$ and $\theta = \arg(\f{z-1}{z+1})$.
An evaluation of the PDF $f_z(z)$ is given by the intensity plot in \Cref{fig:zPdfMaxEntropic}. It exhibits a concentration of probability mass near $z=0$ but also heavy tails in both resistance and reactance,  which are hard to discern here. We note that the PDF is a continuous approximation of the high-SNR capacity-achieving PDFs in \Cref{fig:CapDistr_z18dB,fig:CapDistr_z24dB}.

\ifdefined\SingleColumnDraft
\begin{figure}[t]\centering
\else 
\renewcommand\myFigHeight{33mm}
\begin{figure}[!ht]\centering
\fi
\includegraphics[width=76mm,trim=0 15 0 0]{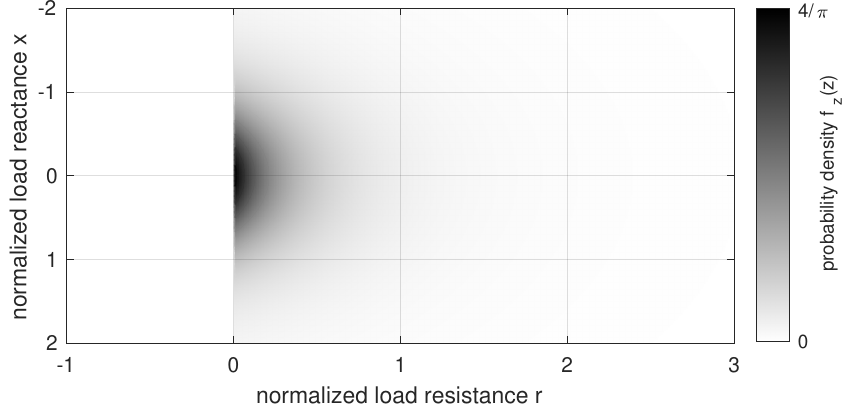}
\caption{Distribution of normalized load impedance $\r z \in \bbC$ with maximum-entropy transmit signaling (reflection coefficient $\r\Gamma$ uniform over unit disk).
}
\label{fig:zPdfMaxEntropic}
\end{figure}

\subsubsection{Purely Reactive Load Modulation}
\label{sec:zStatsReactive}

We found that $\r\Gamma = e^{j\r\theta}$ with
$\r\theta \sim \calU(0,2\pi)$
achieves channel capacity for a purely reactive load (\Cref{sec:CapacityReact}) or for the low-SNR case of a general passive load (\Cref{sec:CapacityGeneral}).
To determine the associated impedance statistics, we note that
$\r z = \f{1 + e^{j\r\theta}}{1 - e^{j\r\theta}} = j\r x$
is on the imaginary axis, with the normalized load reactance
$\r x = \cot(\r\theta / 2)$.
By a change of variables we find that $\r x$ has the PDF
$f_{\r x}(x) = \f{1}{\pi (1 + x^2)}$,
$x \in \bbR$,
which is a standard Cauchy distribution (a.k.a. Lorentz distribution).
This Cauchy PDF can be seen along the imaginary axes of all plots in \Cref{fig:CapDistr_z}.
It has significant tails, i.e. large positive and negative $x$-values are chosen with significant probability.

\subsubsection{Purely Resistive Load Modulation}
\label{sec:zStatsResistive}

Consider a real-valued
$\r\Gamma = \r\Gamma\reSub \in [-1,1]$,
associated with the resistance
$\r z  = \r r = \f{1+\r\Gamma\reSub}{1-\r\Gamma\reSub} \in \bbR_{\geq 0}$.
In \Cref{sec:CapacityResist} we noted that the capacity-achieving distribution is discrete, with mass points 
$\Gamma_m$
and resistances
$r_m = \f{1 + \Gamma_m}{1 - \Gamma_m}$.
If instead real-valued maximum-entropy signaling $\r\Gamma\reSub \sim \calU(-1,1)$ is used to approach capacity at high SNR, then $\r r$ is distributed according to the PDF
$f_{\r r}(r) = \f{1}{(1 + r)^2}$
and CDF
$F_{\r r}(r) = \f{r}{1 + r}$ with
$r \in \bbR_{\geq 0}$.
This is an instance of various established distributions:
standard Beta prime $\r r \sim \beta'(1,1)$,
Pareto type II $\r r \sim \mathrm{Lomax}(1,1)$,
and log-logistic $\r r \sim \mathrm{LL}(1,1)$.
It exhibits a heavy tail but also a tendency towards small $r$, i.e. towards resonance.

%% file: AP-RateCalc.tex
\ifdefined\SingleColumnDraft
\newcommand\mySubsecSeparator{}
\else
\newcommand\mySubsecSeparator{}
\fi
\subsubsection{Numerical Calculation: General-Case Capacity}
\label{apdx:CapOptProb}

We describe an iterative numerical procedure to solve the optimization problem \Cref{eq:Capacity}, yielding the channel capacity evolution $\RateMaxGP(\SNR)$ in \Cref{fig:Rates_Capacity} and the optimal parameter evolution in \Cref{fig:Radii_PMF_evolution}. A similar scheme was given in \cite[Sec.~III-B]{ShamaiTIT1995}.

Choose an initial SNR value $\SNR < 4.8\dB$ where the optimality of $K = 1$ is formally guaranteed. Increase $\SNR$ iteratively with a small increment (we choose $0.1\dB$). For each iteration, run the following procedure.
\begin{enumerate}[i]
\item Use the same number of circles $K$ as the previous iteration.
\item Optimize all $\Radius_2 , \ldots , \Radius_K$ and $\CircProb_1 , \ldots , \CircProb_K$ jointly according to \Cref{eq:Capacity}. For this purpose we use the interior-point algorithm for constrained nonlinear optimization \cite{InteriorPointAlgorithm} with carefully tuned stopping criteria and initialized at the optimal parameter values of the previous iteration. Remember the optimized parameter values and $I(\r y;\r\Gamma)$.
\label{AlgoStepTwo}
\item Add a new circle on trial (i.e. increment $K$ by $1$) and set its radius $\Radius_K = 0$ and its probability $\CircProb_K$ to a sensible nonzero value (we use $\CircProb_K = \f{1}{100}\,\CircProb_1$).
\item Optimize all circle parameters with the aforementioned interior-point algorithm.
\item If the addition of the trial circle increased $I(\r y;\r\Gamma)$ by at least $1$ part per million, then keep the trial circle. Else discard the trial circle and associated parameter adaptations and roll back to the parameter values from step \ref{AlgoStepTwo}.
\item Increment $\SNR$ and repeat the procedure, unless a predefined terminal SNR $\rho$ has been reached.
\end{enumerate}

It shall be noted that the numerical threshold choices have a noticeable effect on the results at high-SNR. Because there, parameter fine tuning for the innermost circles causes only tiny rate changes near the floating point accuracy.
These numerical issues have also been indicated in \cite{ShamaiTIT1995} and cause the rather shaky high-SNR behavior of the smaller radii in \Cref{fig:Radii_evolution}.

\mySubsecSeparator

\subsubsection{Rate Calculation for DAUIP Signaling}
\label{apdx:MutInfCalc}

We derive the evaluable expression \Cref{eq:MutualInformationUP} for the mutual  $I(\r y;\r\Gamma)$ with DAUIP $\r\Gamma$. We do so in a fashion that should be accessible to anyone with basic information theory knowledge. In the process, we prepare important statements for subsequent derivations.

The additive noise channel \Cref{eq:SignalModelRV} is continuous-valued and memoryless. Thus, the mutual information over the channel is a difference of differential entropies,
$I(\r{y};\r\Gamma) = h(\r y) - h(\r y | \r \Gamma) = h(\r y) - h(\r w)$.
The Gaussian noise entropy is given by
$h(\r w) = \log_2(\f{\pi e}{\SNR})$
\cite[Eq.~(B.40)]{Tse2005}. Hence,
\begin{align}
I(\r{y};\r\Gamma)
= h(\r{y}) + \log_2\left(\f{\SNR}{\pi e}\right)
\label{eq:MutualInformation} \, .
\end{align}
It remains to compute the differential entropy $h(\r{y})$ of the received signal. By definition it is given by \cite[Cpt.~8]{Cover2006}
\begin{align}
h(\r y) 
= -\!\int_\bbC f_\r{y}(y) \, \log_2\big( f_\r{y}(y) \big) \,dy
\label{eq:SignalEntropyGeneral}
\end{align}
which is shorthand notation for a double integral of $\Re(y)$ and $\Im(y)$ over $\bbR^2$.
A more specific formula can be given if $\r y$ has uniform independent phase (UIP).
We note that $\r y = \r\Gamma + \r w$ inherits the UIP property from $\r\Gamma$ due to the circularly-symmetric $\r w$.
This is formalized as
$\r\theta , \r\phi \sim \calU(0,2\pi)$
for the polar angles in \Cref{eq:PolarDef}.
Therewith, one can easily derive\footnotemark{} the UIP-specific formula \cite[Eq.~(13)]{ShamaiTIT1995}
\begin{align}
h(\r y) &= \log_2( 2\pi ) -\!\int_0^\infty \!
f_\r{b}(b) \log_2\!\Big( \f{f_\r{b}(b)}{b} \Big) \, db
\, . \label{eq:SignalEntropyUP}
\end{align}
\footnotetext{%
To derive \Cref{eq:SignalEntropyUP}, write \Cref{eq:SignalEntropyGeneral} as double integral of $b,\phi$ with Jacobian determinant $b$, note that
$b \cdot f_\r{y} = f_{\r{b},\r\phi} = \f{1}{2\pi} f_\r{b}$
for UIP $\Rightarrow\ f_\r{y} = \f{1}{2\pi} \f{1}{b} f_\r{b}$. Note that the integrand is constant w.r.t. $\phi$ and compute the trivial integral.%
}%
Used in \Cref{eq:MutualInformation} this directly yields the UIP-case mutual information \Cref{eq:MutualInformationUP}.
Evaluation requires the PDF of the noisy radius $\r b = |\r a e^{j\r\theta} + \r w|$. It has the same statistics as $|\r a + \r w|$: a Rice  distribution
$\r b|\r a \sim \mathrm{Rice}(a,1/\sqrt{2\SNR})$.
Marginalization
$f_\r{b}(b) = \int_0^1 f_\r{a}(a) f_{\r b|\r a}(b|a) da$
yields the receive amplitude PDF\footnotemark{}
\begin{align}
&f_\r{b}(b) = 2\SNR b
\int_0^1 \! f_\r{a}(a)\,e^{-\SNR(b - a)^2} g( 2\SNR ab ) \, da \, .
\label{eq:NoisyRadiusPDF} 
\end{align}
\footnotetext{Expression \Cref{eq:NoisyRadiusPDF} is equivalent to \cite[Eq.~(11)]{ShamaiTIT1995}. A conversion from our formalism to that of Shamai and Bar-David \cite{ShamaiTIT1995} is achieved as follows. Set the peak-power to $\rho_\mathrm{p} := 2\SNR$ and multiply \Cref{eq:SignalModelRV} with $\sqrt{\rho_\mathrm{p}}$ to obtain a signal model
$\r{\tilde y} = \r x + \r\omega$
with
$\r{\tilde y} = \r y \sqrt{\rho_\mathrm{p}}$,
$\r x = \r\Gamma \sqrt{\rho_\mathrm{p}}$, and
$\r\omega \sim \mathcal{CN}(0,2)$.
The peak-power constraint is
$|\r x|^2 = |\r\Gamma|^2 \rho_\mathrm{p} \leq \rho_\mathrm{p}$.
The average-power constraint
$\EV{|\r x|^2} \leq \rho_\mathrm{a}$
in \cite[Eq.~(3)]{ShamaiTIT1995} is not relevant to the backscatter problem; it is deactivated by setting
$\rho_\mathrm{a} = \rho_\mathrm{p}$.
In $\r x = \r r e^{\r\theta}$, $\r{\tilde y} = \r R e^{\r\phi}$ the polar radii fulfill
$\r a = \r r / \sqrt{2\SNR}$, $\r b = \r R / \sqrt{2\SNR}$
and
$f_\r{a}(a) = f_\r{r}(r) \sqrt{2\SNR}$,
$f_\r{b}(b) = f_\r{R}(R) \sqrt{2\SNR}$.
With these substitutions, the equations \eqref{eq:MutualInformation},\eqref{eq:SignalEntropyUP},\eqref{eq:NoisyRadiusPDF} become \cite[Eq.~(4),(13),(11)]{ShamaiTIT1995}. Finally, $\mathrm{bits}$ are converted to $\mathrm{nats}$ by replacing each $\log_2$ with $\mathrm{ln}$.}
These formulas allow to calculate $I(\r{y};\r\Gamma)$ for UIP $\r\Gamma$ given $f_\r{a}(a)$ and $\SNR$: use \Cref{eq:NoisyRadiusPDF} in \Cref{eq:MutualInformationUP} and do numerical integration. We note that a finite integration interval $b \in [0,1+5/\sqrt{\SNR}]$ suffices for accurate results in \Cref{eq:MutualInformationUP}.

Consider now that $\r\Gamma$ has DAUIP, i.e. discrete amplitude $\r a$ and UIP $\r\theta$. This comprises the capacity-achieving distribution. The radius $\r a$ assumes a discrete distribution with a finite number of mass points; the PDF is of the form
$f_\r{a}(a) = \sum_{k=1}^K \CircProb_k\,\delta(\Radius - \Radius_k)$ where $\delta(.)$ is the Dirac delta distribution.
Consequently, the integral \Cref{eq:NoisyRadiusPDF} simplifies to the Rician mixture sum \Cref{eq:NoisyRadiusPDFCircleSum}. The subsequent computation of the mutual information \Cref{eq:MutualInformationUP} still requires numerical integration.

\mySubsecSeparator

\subsubsection{Derivation of Lower Bound $\log_2(1 + \SNR/e)$}
\label{apdx:LowerBoundSNReDeriv}

In \Cref{sec:MaxEntropyRate} we argued that uniform signaling allows for near-capacity rates at high SNR. We shall analyze this in more detail.
For the complex-valued channel
$\r y = \r\Gamma + \r w$,
the two-dimensional entropy power inequality
$2^{h(\r y)} 
\geq 2^{h(\r\Gamma)} + 2^{h(\r w)}$ holds \cite{BlachmanTIT1965},\cite[Eq.~(36)]{ShamaiTIT1995}.
Application to $h(\r y)$ in \Cref{eq:MutualInformation} and rearrangements yield
\begin{align}
\log_2\left(1 + \f{2^{h(\r\Gamma)}}{\pi} \cdot \f{\SNR}{e} \right) \leq I(\r y;\r\Gamma)
\label{eq:GeneralResultFromComplexEPI}
\end{align}
for the AWGN channel, for any distribution of $\r\Gamma$.
On the other hand, in \Cref{apdx:MaxEntropy} we show that $h(\r\Gamma) \leq \log_2(\pi)$ holds and is achieved with equality through UD signaling. \QEDB

\mySubsecSeparator

\subsubsection{Rate Calculation for UD Signaling}
\label{apdx:UDRateCalc}

We consider a UD transmit signal $\r\Gamma$ and derive the mutual information over a complex AWGN channel with SNR $\SNR$. Here $\r\Gamma = \r a e^{j\r\theta}$ has UIP (but not DAUIP) and a linear radius PDF $f_a(a) = 2a$ for $a \in [0,1]$. The resulting PDF of the receive-signal radius $\r b$ is described by the integral \Cref{eq:NoisyRadiusPDF}. From a special case of \cite[Eq.~(33)]{ShamaiTIT1995} and a change of variables, we obtain the solution
$f_\r{b}(b) = 2b (
1 - Q_1(b\sqrt{2\SNR},\sqrt{2\SNR}) )$
where $Q_1(\cdot,\cdot)$ is the Marcum Q-function of order $1$ \cite{MarcumQ}.
The information rate
$\RateME(\SNR) := h(\r{y}) + \log_2(\f{\SNR}{\pi e})$
follows via \Cref{eq:MutualInformationUP} and numerical integration.

The $3\dB$ loss in the power-limited regime as compared to channel capacity is due to the mean squared amplitude
$\EV{\r a^2}
= \int_0^1 a^2 f_{\r a}(a)\,da
= \int_0^1 2 a^3\,da
= \f{1}{2}$.

\mySubsecSeparator

\subsubsection{Derivation, $\infty$-PSK Asymptote}
\label{apdx:ReactAsymptote}

With constant amplitude and UIP, the receive-amplitude has Rice distribution $\r b \sim \mathrm{Rice}(1,1/\sqrt{2\SNR})$. At high $\SNR$, this is closely resembled by a Gaussian $\r b \sim \calN(1,1/(2\SNR))$.
We write \Cref{eq:MutualInformationUP} as
$I(\r{y};\r\Gamma) = \log_2( 2\SNR / e ) + h(\r b) + \EV{\log_2(\r b)}$
and note that
$h(\r b) \approx \f{1}{2}\log_2(\f{\pi e}{\rho})$
and
$\EV{\log_2(\r b)} \approx 0$.
This yields the very accurate approximation
$I(\r{y};\r\Gamma) \approx \f{1}{2} \log_2( \f{4\pi\SNR}{e} )$,
also found in \cite{WynerBSTJ1966},\cite[Table~II]{Blachman1953},\cite[Eq.~(42)]{ShamaiTIT1995}.

\mySubsecSeparator

\subsubsection{Rate Calculation, Finite Constellations}
\label{apdx:RateFinte}

For symbols $\Gamma_m$ from a finite constellation, chosen with odds $\SymbProb_m$,
the receive-signal PDF is the Gaussian mixture
$f_\r{y}(y) = \f{\SNR}{\pi} \sum_{m=1}^M
\SymbProb_m \exp\left( -\SNR\,|y-\Gamma_m|^2 \right)$
for SNR $\SNR$.
Therewith, calculate $I(\r{y};\r\Gamma)$ with \Cref{eq:MutualInformation,eq:SignalEntropyGeneral} and numerical integration.

\mySubsecSeparator

\subsubsection{Capacity Calculation, Purely Resistive Load Modulation}
\label{apdx:CapCalcRes}

The number of mass points, their positions and probabilities must be found with optimization for the target SNR, analogous to the circle parameters in \Cref{eq:Capacity}. The UIP-specific \Cref{eq:Capacity} and \Cref{eq:SignalEntropyUP}-\Cref{eq:NoisyRadiusPDFCircleSum} however do not apply here. 

The mutual information is calculated by numerical integration in
$I(\r{y}\reSub;\r\Gamma\reSub) \! = \! \f{1}{2} \log_2(\f{\SNR}{\pi e}) - \!\int_{-\infty}^\infty f_{\r{y}\reSub}(y) \log_2( f_{\r{y}\reSub}(y) ) dy$
with the PDF
$f_{\r{y}\reSub}(y) = \sqrt{\SNR / \pi} \sum_{m=1}^M \SymbProb_m \exp( -\SNR(y-\Gamma_m)^2  )$.
The capacity $\RateResist$ is obtained by numerical maximization of
$I(\r{y}\reSub;\r\Gamma\reSub)$
with respect to $M \in \bbN$ and all free parameters $\Gamma_m \in [-1,1]$ and $\SymbProb_m  \in [0,1]$ subject to $\sum_{m=1}^M \SymbProb_m = 1$.
Many parameters are fixed because: (i) the outmost points $\Gamma_m = \pm 1$ always occur and (ii) for any mass point $\Gamma_m > 0$, $-\Gamma_m$ is another mass point and has equal probability.

\mySubsecSeparator

\subsubsection{Lower Bounds, Purely Resistive Load Modulation}
\label{apdx:ResBounds}


Another lower bound is found from the rate of the suboptimal transmit signaling $\r\Gamma\reSub \sim \mathcal{U}(-1,1)$.
The real-valued AWGN channel
$\r y\reSub = \r\Gamma\reSub + \r w\reSub$ fulfills the entropy power inequality
$2^{2\,h(\r y\reSub)} 
\geq 2^{2\,h(\r\Gamma\reSub)} + 2^{2\,h(\r w\reSub)}$ \cite{BlachmanTIT1965}.
Now $h(\r w\reSub) = \f{1}{2}\log_2(\f{\pi e}{\SNR})$ and rearrangements give
$\f{1}{2}\log_2(1 + \f{\SNR}{\pi e} 2^{2h(\r\Gamma\reSub)}) \leq I(\r y\reSub;\r\Gamma\reSub)$.
If $\r\Gamma\reSub \in [-1,1]$, then $h(\r\Gamma\reSub) \leq \log_2(2) = 1$ holds and is achieved with equality by $\r\Gamma\reSub \sim \calU(-1,1)$. 
This yields the lower bound
$\f{1}{2} \log_2( 1 + \f{4\,\SNR}{\pi e} ) < I(\r y\reSub;\r\Gamma\reSub)$.

%% file: AP-SubareaEntropy.tex
Consider a continuous complex-valued random variable $\r x$ whose realizations are constrained to a compact set $\mathcal{A} \subset \bbC$ with non-empty interior, i.e.
$0 < \mathrm{area}(\mathcal{A}) < \infty$ with
$\mathrm{area}(\mathcal{A}) := \int_\mathcal{A} dx$.
Then the differential entropy is upper-bounded by $h(\r x) \leq \log_2(\mathrm{area}(\mathcal{A}))$. Equality holds for a uniform distribution $\r x \sim \calU(\mathcal{A})$.

\subsubsection*{Proof Sketch} 
The optimality of $\r x \sim \calU(\mathcal{A})$ follows from an $\bbR^2$ description $\Re(\r x), \Im(\r x)$ and \cite[Thm.~12.1.1]{Cover2006} which dictates that,
in the absence of further constraints, the entropy-maximizing PDF must be constant inside the support set, i.e. $f_\r{x}(x) = 1/\mathrm{area}(\mathcal{A})$.
Then 
$h(\r x)
:= -\int_\mathcal{A} f_\r{x}(x) \, \log_2( f_\r{x}(x) ) \,dx$
$= -\log_2(1/\mathrm{area}(\mathcal{A})) \int_\mathcal{A} f_\r{x}(x) dx$
$= \log_2(\mathrm{area}(\mathcal{A}))$.